\documentclass[reprint,aps,prd,showpacs,footinbib,amsmath,amssymb,amsfonts,superscriptaddress]{revtex4-2}

\usepackage{bm}
\usepackage{graphicx}
\usepackage{tikz}
\usepackage{color}
\usepackage{dsfont}
\usepackage[normalem]{ulem}
\usepackage{hyperref}
\usepackage{mathrsfs}
\usepackage{subfig}
\usepackage{float}
\usepackage{varioref}
\usepackage[active]{srcltx}
\usepackage{lmodern}
\usetikzlibrary{decorations.pathmorphing}
\tikzset{snake it/.style={decorate, decoration=snake}}
\expandafter\ifx\csname package@font\endcsname\relax\else
\expandafter\expandafter
\expandafter\usepackage
\expandafter\expandafter
\expandafter{\csname package@font\endcsname}%
\fi
\hypersetup{
colorlinks=true,       % false: boxed links; true: colored links
linkcolor=red,          % color of internal links
citecolor=blue,        % color of links to bibliography
}

\usepackage[section]{placeins}

\usepackage{fancybox}
\labelformat{equation}{(#1)} 
\labelformat{figure}{Fig.~#1} 
\labelformat{subfigure}{Fig.~\thefigure#1} 
\labelformat{table}{Tab.~#1} 
\newcommand{\be}{\begin{equation}}
\newcommand{\ee}{\end{equation}}
\newcommand{\bea}{\begin{eqnarray}}
\newcommand{\eea}{\end{eqnarray}}

\usepackage{subfig}
\newcommand{\ket}[1]{|#1\rangle}
\newcommand{\bra}[1]{\langle#1|}

\begin{document}

\title{Quenched Kitaev Chain : Analogous model of Gravitational Collapse}

\author{Sandra Byju} \email{byju.s@husky.neu.edu}
%\affiliation{ School of Physics, Indian Institute of Science Education and Research, Vithura, Trivandrum 695551, India}
\affiliation{Department of Physics, Northeastern University, Boston, Massachusetts 02115, USA}
\author{Kinjalk Lochan}%
\email{kinjalk@iisermohali.ac.in}
%\affiliation{School of Physics, Indian Institute of Science Education and Research, Thiruvananthapuram (IISER-TVM), Trivandrum 695016, India}
\affiliation{Department of Physical Sciences, IISER Mohali, Manauli 140306, India}
\author{S. Shankaranarayanan}%
\email{shanki@phy.iitb.ac.in}
%\affiliation{School of Physics, Indian Institute of Science Education and Research, Thiruvananthapuram (IISER-TVM), Trivandrum 695016, India}
\affiliation{Department of Physics, Indian Institute of Technology Bombay, Mumbai 400076, India}

\begin{abstract}
	We investigate generalized thermalization in an isolated free Fermionic chain evolving from an out of equilibrium initial state through a sudden quench. We consider the quench where a Fermionic chain is broken into two disjoint chains. We focus on the evolution of the local observables namely, occupation number, nearest neighbor hopping, information sharing and out-of-time-order correlations after the quench and study the relaxation of the observable, leading to generalized Gibbs ensemble for the system in the thermodynamic limit  though it has been argued that non-interacting or free Fermionic models in general do not relax to GGE. We obtain the light cone formed by the evolution of the observables along the Fermionic lattice chain due to the sudden quench which abides by the Lieb-Robinson bound in quantum systems. We also analytically study a simpler model which captures the essential features of the system. Our analysis strongly suggest that the internal interactions within the system do not remain of much importance once the quench is sufficiently strong.	
\end{abstract}

\maketitle 
% Keywords
%\keyword{Generalized thermalization; quantum quench; Fermionic chain}

\section{Introduction}
Numerous fascinating processes proposed within the context of relativistic quantum field theory in curved space-time --- such as the particle creation in the early universe~\cite{1969-Parker-PR}, Hawking radiation~\cite{1975-Hawking-CMP}, and Unruh effect~\cite{1976-Unruh-PRD} --- are currently inaccessible to direct observation. Since these processes occur at extreme conditions, creating such conditions in a laboratory environment is a formidable challenge. However, analog models can be established as an alternative method for gaining insight into this exotic process~\cite{1981-Unruh-PRL}. Since Unruh's seminal work, a multitude of analogues were studied, including flowing water, Bose-Einstein condensates (BEC), dilute gases, fiber optics, and nonlinear dielectrics~(for a review, see~\cite{2011-Barcelo.etal-LRR}). In addition, a number of experimental tests have been conducted, despite the lack of conclusive confirmation of the phenomenon~(for a review, see~\cite{2019-Belgiorno.etal-Book}).

One fundamental difference between these exotic phenomena and the experimental set-up is the system size. In the case of Hawking radiation or particle creation in the early Universe, the field theoretic assumption is valid, however, in the low-temperature \emph{quantum gas} experiments this is not the case. To understand how far these analogue models mimic the exotic phenomena, one need to understand some of the fundamental questions in quantum many-body systems: If the many-body 
system starts out in a ground state at early times what is the fate of the system at late times? Does the system evolve into a steady state at late times ? If so does the state resemble a thermal state ? What is the characteristic time scale for this to happen? An important setting where these problems come to forefront is when matter collapse leads to a black hole formation. Not only the thermality of exterior modes post the horizon formation, is an open problem but there also have been arguments suggesting a possible setting up of chaos as a result of thermality and protection of monogamy of entanglement \cite{Hawking:2014tga, Magan:2018nmu, Roberts:2018mnp}. In this work we model the black hole formation by an analogue coupled atomic chain system undergoing quench action in terms of getting disjoint at a particular time. In terms of gravitational collapse this resembles formation of a black hole event horizon where certain modes get causally disconnected \footnote{Note that the ingoing and out moving modes {\it do not} get causally disconnected, but {\it left} moving and {\it right} moving modes do !}. Such atomic  Kitaev chain models have extensively been studied for undergoing quench action and post quench relaxation, see \cite{Magan:2016ehs} and references therein.

In recent years the relaxation of such atomic chains has received attention as these can now be experimentally measured in cold atom experiments~\cite{2007-Lewenstein.etal-AP}.
Large enough systems left on their own, over large times seem to settle into a thermodynamic configuration. However, the ideas of thermalization and unitary evolution of the unnderlyinng quantum theory do not go hand in hand ~\cite{maldacena2016bound}. A quantum evolution keeps a pure state as pure throughout, whereas thermalization demands a {\it mixed state description}. The initial contributions to resolve this apparent conflict in this area came in the 1920's from von Neumann about thermalization in isolated many-body quantum systems, proposing that demand on thermalization on large systems may be relaxed to the demand that only the expectation values of macroscopic observables need to thermalize ~\cite{neumann1929beweis,von1956probabilistic}. Thus, for large enough systems the late time expectations should closely resemble those of a thermalized system, and that is about it! The system then {\it thermalizes without really thermalizing} \footnote{The actual information that the system has not thermalized will be reflected in correlators and not in expectations.}.

However the problem of explicit verification of this idea remained dormant for nearly eight decades because of the analytical complexity; as the thermalization is supposed to work for {\it large systems} and the Hilbert space dimension increases exponentially as the number of degrees of freedom increases, making the analytic handling almost intractable.

When a system thermalizes, we expect the macroscopic properties of the system to equilibrate to its corresponding statistical ensemble predictions. Classically a system is called integrable if it has $N$ independent constants of motion in a $2N$ dimensional phase space; by doing a canonical transformation to its corresponding action-angle coordinates, the action is conserved, and the angle evolves linearly in time. Hence for an integrable system, its dynamics can be predicted at all times, and it will never be ergodic. However, dynamics for generic non-integrable systems sufficiently away from any integrable limit is governed by non-linearity and chaos, making the evolution ergodic thereby, resulting in thermalization ~\cite{gogolin2016equilibration,polkovnikov2011colloquium,d2016quantum}. See references for some interesting exceptions to this dictum ~\cite{kolmogorov1954conservation,moser1962invariant,jain2017nodal}.

The isolated quantum many-body systems have a different mechanism for thermalization owing to its unitary time evolution. A pure initial state would never evolve into a mixed thermal state density matrix through unitary evolution in an isolated system. However as pointed out by von Neumann, one needs to compare the expectation value of macroscopic observables in the thermodynamic limit, and not the density matrix themselves ~\cite{neumann1929beweis}. Isolated quantum systems with short-range interactions are said to thermalize if, after a long time, the expectation values of few-body observables equilibrate to a steady state predicted by statistical mechanics ~\cite{d2016quantum}. Srednicki and Deutsch proposed a mechanism suggesting that the thermalization occurs at the level of eigenstates ~\cite{srednicki1994chaos,deutsch1991quantum}. This mechanism is referred to as the Eigenstate Thermalization Hypothesis. Eigenstate thermalization holds for non-integrable systems where the expectation value of observables settles down to the value given by the ensemble description. 

In order to look at how unitary time evolution generated by an arbitrary Hamiltonian $\hat{H}$ in an isolated quantum system could lead to thermalization, let us consider the dynamics of an initial state $\rho_{I}=\ket{\psi(0)}\bra{\psi(0)}$ (where $[\hat{\rho}_{I}, \hat{H}] \neq 0$) and compare the expectation value of the macroscopic observables with the value given by statistical ensemble. For now, let the Hamiltonian be the only physically relevant conserved quantity for the system. 

Let $E_{n}$ and $\ket{n}$ be the eigenvalues and eigenvectors of $\hat{H}$. The energy of the system is a conserved quantity and is set by the initial state given by $\bar{E}=\text{Tr}(\rho_{I}H)$. The fluctuations in energy is given by $\delta E=\sqrt{\text{Tr}(\rho_{I}H^{2})-\bar{E}^{2}}$. We choose $\rho_{I}$ such that $\delta E$ is subextensive. The time evolution of the state $\ket{\psi(t)}$ is given by 
\begin{equation}
\ket{\psi(t)}= \exp{(-i\hat{H}t)}\ket{\psi(0)}=\sum_{n=1}^{D}C_{n} e^{(-iE_{n}t)}\ket{n}
\end{equation}
where $C_{n}=\langle{n}|\psi(0)\rangle$ and D is the dimension of the Hilbert space. (Throughout this work, we set Planck constant $\hbar=1$ and Boltzmann constant $\text{k}_{B}=1$.)

The expectation value of an observable is given by
\begin{eqnarray}
\langle \hat{O}(t) \rangle =\bra{\psi(t)}\hat{O}\ket{\psi(t)}=\sum_{n}|C_{n}|^{2}\bra{n}\hat{O}\ket{n}\nonumber\\
+\sum_{n,m,n\neq m}C_{m}^{*} C_{n}e^{(-i(E_{n}-E_{m})t)}\bra{m}\hat{O}\ket{n}
\end{eqnarray}
For non-integrable systems, after a reasonably long time, it is phenomenologically observed that $\langle\hat{O}(t)\rangle$ equilibrates to a steady state given by the Gibbs (microcanonical) ensemble i.e.,
\begin{equation}
\lim_{t \to \infty, L \to \infty}\langle\hat{O}(t)\rangle \approx O(\bar{E})
\end{equation}
where ${O}(\bar{E})= \frac{1}{\Omega}\sum_{n}O_{nn}$ where $\Omega$ is the number of energy eigenstates with energies within the window $[\bar{E}-\Delta E,\bar{E}+\Delta E]$ with $\Delta E<< \bar{E},$ ~\cite{rigol2008thermalization,rigol2009quantum,d2016quantum}. It is important to note that this is not a mathematically proven result, but is motivated from the ideas of random matrix theory and quantum chaos,~\cite{jain1997quantum,alonso1996random,magan2016random} and is referred to as 
Eigenstate Thermalization Hypothesis(ETH)~\cite{srednicki1994chaos,deutsch1991quantum}.  However, it has been studied and verified in a variety of non-integrable systems ~\cite{d2016quantum,rigol2008thermalization,rigol2012alternatives}. 

In the case of integrable systems --- which is the focus of this work --- the expectation of observables do not thermalize to Gibbs ensemble. This is because such systems have other 
conserved quantities (totaling $N$) and can relax only to a steady state predicted by the generalized Gibbs ensemble (GGE)~\cite{rigol2007relaxation,cazalilla2006effect,calabrese2011quantum,ilievski2015complete,langen2015experimental,sirker2014locality,gebhard2012particle}. The notion of generalized thermalization was obtained in integrable systems by generalizing the statistical ensemble description for integrable systems by including all the integrals of motion. We can arrive at GGE by maximizing entropy subject to constraints imposed by the integrals of motion,i.e.,
\begin{equation}
\hat{\rho}_{GGE}=\frac{e^{(-\sum_{m}\lambda_{m}\hat{I}_{m})}}{\text{Tr}[e^{(-\sum_{m}\lambda_{m}\hat{I}_{m})}]},
\end{equation}
where $\hat{I}_{m}$ is the set of all integrals of motion and $\lambda_{m}$ are the Lagrange multipliers which are fixed using initial condition $\text{Tr}[\hat{\rho}_{GGE}\hat{I}_{m}]=\langle \hat{I}_{m} \rangle(t=0)$.
Using which we arrive at the state to which the diagonal ensemble settle down. Just like ETH, there is no general proof for generalized thermalization~\cite{vidmar_rigol_16,cassidy2011generalized}.

Though a complete analytical understanding of the generalized thermalization is still lacking, recent advances in ultracold atom experiments and computational techniques made it possible to simulate dynamics of nearly isolated quantum systems and study out of equilibrium dynamics~\cite{trotzky2012probing,kinoshita2006quantum,bloch2005ultracold,giamarchi2016strongly,langen2015ultracold,greiner2002collapse,will2010time,will2015observation,gring2012m,kaufman2016quantum,tang2018thermalization,langen2015experimental,onofrio2017physics}.
Also, there have been various studies in the literature for Fermionic and Bosonic integrable systems to understand thermalization for integrable systems ~\cite{vidmar_rigol_16,essler2016quench,cazalilla2016ma,caux2016quench,ilievski2016quasilocal}. Most of the Fermionic system's studies have focused on the 1-dimensional Fermionic chain that makes a transition from a non-integrable to an integrable configuration. 

The primary goal of this work is to study the equilibration of observables in isolated integrable 1-D free non-number conserving Fermionic lattice (Kitaev) chain~\cite{2001-Kitaev-PhysUsp,2012-Alicea-RPP,2021-Borla.etal-SciPost} and compare it with the GGE prediction. We study the quench ~\cite{greiner2002collapse,will2010time,will2015observation,trotzky2012probing,porta2018nonmonotonic,porta2016out} where a Fermionic lattice chain is broken into two smaller disjoint chains which result in moving the initial system out of equilibrium. Thus a Fermionic system jumps from one integrable set up to another integrable setting.  We verify whether the systems lands into a GGE owing to this quench. We calculate expectation values for observables which play the role of conserved charges. To visualize the dynamical evolution of the system into a GGE description, we calculate the information content in bits per Fermion before and after the quench~\cite{magan2016random} and out of time ordered correlators (OTOC)~\cite{maldacena2016bound}. We show that information content per Fermion provides crucial information about thermalization of the isolated system under quench.

It is essential to compare and contrast the current work with the earlier works:   
To visualize the thermalization in integrable systems, one of the most analyzed models is the 1-dimensional Fermionic chain that makes a transition from non-integrable to integrable configuration~\cite{Luttinger} (see also, \cite{Cstro, Bertini,caux2013time,Luca2015,viti2016,Rentrop,Guan, Essler, Vidmar, Pozsgay, Fagotti, Fagotti2, Pozsgay2,Aschbacher2003, Aschbacher2006, Luca2015, lai2015entanglement, riddell2018generalized}). In our case, the Fermionic system jumps from one integrable set up to another integrable setting.

As mentioned earlier, the current framework mirrors the gravitational setting, where the thermalization is related to the formation of black-hole. Once a black hole forms, the Hilbert space of the initial data gets bifurcated into Hilbert spaces of interior and exterior, where the exterior appears to be put (at late times) in the thermal environment. Further, as shown in \ref{fig:penroseBF}, the exterior region of the hole is made up of two sets of modes that disjoint after the formation of the so-called bifurcation horizon, mimicking the system, which undergoes quench at a particular time. Thus, the current model helps us get insights into the settings where one part of the system is dynamically decoupled from another part in a thermal backdrop. In Ref.~\cite{Sushruth2018}, two of the present authors studied a Bosonic system jumping from an integrable to another integrable setting. However, the effect of quench in that system was the joining of two disjoint chains, the inverse of the present system of study. It was shown that the system tends towards the GGE at large times as the system size increases. In the Bosonic case, the physical quantities can be computed only up to the leading order of the product of the creation/annihilation operator. Post quench state can be expressed in terms of excitations of various normal modes. This work analyzes the reverse scenario (disjointing of a chain) for fermionic degrees of freedom. In a model set-up for fermionic degrees of freedom, the fermionic nature of the system controls the dimension of the Hilbert space involved.   In Fermionic systems, due to the anticommutator structure, the expansion of the ground state in terms of post-quench operators truncates at the leading order of the product of creation/annihilation operators. Therefore, in the present case, we can compute the quantities precisely for all orders. 

The rest of the paper is organized as follows: In section \ref{Model}, we introduce the model Hamiltonian, the quench protocol and the observables of interest. In section \ref{Results}, we compare the long time evolution of the observable quantities against the corresponding GGE value by varying the system size and the time of evolution. Through various estimators, we demonstrate that the system quickly settles into a GGE configuration with increasing size. In section \ref{analytic}, we provide the analytical calculations for the observable quantities of interest in a similar yet simpler model to attain a better understanding of the general characters observed. Finally, section \ref{Conclusion} sums up the findings and discuss the implications for the field theoretic set up.

%%%%%%%%%%%%%%%%%%%%%%%%%%%%%%%%%%%%%%%%%%
\section{Model and setup} \label{Model}

\begin{figure}[h]
    \begin{center}
        \begin{tikzpicture}[scale=1.25]
     
            \draw[blue] (-1.5,-1.5) -- (0,0) -- (1.5,1.5);
        
            \draw[orange, snake it] (-1.5,1.5) -- (1.5,1.5);
              \draw[black] (0,-3) -- (3,0) -- (1.5,1.5);
            \draw[black] (0,-3) -- (-3,0) -- (-1.5,1.5);
            \draw[blue] (1.5,-1.5) -- (0,0) -- (-1.5,1.5); 
            % \draw[blue] (1.60,-1.40) -- (-1.3,1.50);
             
           % \draw[magenta, bend left=15] (0,-3) to (-1.5,1.5);
            \draw[magenta,line width=0.25mm,bend right=10] (0,-3) to (1.5,1.5);
             \draw[magenta,line width=0.25mm,bend right=25] (0,-3) to (1.5,1.5);
            \draw[magenta,line width=0.25mm,bend left=15] (0,-3) to (1.5,1.5);
            
             \draw[cyan,line width=0.25mm,bend right=10] (0,-3) to (-1.5,1.5);
             \draw[cyan,line width=0.25mm,bend right=25] (0,-3) to (-1.5,1.5);
            \draw[cyan,line width=0.25mm,bend left=15] (0,-3) to (-1.5,1.5);

             \draw[green,line width=0.25mm,bend right=40] (-3,0) to (3,0);
             
             \draw[green,line width=0.25mm,bend left=10] (-3,0) to (0,0);
              \draw[green,line width=0.25mm,bend right=10] (0,0) to (3,0);
            
           % \node[label=right:$i^+$] (1) at (1.2,1.8) {};
            %\node[label=right:$i^-$] (2) at (0,-3) {};
            %\node[label=above:$i^0$] (3) at (3,0) {};

            \node[label=right:$\mathcal{J}_R^+$] (4) at (2.3,0.8) {};
            \node[label=right:$\mathcal{J}_R^-$] (5) at (1.5,-2) {};
       %     \node[label=below: $x_i^+$](6) at (1.5,-1.4) {};
        %   \node[label=below: $x^-$](6) at (-0.75,-2.2) {};
            \node[label=left:$\mathcal{J}_L^+$] (7) at  (-2.3,0.8) {};
            \node[label=left:$\mathcal{J}_L^-$] (8) at (-1.5,-2) {};
        %     \node[label=below: $x_f^+$](9) at (2.1,-0.9) {};
       %     \node[label=below: $x^+$](10) at (0.75,-2.2) {};
              \node[label=above: $i_L^{+}$] at (-1.5,1.5) {};
            \node[label=above: $i_R^{+}$] at (1.5,1.5) {};
             \node[label=below: $i^{-}$] at (0,-3) {};
     %        \node[label=below: $\mathcal{H}_{R}$] at (-0.5,-0.5) {};    
             \node[label=right:$t_{\text{cross}}$] (9) at (-1.4,0) {};
 	\node[label=right:$t_{\text{initial}}$] (10) at (-1,-1.2) {};

 %           \draw[->,bend right=35] (4) to (2.3,0.8);
  %          \draw[->,bend left=35] (5) to (2.5,-1.0);
   %         \draw[->,bend left=35] (7) to (-2.3,0.8);
     %       \draw[->,bend right=35] (8) to (-2.5,-0.8);

        \end{tikzpicture}
    \end{center}
    \caption{ Black hole formation in isotropic co-ordinates}
    \label{fig:penroseBF}
\end{figure}  
In this work, we wish to study the effect of horizon formation in a gravitational collapse culminating into a black hole on the matter escaping the black hole. Standard consideration explores the interior and the exterior region of the black hole and the entanglement between them. However, there exists another class of disjoint Hilbert space between outgoing left-moving modes and outgoing right-moving modes which become decoupled once they move sufficiently apart, i.e., $ t> t_{\text{cross}}$, a configuration best viewed in the isotropic coordinates, as shown in \ref{fig:penroseBF}.

 For matter moving in a black hole background, the particles entering the horizon get causally disconnected (in one direction) from the particles in the exterior. However, they do not become decoupled systems as such. However, the left moving and right moving modes (particles) which do not cross the horizon become decoupled as there is no causal communication (to and fro) between them possible after a time ($t_{\text{cross}}$). Thus, the data on the initial constant time slice $t_{\text{initial}}$, after $ t> t_{\text{cross}}$, become two disjoint systems. In \ref{fig:penroseBF}, the right-moving modes (magenta curves) and the left-moving modes (cyan curves) depict the decoupled modes in black hole spacetime.
 
Such a system where an initial set of particles breaks into two disjoint sets can be considered in the purview of the Kitaev chain~\cite{2001-Kitaev-PhysUsp,2012-Alicea-RPP,2021-Borla.etal-SciPost} undergoing a sudden quench. We wish to study if this kind of sudden quench,  introduced naturally in the exterior of a black hole through the bifurcate horizon formation, leads to any instabilities instead of thermalization as supposed in the black hole exterior. For this purpose, we consider a Kitaev chain ~ \cite{2001-Kitaev-PhysUsp,2012-Alicea-RPP,2021-Borla.etal-SciPost} which undergoes a break at a given time in a thermal setting (accounting for the black hole mass).

\subsection{Model Hamiltonian}
In this subsection we study the dynamics of local quench in analytically solvable one dimensional spinless Fermionic system. The model we consider is the non-number conserving free Fermionic model whose Hamiltonian is ~\cite{kogut1979introduction}:
\begin{eqnarray}\label{ham}
H=- \frac{J}{2}\sum_{j=1}^{2N}(\hat{a}_{j+1}^{\dagger}\hat{a}_{j}+\hat{a}_{j}^{\dagger}\hat{a}_{j+1})-h\sum_{j=1}^{2N}\hat{a}_{j}^{\dagger}\hat{a}_{j}\nonumber\\
-\frac{J}{2}\sum_{j=1}^{2N}(\hat{a}_{j}^{\dagger}\hat{a}_{j+1}^{\dagger}+\hat{a}_{j+1}\hat{a}_{j})
\end{eqnarray}
where $a_{j}^{\dagger}(a_{j})$ creates(annihilates) Fermion at lattice site j.As mentioned earlier, this is an integrable model. For the ease of computations, we assume a periodic boundary condition for the Fermionic chain. In Sec. \ref{analytic}, we provide an analytic study of a simpler yet similar number conserving system, tight binding model with Fermions.

The Hamiltonian \ref{ham} can be diagonalized to normal modes by a Fourier transformation followed by the Bogoliubov transformation. Under the Fourier transformation,
$$\hat{b}_k=\frac{1}{\sqrt{2N}}\sum_{j=1}^{2N}\hat{a}_{j} e^{(\frac{2 \pi i j k}{2N})}; ~\hat{b}^{\dagger}_k=\frac{1}{\sqrt{2N}}\sum_{j=1}^{2N}\hat{a}^{\dagger}_{j} e^{(\frac{-2 \pi i j k}{2N})}$$
the Hamiltonian \ref{ham} gets transformed to
\begin{equation}
H=\sum_{k=1}^{N}\omega_k(\hat{b}_k^\dagger \hat{b}_k +\hat{b}_{-k}^\dagger \hat{b}_{-k})+\sum_{k=1}^{N} i \Delta_k(\hat{b}_k^\dagger \hat{b}_{-k}^\dagger -\hat{b}_{-k} \hat{b}_{k})
\end{equation}
where 
$$ \omega_k=-h-J\cos\left({\frac{2 \pi k}{2N}}\right) \mbox{ and } 
\Delta_k=J\sin\left({\frac{2 \pi k}{2N}}\right).$$

Thereafter, performing the Bogoliubov transformation:
 $$ \hat{\gamma}_{k1}=\alpha_{k}\hat{b}_k +i \beta_k \hat{b}_{-k}^\dagger ;~~\hat{\gamma}_{k2}=\alpha_{k}\hat{b}_{-k} -i \beta_k \hat{b}_{k}^\dagger \, ,$$ 
the Hamiltonian \ref{ham} becomes

\begin{equation}
\label{eq:daigham}
H=\sum_{k=1}^{N}E_k(\hat{\gamma}_{k1}^\dagger \hat{\gamma}_{k1}+\hat{\gamma}_{k2}^\dagger\hat{\gamma}_{k2})-(E_k -\omega_k)
\end{equation}
where 
$$ \alpha_k^2=\frac{1}{2}\left(1+\frac{\omega_k}{E_k} \right) \, ,  \beta_k^2=\frac{1}{2}\left(1-\frac{\omega_k}{E_k} \right) \, , 
 $$
and $E_k=\sqrt{J^2 +h^{2}+2 \, J \, h\cos{\frac{2 \pi k}{2N}}}$, $k=1,2,...,N$ are the normal mode frequencies.

\subsection{The quench and covariance matrix}

The initial Hamiltonian $H_{I}$ is $H_{I}=H_{2N+2M}$ where $H_{2N+2M}$ describes the Fermionic lattice of size $(2N+2M)$ with periodic boundary condition. The effect of quench corresponds to (i) simultaneously switching off the hopping term between $1^{st}$, and $(2N+2M)^{th}$ sites and $2N^{th}$ and $(2N+1)^{th}$ sites of $H_{2N+2M}$ and (ii) introducing the hopping term (with coupling constant $J$) between $1^{st}$ and $2N^{th}$ site resulting in $H_{2N}$ and $(2N+1)^{th}$ and $(2N+2M)^{th}$ site resulting in $H_{2M}$. In other words, we break the chain of lattice size $(2N+2M)$ into two independent chains of sizes $(2N)$ and $(2M)$, resulting in the quenched Hamiltonian $H_{f}=H_{2N}\oplus H_{2M}$. 
\begin{figure}[H]
    \includegraphics[width=1\columnwidth]{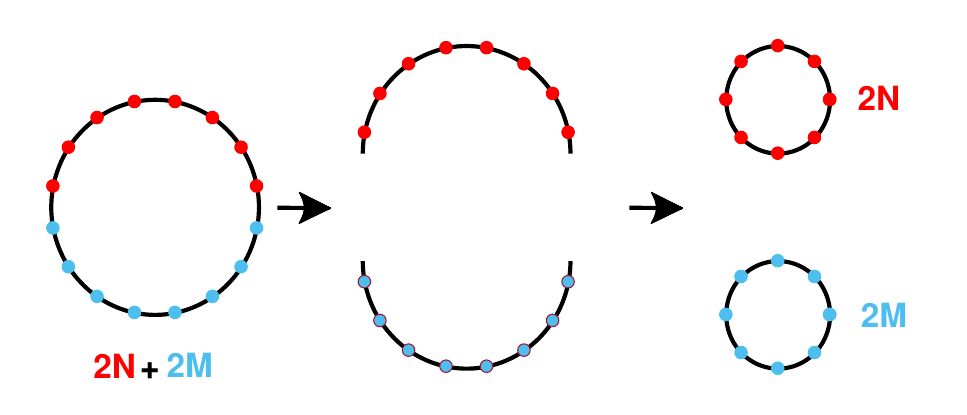}
    \caption[An illustration for the quench]{An illustration of the quench that breaks an initial chain of size $2N+2M$ into two disjoint chains with periodic boundary condition for all the chains.}\label{illustration}
\end{figure}
Since the initial and the final Hamiltonian can be diagonalized, the system is described by non-interacting quasi-particles. The non-interacting nature implies that all the information about the system can be obtained from the expectation value of the two-point correlators between various lattice points which can be compactly arranged in the covariance matrix $\hat{G}$  defined as the outer product of $\vec{A}$ and $\vec{A}^{T}$ i.e., $\hat{G}=\vec{A}\vec{A}^{T}$ where the column vector $\vec{A}=[a_{1} a_{1}^{\dagger} a_{2} a_{2}^{\dagger}...a_{N} a_{N}^{\dagger}]$, leading to 
%Thus, the covariance matrix $G_{ij}$ can be defined as 
%
%\begin{equation}
%G_{ij}=\langle\hat{G}_{ij}\rangle=Tr[\rho \hat{G}_{ij}]\% end{equation}
\begin{eqnarray}
G = \left[ \begin{array}{ccccc}
\langle a_1 a_1\rangle &\langle a_1a_1^{\dagger}\rangle  &\dots &\langle a_1 a_N\rangle &\langle a_1a_N^{\dagger}\rangle\\
\langle a_1^{\dagger} a_1\rangle &\langle a_1^{\dagger}a_1^{\dagger}\rangle  &\dots &\langle a_1^{\dagger} a_N\rangle &\langle a_1^{\dagger}a_N^{\dagger}\rangle\\
\vdots &\vdots&\ddots&\vdots&\vdots\\
\langle a_N a_1\rangle &\langle a_Na_1^{\dagger}\rangle  &\dots &\langle a_N a_N\rangle &\langle a_Na_N^{\dagger}\rangle\\
\langle a_N^{\dagger} a_1\rangle &\langle a_N^{\dagger}a_1^{\dagger}\rangle  &\dots &\langle a_N^{\dagger} a_N\rangle &\langle a_N^{\dagger}a_N^{\dagger}\rangle
\end{array}\right]
\end{eqnarray}

Hence a symplectic transformation (linear transformations that preserve Fermionic anti-commutation relation) of the creation and annihilation operators like, $\vec{\gamma}=U \vec{a}$ would cause the covariance matrix to transform as 
\begin{equation}
G'=UGU^{T}
\end{equation}
and one can use the transformed covariance matrix to obtain correlators.
The creation and annihilation operators for Fermions in the energy eigenstates of the quenched Hamiltonian evolves in time as $\gamma_{k}(t)=e^{-iE_{k}t}\gamma_{k}$ and $\gamma_{k}^{\dagger}(t)=e^{iE_{k}t}\gamma^{\dagger}_{k}$ where $E_{k}$ are the energy eigenvalues of $H_{f}$. 
    
\subsection{Observables and GGE}

In this work, we have used three estimators to quantify thermalization. 
As we will show in the next section, these three estimators provide complementary information about how the system drives to a generalized Gibbs ensemble. 

\begin{enumerate}

\item {\bf Occupancy of a site} {One of the macroscopic observables of our interest is the number density per lattice site in real space. The expectation value of the time evolved number operator can be obtained from the time evolved covariance matrix $G$, given by $\langle n_{i}(t)\rangle=\langle a_{i}^{\dagger}(t) a_{i}(t)\rangle$. The main aim would be to verify if the longtime expectation value of the number operator per lattice site in the real space would converge to GGE.}

In order to find the GGE ensemble, we need to find the conserved quantities of the system. Since the system is non-interacting in the normal modes, the occupation number of each normal mode after the quench is conserved. Hence the independent conserved quantities are  $n_{k}= \hat{\gamma}^{\dagger}_{k1(2)}\hat{\gamma}_{k1(2)}$ where $k=1,2\cdots (N+M)$. The GGE density matrix is given by
\begin{equation}
\hat{\rho}=\frac{\exp{(-\sum_{k=1}^{N+M}\lambda_{k}\hat{n}_{k})}}{\text{Tr}[\exp{(-\sum_{k=1}^{N+M}\lambda_{k}\hat{n}_{k})}]}
\end{equation}
where $\lambda_{k}$ are the Lagrange multipliers which are fixed using the initial condition $\text{Tr}[\hat{\rho} \, \hat{n}_{k}]=\langle \hat{n}_{k}(0) \rangle$.
Using the initial condition, we obtain the Lagrange multipliers to be 
\begin{equation}
\lambda_{k}=\text{ln}\left(\frac{1-\langle \hat{n}_{k}(0) \rangle}{\langle \hat{n}_{k}(0) \rangle}\right)
\end{equation}
where $\langle \hat{n}_{k}(0) \rangle$ can be obtained from the covariance matrix in the normal modes.

\item{\bf Nearest neighbor correlation} We also study the nearest neighbor correlation in the real lattice. The nearest neighbor correlation is defined as $\langle a_{i}^{\dagger}(t) a_{i+1}(t) + a_{i+1}^{\dagger}(t) a_{i}(t) \rangle$. Here again we verify if the longtime expectation value of the operator in the real space would converge to its corresponding GGE. 

\item {\bf Bits per Fermion} Another important quantity we calculate is the information content in bits per Fermion in each normal mode and compare its profile before and after the quench. The von Neumann entropy for the density matrix $\rho$ is given by $S(\rho)=-Tr(\rho \ln \rho)$. The von Neumann entropy for the GGE density matrix, for a normal mode $k$, is given by 
\begin{equation}
\label{eq:BitsperF-Entropy}
~~~~~~S(k)=-\langle \hat{n}_{k} \rangle\ln \langle \hat{n}_{k} \rangle-(1-\langle \hat{n}_{k} \rangle)\ln(1-\langle \hat{n}_{k} \rangle),
\end{equation}
where {$\langle \hat{n}_{k} \rangle$ is the expectation value of the number operator in the corresponding mode.} One can define the information content in bits per Fermion per normal mode as~\cite{magan2016random}
\begin{equation}
\label{eq:BitsperF}
I(k)=\dfrac{S(k)}{\langle \hat{n}_{k} \rangle\log(2)}
\end{equation}
\item {\bf Out-of-time-order correlator}  Out of time order correlator (OTOC) corresponding to  $[\hat{x}(t),\hat{p}(0)]$ measures the quantum analog of the classical quantity $\delta x(t)/\delta x(0)$ for Bosonic systems, identifying the measure of chaos in the system. If the system turns chaotic, this quantity should gradually rise with time, while for a system landing in a pre-ascribed configuration, the strength of OTOC should remain within bounds for large times \cite{lin2018out}, whereas for Fermionic systems it shows a tendency of flattening out~\cite{tsuji2017exact}.  An OTOC can be constructed for any two non commuting observables. We will be considering an  out-of-time-order correlator for the chain, given by 
\begin{equation}
\label{eq:OTOC-def}
F_{ij}(t)=\frac{1}{2}\langle[\hat{x}_i(t),\hat{p}_j(0)]^2\rangle
\end{equation}
where we define Hermitean observables
\begin{equation}
\label{eq:OTOC-def2}
\hat{x}_i(t)=\frac{\hat{a}_i^{\dagger}(t)+\hat{a}_i(t)}{\sqrt{2}};~~ \hat{p}_j(0)=i \frac{\hat{a}_j^{\dagger}(0)-\hat{a}_j(0)}{\sqrt{2}}
\end{equation}
\end{enumerate}
in analogy to the Bosonic case (but keeping in mind  that in Fermionic systems they satisfy anticommutation relation).
%%%%%%%%%%%%%%%%%%%%%%%%%%%%%%%%%%%%%%%%%%
We will now calculate these estimators in order to test the robustnness of genaralized thermalization and confirm whether the  jump of a system  from one integral configuration to another integral configuration  with causal disruption does not make it chaotic~\cite{lewkowycz2013generalized}.

\begin{figure*}
\includegraphics[width=0.9\textwidth]{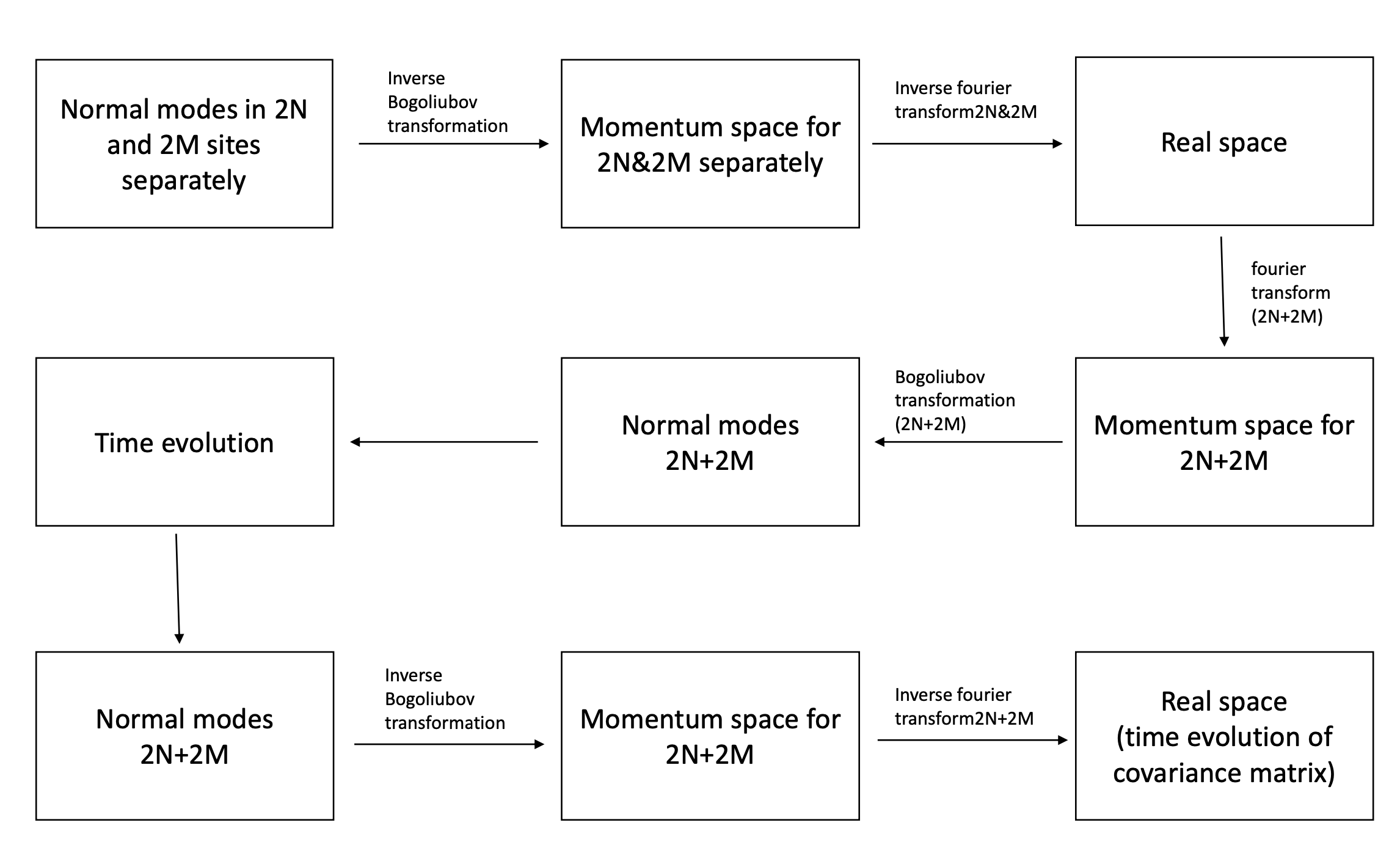}
\caption{Flowchart of the procedure used in numerical evaluation of the physical quantities.}
\label{fig:flowchart}
\end{figure*}

\section{Results} \label{Results}

In this section, we present semi-analytical results for the model Hamiltonian \ref{ham} with the effect of quench. As mentioned earlier, we use the following estimators--- occupancy at a site, nearest neighbor hopping, Bits per Fermion and OTOC --- to identify the late-time evolution of the initial state.  

{The flowchart of the procedure used in numerical evaluation of the physical quantities can be seen in \ref{fig:flowchart}. Numerics is done in MATLAB.}

\subsection{Occupation Number}

To study equilibration of a local observable, we look at the evolution of expectation value of the number operator at a particular lattice site in real space. For the verification, we have plotted the mean value of the evolved number operator and the value given by the GGE. For all the plots, we have fixed the parameters $h/J=-2$ and the nearest neighbor interaction $J$ to be $0.5$. The initial state is chosen to be a thermal state with the lattice chain at temperature $T/J=0.5$, i.e., inverse temperature $\beta_{I}=1/T=4$. We calculate the energy in the unit of the onsite coupling constant $h = -1$. This value sets the unit of time to be ${1/h}$.

\begin{figure}[H]
    \centering
    \subfloat[]{
        \includegraphics[width=0.5\columnwidth,height=48mm]{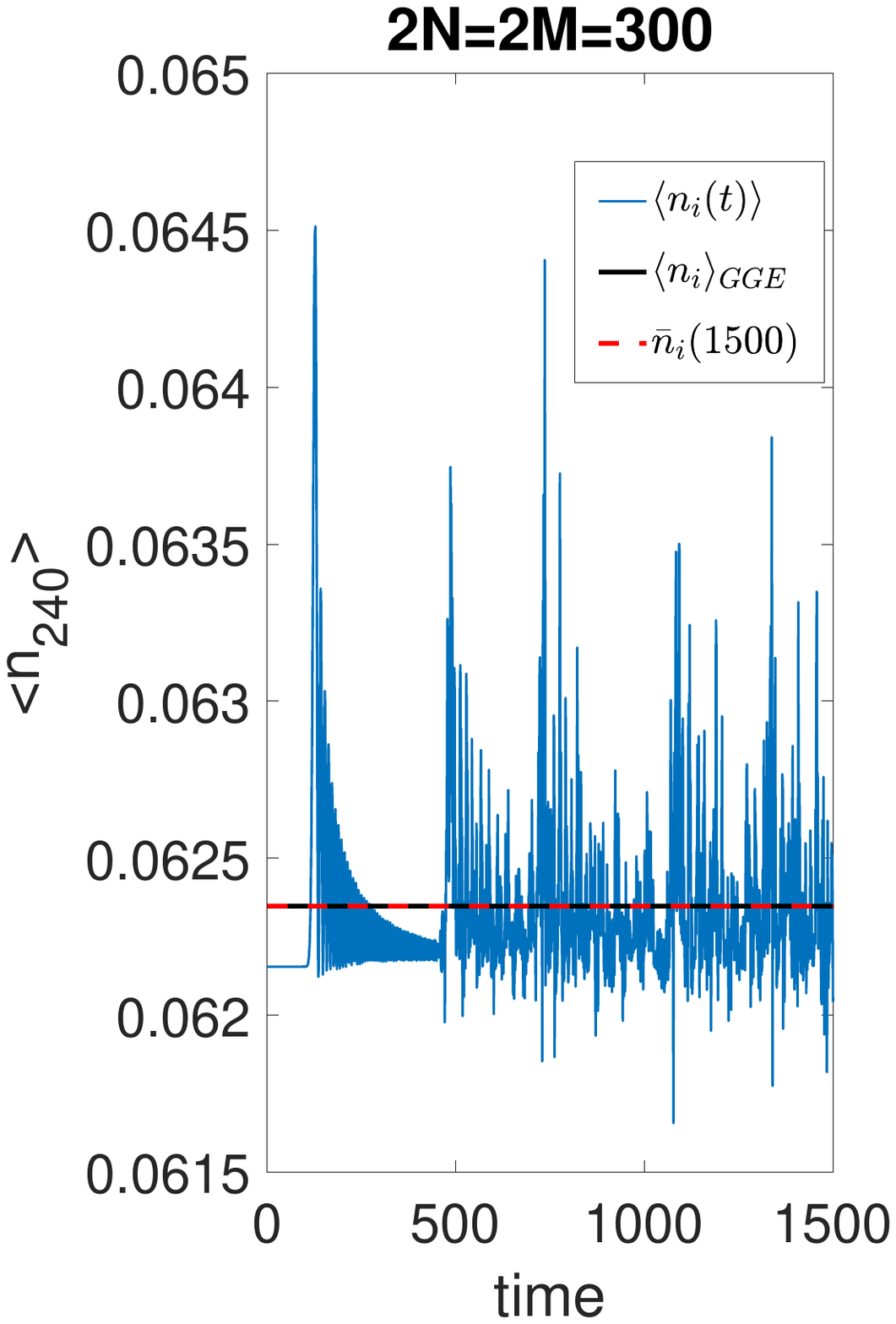}
    }
    \subfloat[]{
        \includegraphics[width=0.5\columnwidth,height=48mm]{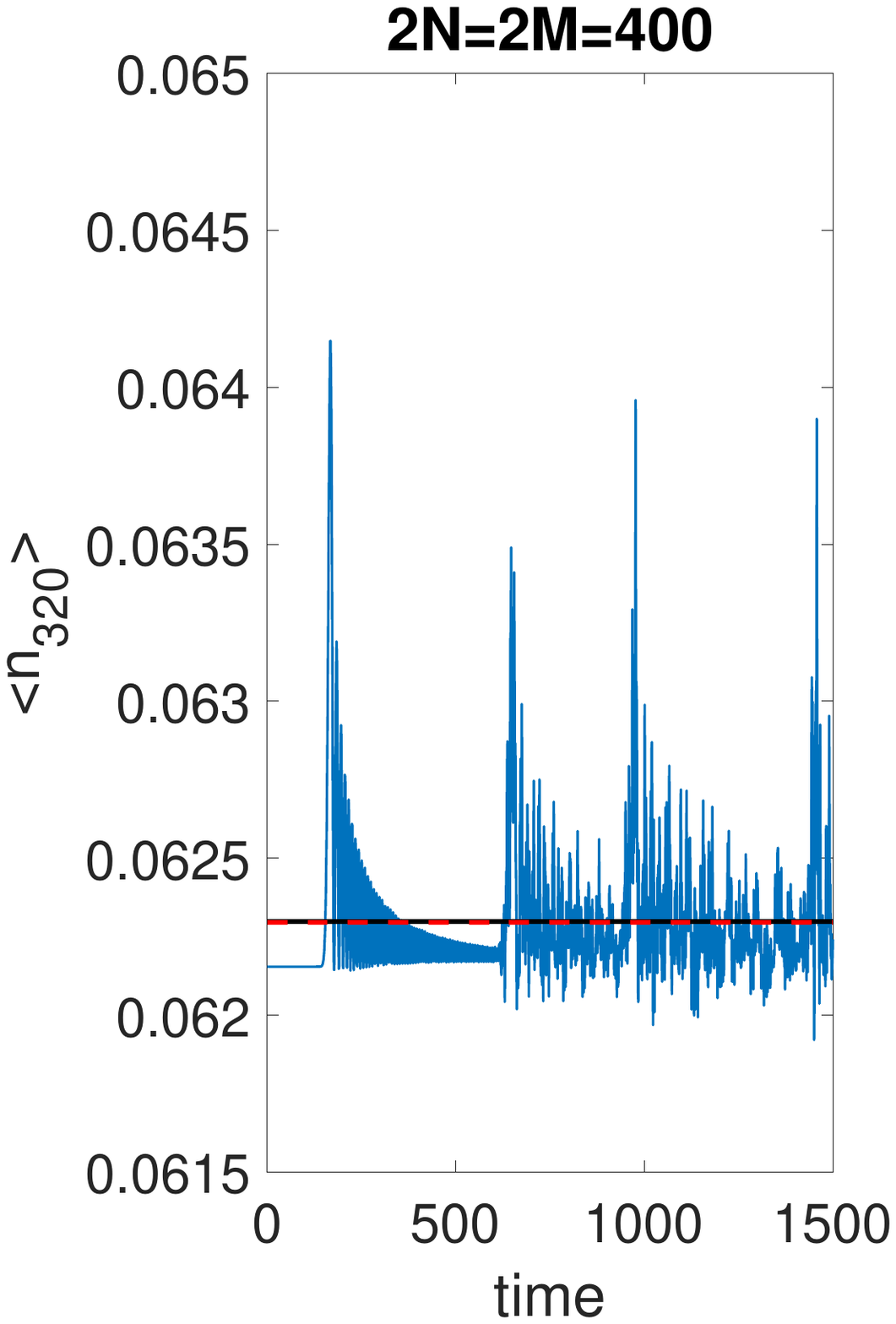}
    }
    \hspace{1mm}
    \subfloat[]{
        \includegraphics[width=0.5\columnwidth,height=48mm]{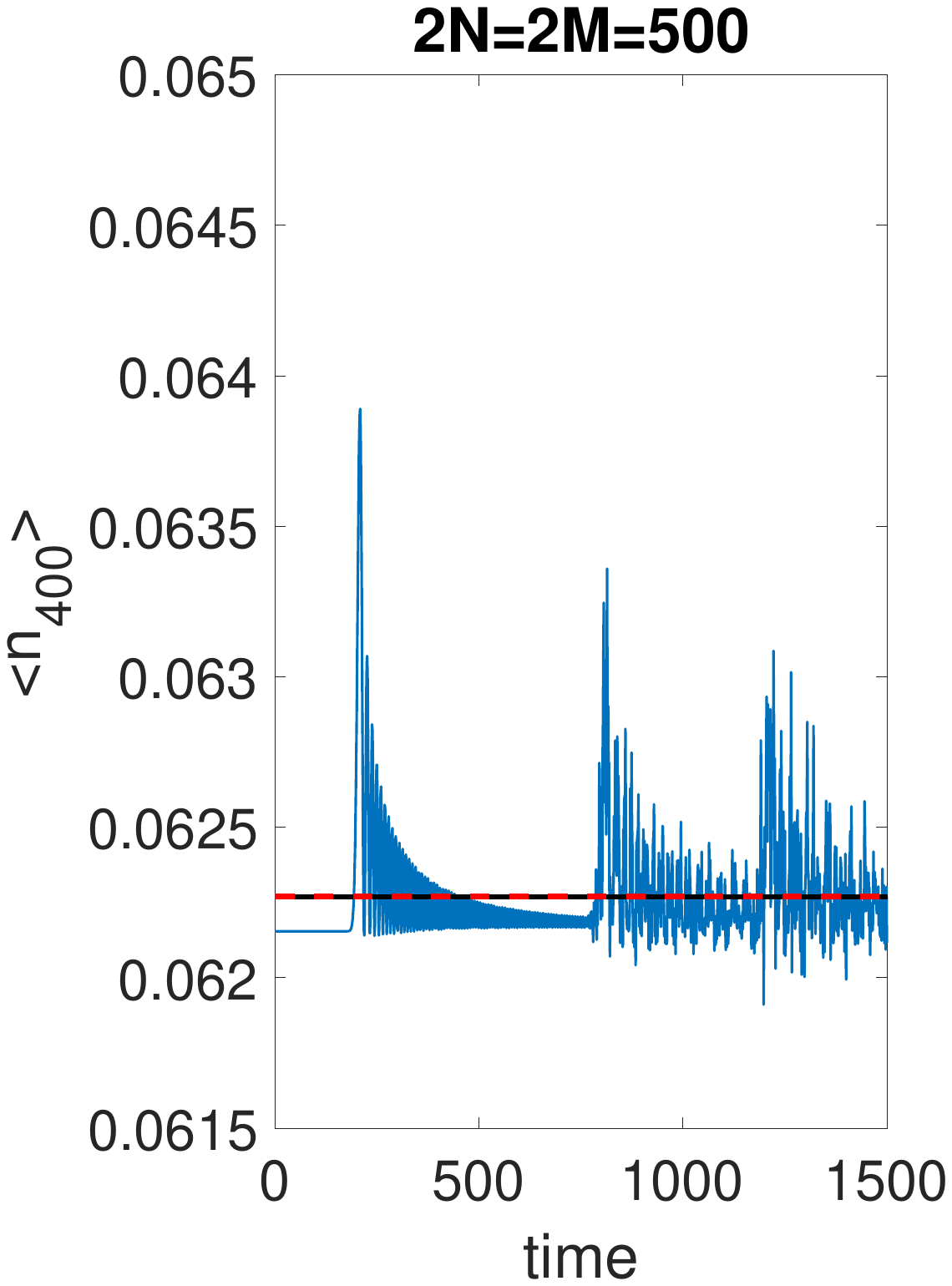}
    }
    \subfloat[]{
        \includegraphics[width=0.5\columnwidth,height=48mm]{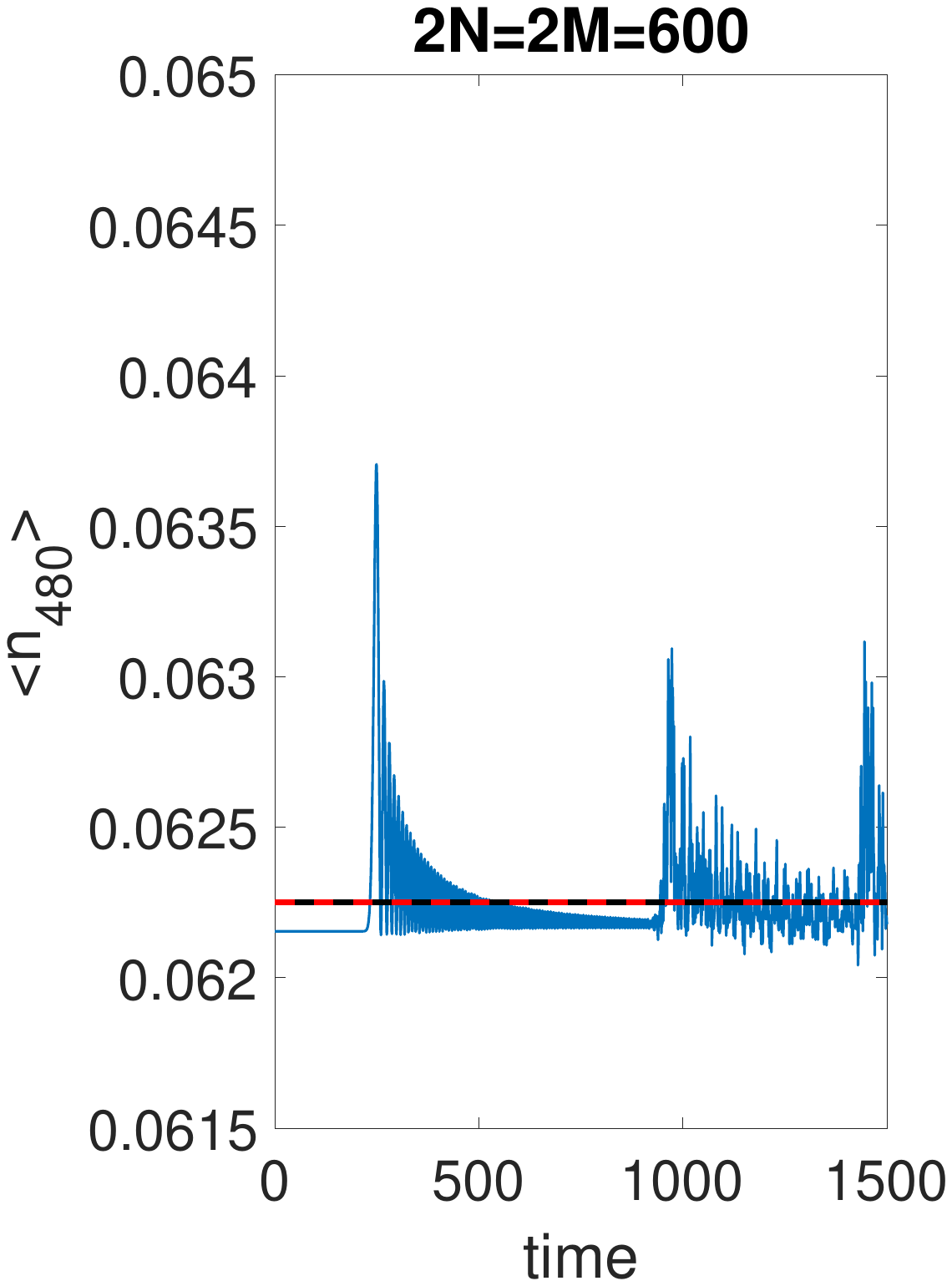}
    }
    \caption[Plot of the expectation value of the number operator at a site (slightly away from the site of quench) as a function of time. Note that the number of sites on both the chains is the same after the quench.]{Plot of the expectation value of the number operator at a site slightly away from the quenching  site $\langle n_{i}(t)\rangle$ as a function of time for the case when $2N = 2M$, i. e., (a) $2N = 2M = 300$, (b) $2N = 2M = 400$, (c) $2N = 2M = 500$, (d) $2N = 2M = 600$. GGE value and time average of $\langle n_{i}(t)\rangle$ is also plotted in each case.}\label{diffbreak}
\end{figure}

In \ref{diffbreak}, we plot the expectation value of the number operator $\langle n_{i}(t)\rangle$ at a lattice site $i$ slightly away from the site of quench as a function of time. 

We infer the following: First, until the effect of quench reaches the particular site of observation, it remains in the initial thermal state. As soon as the effect of quench reaches the site, the value fluctuates, and the system goes out of equilibrium. Second, 
the fluctuations tend to decay in time, and the expectation value of the number operator equilibrates to the GGE value and has a recurrence property. Third, as the lattice size increases, the fluctuations become smaller, and the system tends to come closer to the GGE value, with recurrences becoming sparse. {As $N$ increases, the difference between GGE and $\langle n_{i}(t)\rangle$ decreases. Therefore in large $N$ limit (and hence large $t$ limit before the first recurrence), it can be seen that the system settles to the GGE description. Hence, the GGE description is expected to be valid for large $N$.}

\begin{figure}[h]
    \centering
    \includegraphics[width=.5\columnwidth]{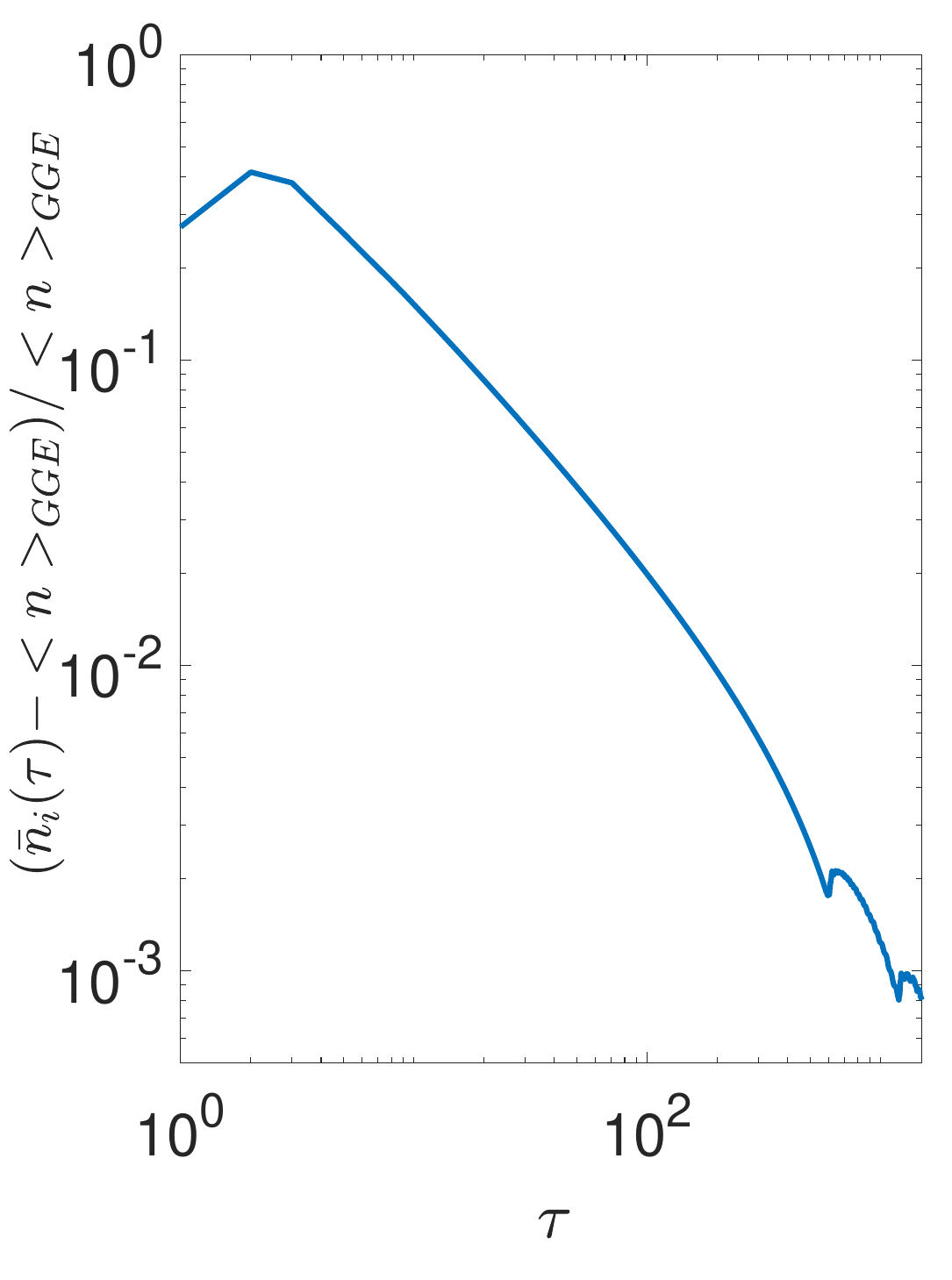}
    \caption[The relative deviation of $\bar{n}(\tau)$ from GGE value plotted against increasing $\tau$]{The relative deviation of $\bar{n}(\tau)$ from GGE value plotted against increasing $\tau$.}\label{dev_longtime}
\end{figure}

As mentioned above, from \ref{diffbreak}, we observe that sufficiently long time average value matches the GGE value. In order to substantiate the same, we evaluate the time average of the observable: 
\begin{equation}
\overline{n}(\tau)=\frac{1}{\tau}\int_{0}^{\tau}\langle \hat{n}_{i}(t) \rangle dt \, .
\end{equation}
and calculate the relative deviation of $\overline{n}(\tau)$  ($\Delta n$)
\begin{equation}
\Delta n = \frac{\left| \overline{n} (\tau) - \langle n \rangle_{GGE} \right|}{\langle n \rangle_{GGE}}
\end{equation} 
from the GGE value as a function of $\tau$. \ref{dev_longtime} contains the plot of $\Delta n(\tau)$ as a function of $\tau$. The figure explicitly shows a power-law decay of the relative deviation. Thus, in the infinite time limit, the relative deviation vanishes.

\begin{figure}[h]
    \centering{
        \subfloat[]{
            \includegraphics[width=0.5\columnwidth]{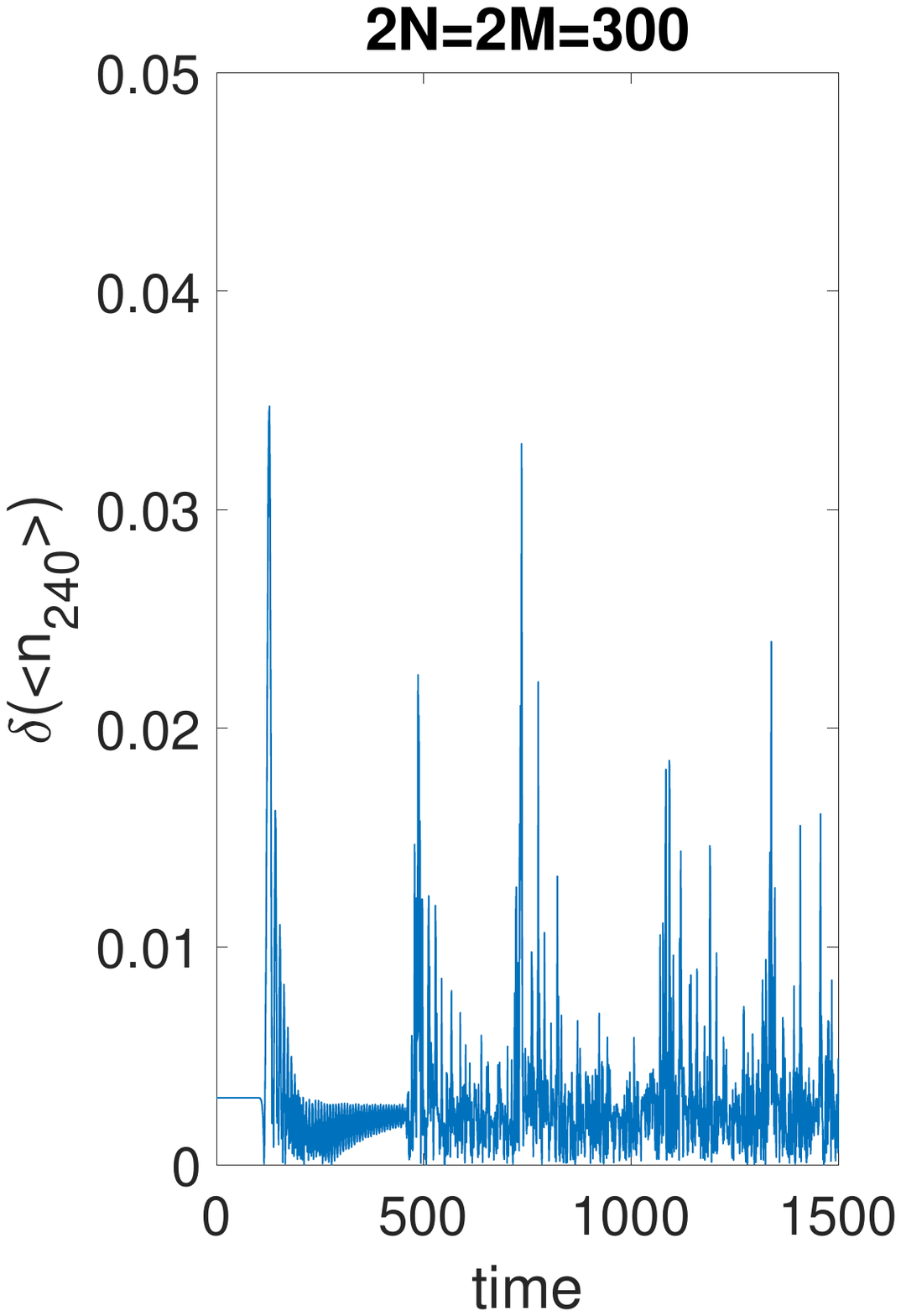}
        }
        \subfloat[]{
            \includegraphics[width=0.5\columnwidth]{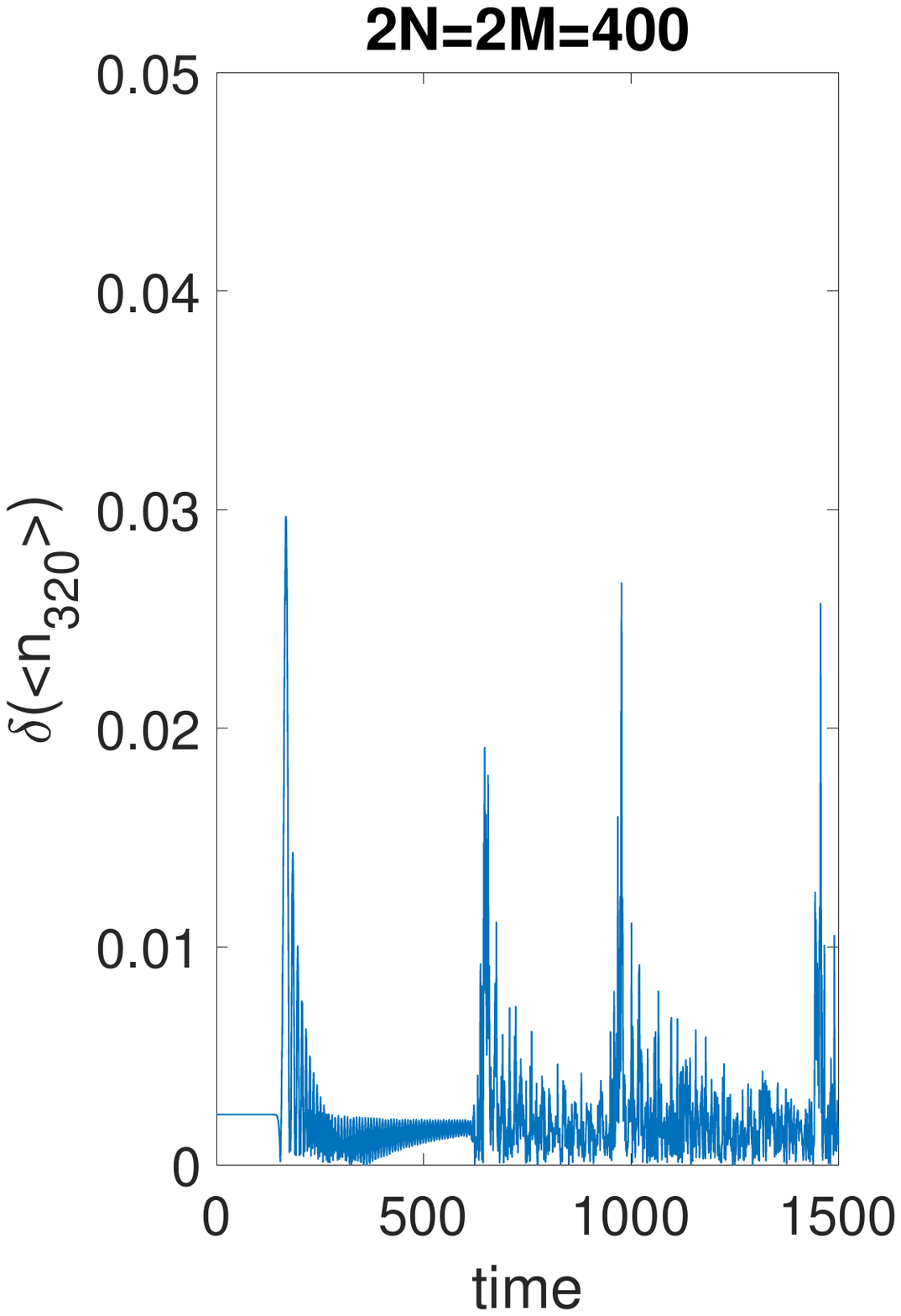}
        }
        \hspace{1mm}
        \subfloat[]{
            \includegraphics[width=0.5\columnwidth]{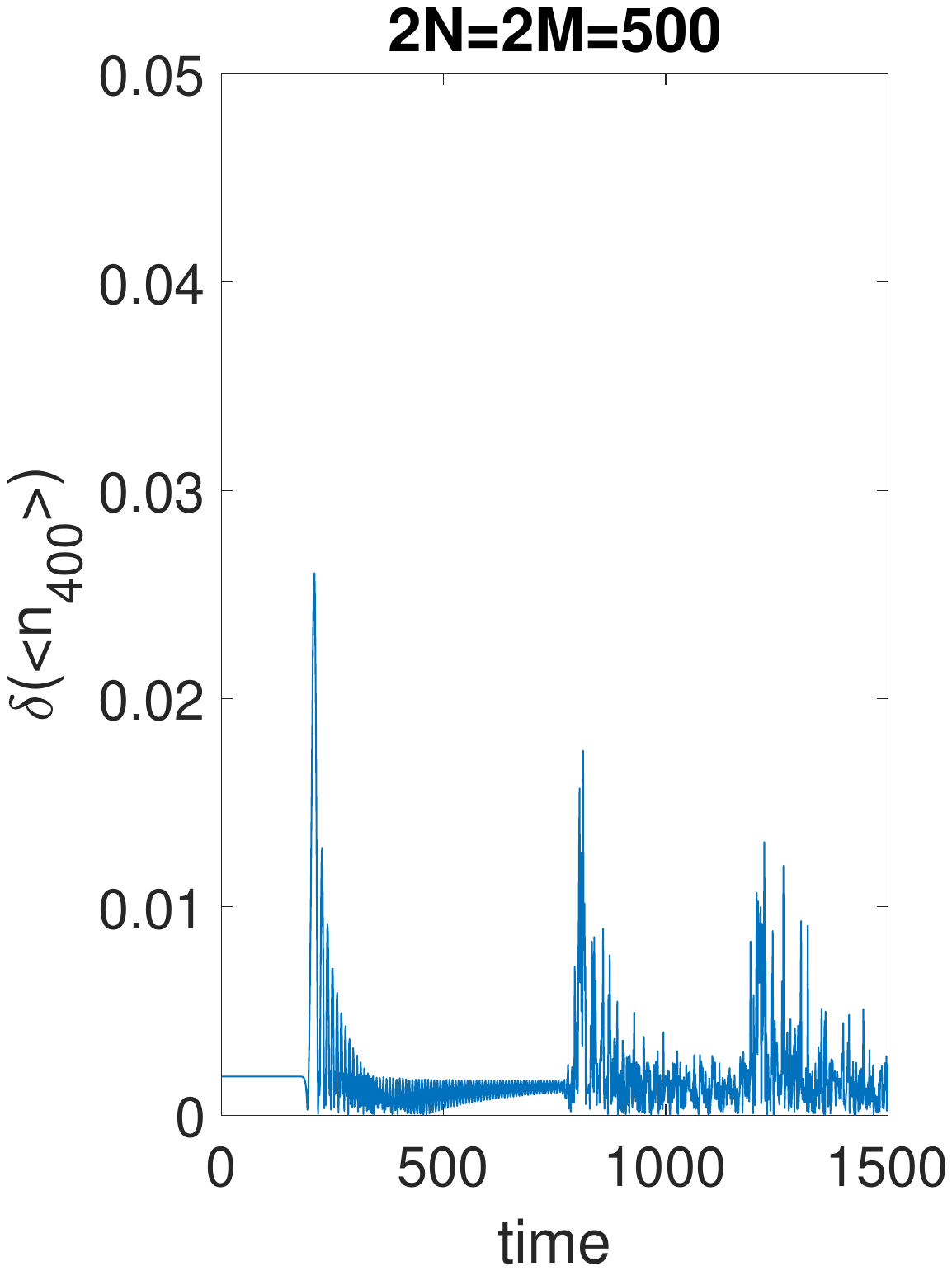}
        }
        \subfloat[]{
            \includegraphics[width=0.5\columnwidth]{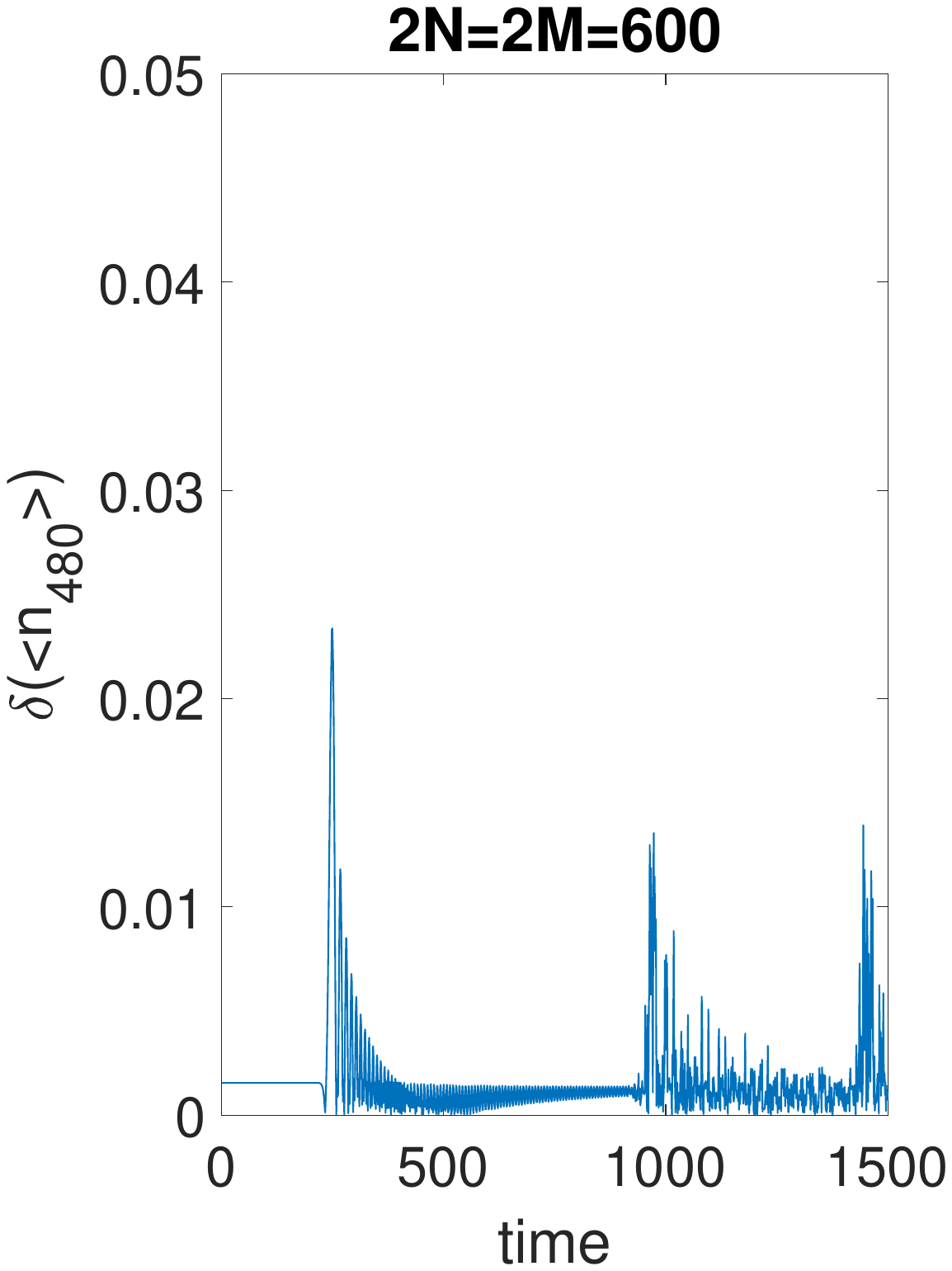}
        }}
        \caption[$\delta(n_{i}(t))=\frac{|<n_{i}(t)>-<n_{i}>_{GGE}|}{<n_{i}>_{GGE}}$ plotted against time]{The deviation of $\langle n_{i}(t)\rangle$ from the GGE value $\delta(n_{i}(t))$ is plotted as a function of time for (a) $2N = 2M = 300$, (b) $2N = 2M = 400$, (c) $2N = 2M = 500$, (d) $2N = 2M = 600$.}\label{diffdevbreak}
    \end{figure}
    
    To further quantify, we evaluate the relative deviation of the lattice occupation number $\langle n_{i}(t)\rangle$ from the GGE value, i. e., 
    \begin{equation}
    \label{eq:rel-deviation}
    \delta n_i(t) = \frac{\left| \langle n_{i}(t)\rangle - \langle n_i \rangle_{GGE} \right|}{\langle n_i \rangle_{GGE}}
    \end{equation} 
    as a function of time. From the \ref{diffdevbreak}, we infer the following: Initially $\delta(n_{i}(t))$  relaxes to zero after quench and at later times, starts showing fluctuations. As the number of lattice sites increases, the time of initiation of the late time fluctuations is delayed in a linear fashion and the magnitude of fluctuation also reduces. 
    \begin{figure}[h]
        \subfloat[]{
            \includegraphics[width=0.5\columnwidth]{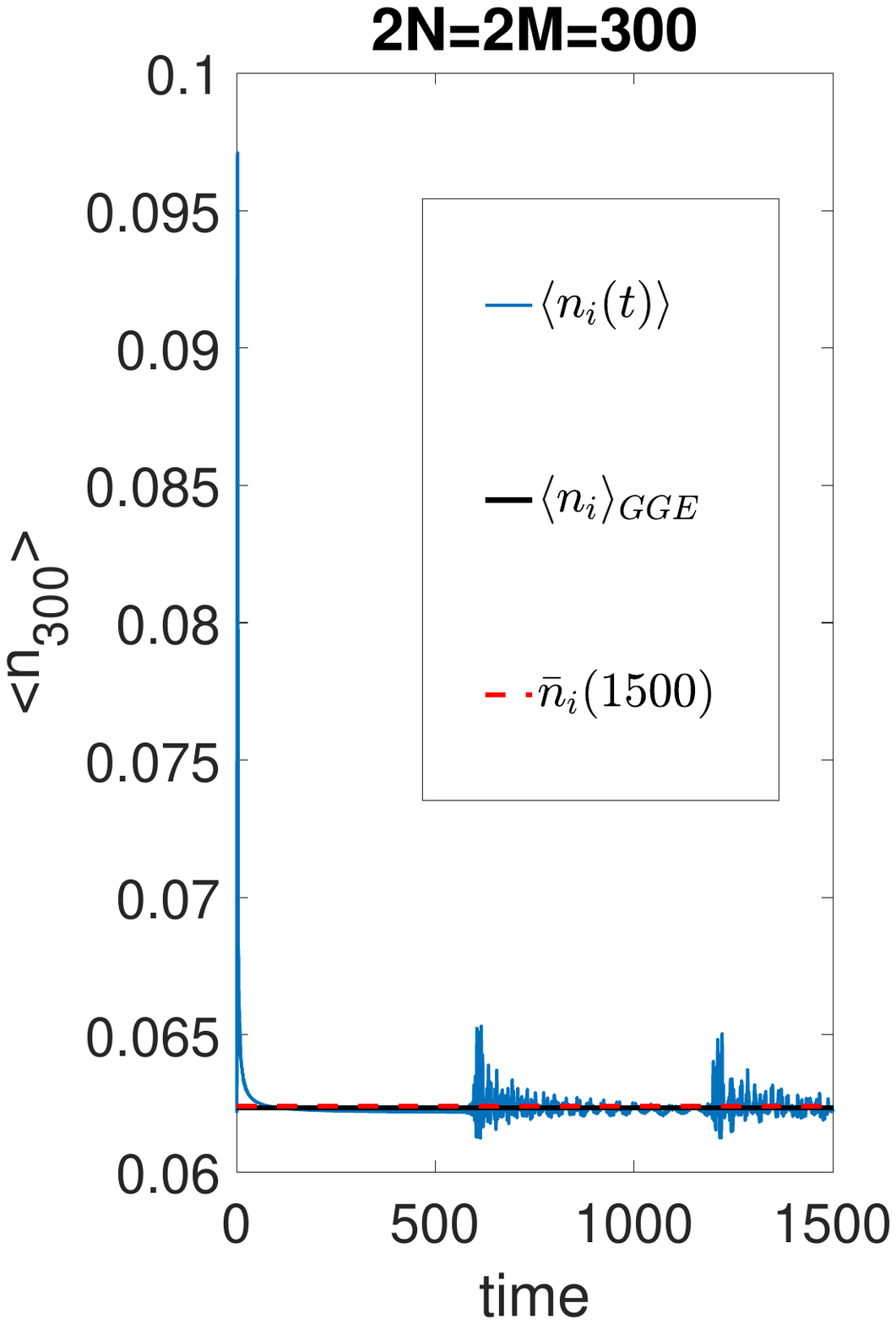}
        }
        \subfloat[]{
            \includegraphics[width=0.5\columnwidth]{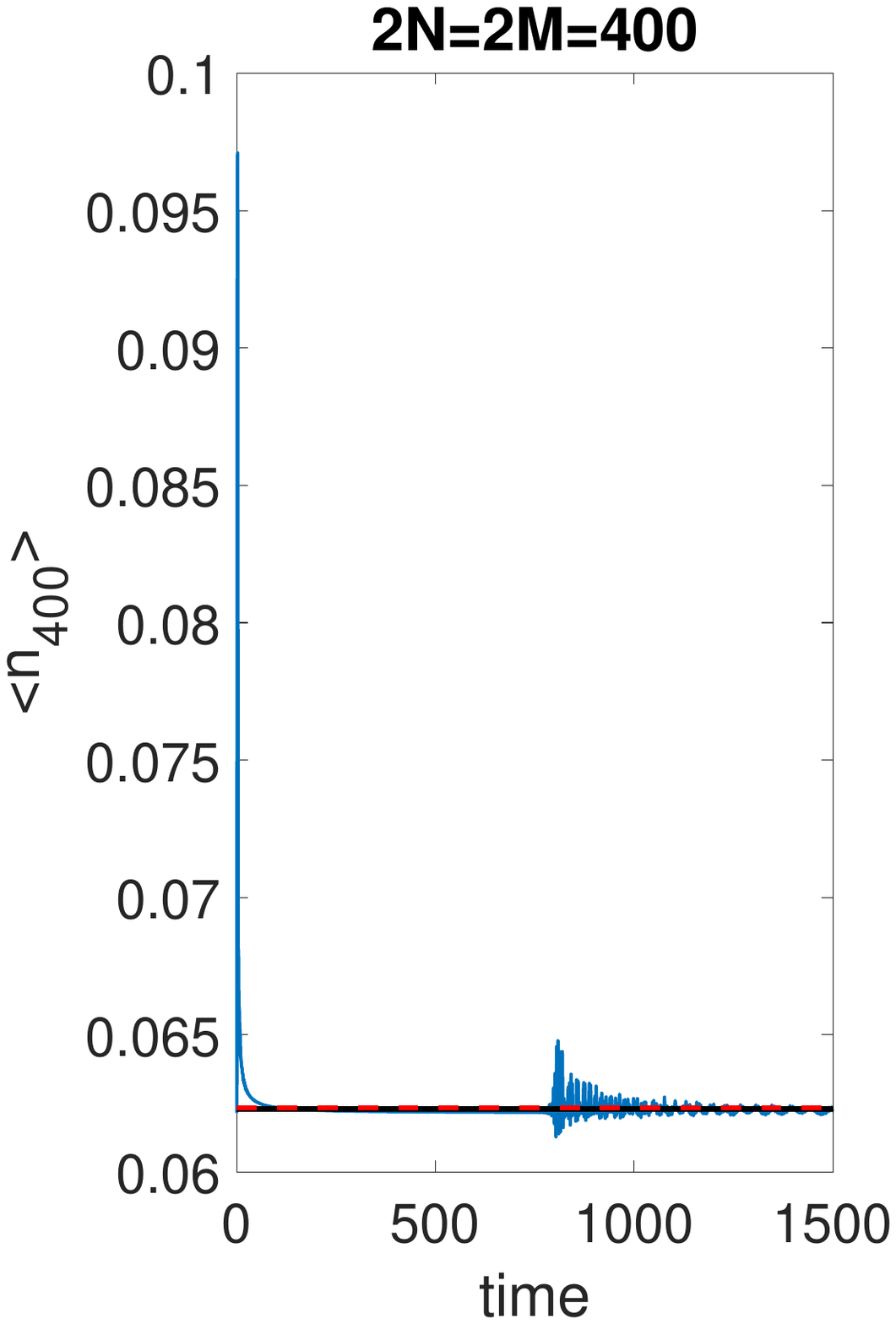}
        }
        \hspace{1mm}
        \subfloat[]{
            \includegraphics[width=0.5\columnwidth]{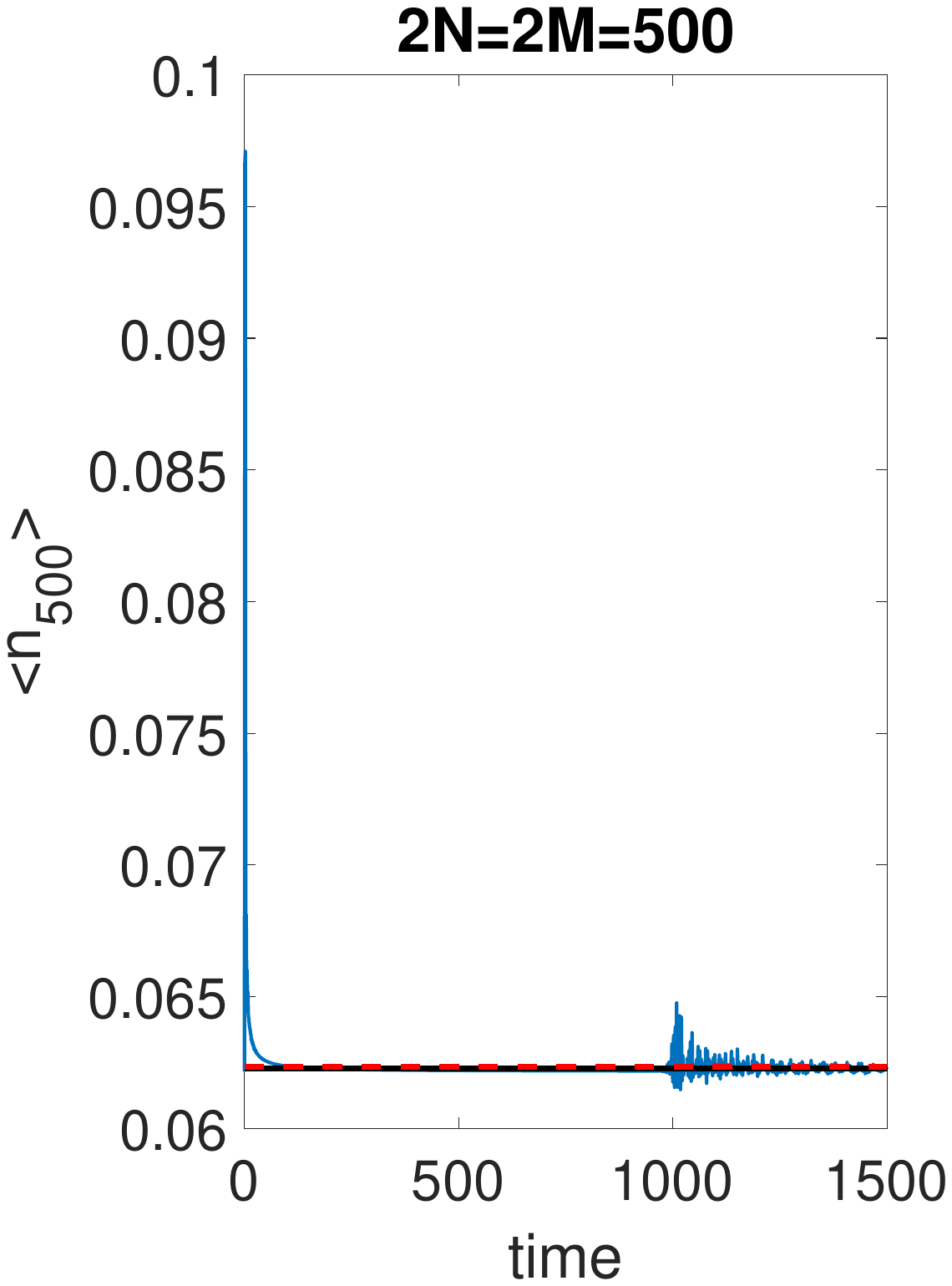}
        }
        \subfloat[]{
            \includegraphics[width=0.5\columnwidth]{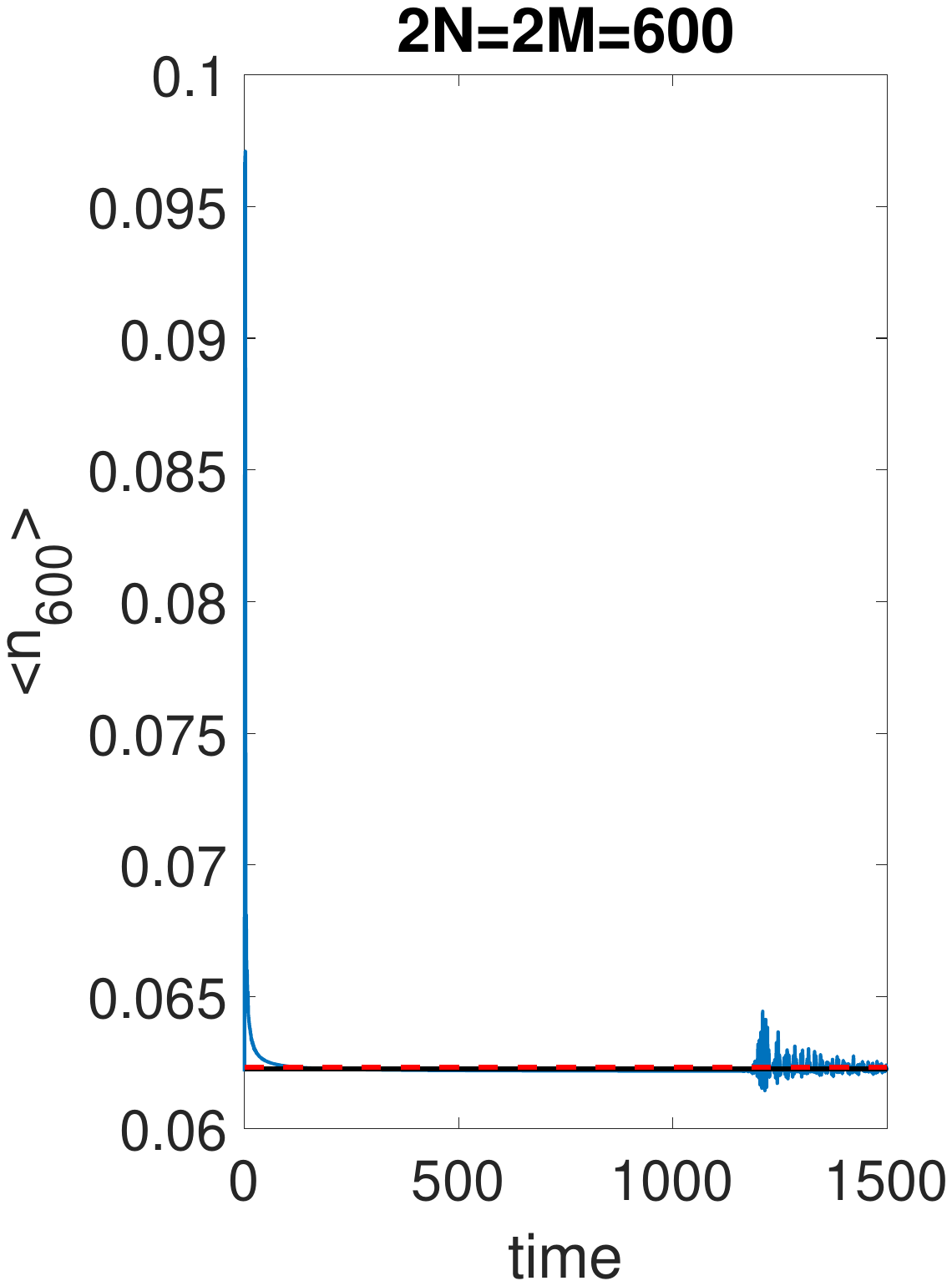}
        }
        \caption[Expectation value of the number operator at a site where quench happens plotted against time.]{The expectation value of the number operator at a site~where quench happens is plotted against time where (a) $2N = 2M = 300$, (b) $2N = 2M = 400$, (c) $2N = 2M = 500$, (d) $2N = 2M = 600$. The GGE value and time average of $\langle n_{i}(t)\rangle$ are also plotted in each case.}
        \label{samebreak}
    \end{figure}
\subsection{Characteristic function of fluctuation}
In order to study the evolution of the system towards the GGE post quenching, one can look at the charactersitic function $\phi_{\mu}$ which gets determined from the probability distribution $P(\Delta n)$ of the fluctuations and  subsequently through the fluctuations themselves in various orders of fluctuation and probe wavelength. The characteristic function
is defined as
\bea
\phi_{\mu} = \int_{-\infty}^{\infty} d \Delta n e^{i \mu \Delta n}P(\Delta n).
\eea
Expanding the exponential in the above expression, we get
\bea
\phi_{\mu} &=& \int_{-\infty}^{\infty} d \Delta n \left(1 + i \mu \Delta n + {\cal O} (\mu \Delta n^2) \right)P(\Delta n) \nonumber\\
&\approx & 1 + i \mu \langle \delta n \rangle.
\eea
   
There are a couple of observations to make. For small "chemical potential" (parameter conjugate to $\delta n$) limit $\mu\rightarrow 0$ the characteristic function goes to unity very quickly as $\langle \delta n \rangle $   decays over time (see \ref{diffdevbreak}). Thus,  such systems are virtually indistinguishable from the GGE configurations soon after quenching. Further on the time scales (within one boundary reflections time cycle), when $\langle \delta n \rangle $ approaches vanishing value, $\phi_\mu$ will approach unity for non-infinitesimal $\mu$ too.  Therefore, generic systems approach GGE configurations on their characteristic time scales.
   
    To investigate further, in \ref{samebreak}, we plot the expectation value of the number operator at the site where quench happens as a function of time. \ref{samebreak} shows the same trend as in \ref{diffbreak}. In other words, 
    as the lattice size increases, the deviation from the GGE is small and better relaxation is observed. The plot indicates that the observable will relax to GGE in the thermodynamic limit. Also, the late time fluctuations get delayed linearly with the increase in lattice size confirming the finite size effect. 
    
    \begin{figure}[h]
        \centering
        \subfloat[]{
            \includegraphics[width=0.5\columnwidth]{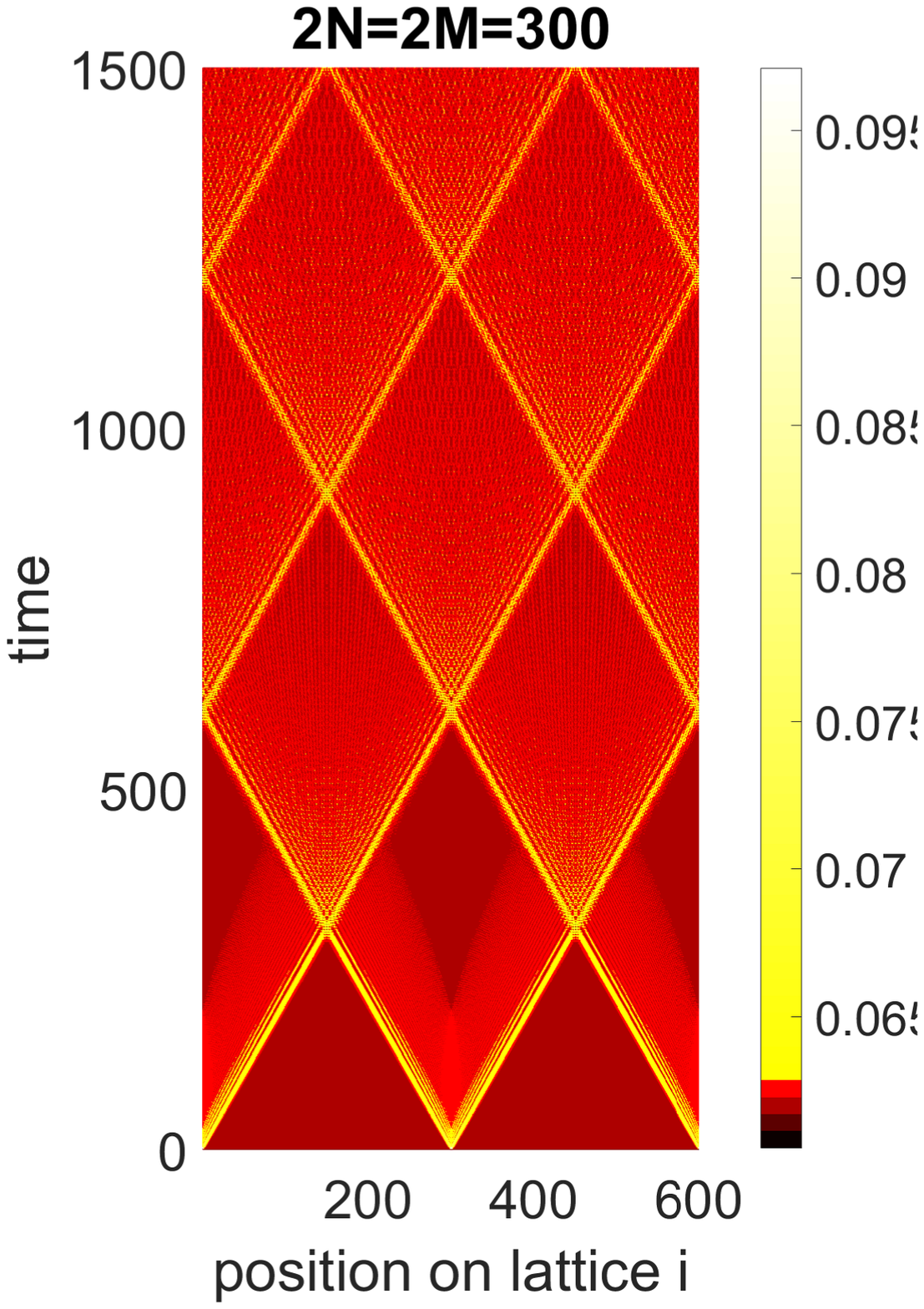}
        }
        \subfloat[]{
            \includegraphics[width=0.5\columnwidth]{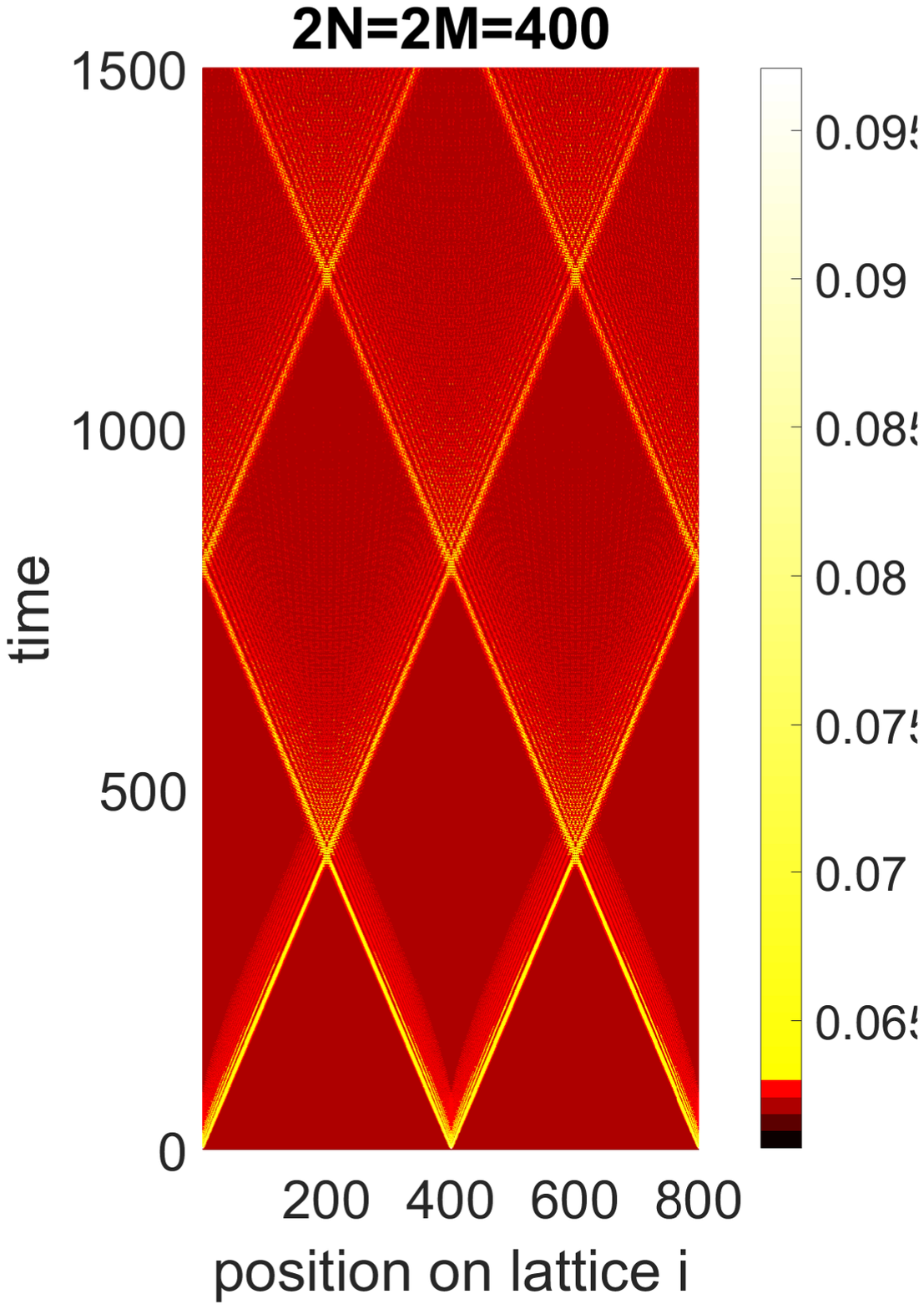}
        }
        \hspace{1mm}
        \subfloat[]{
            \includegraphics[width=0.5\columnwidth]{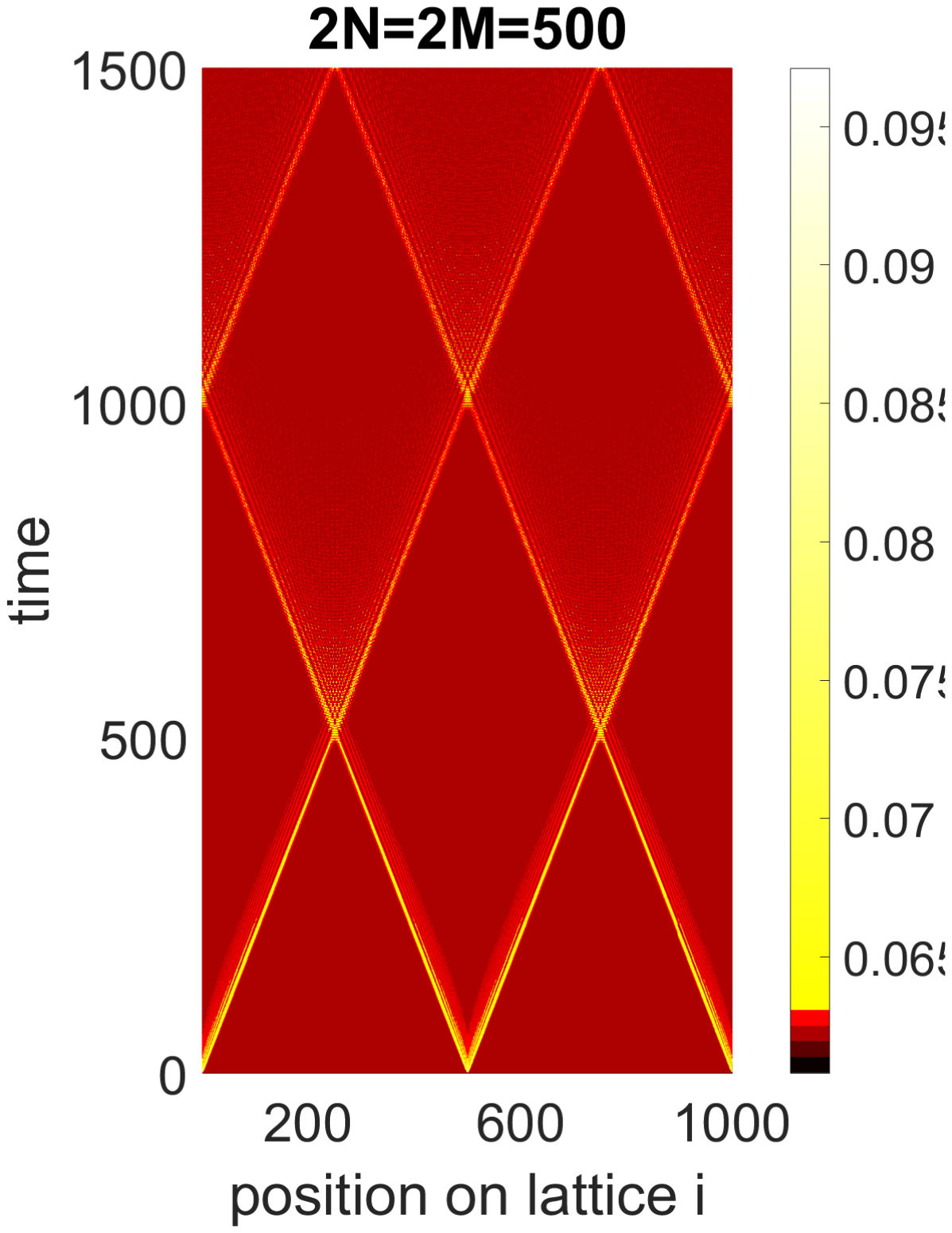}
        }
        \subfloat[]{
            \includegraphics[width=0.5\columnwidth]{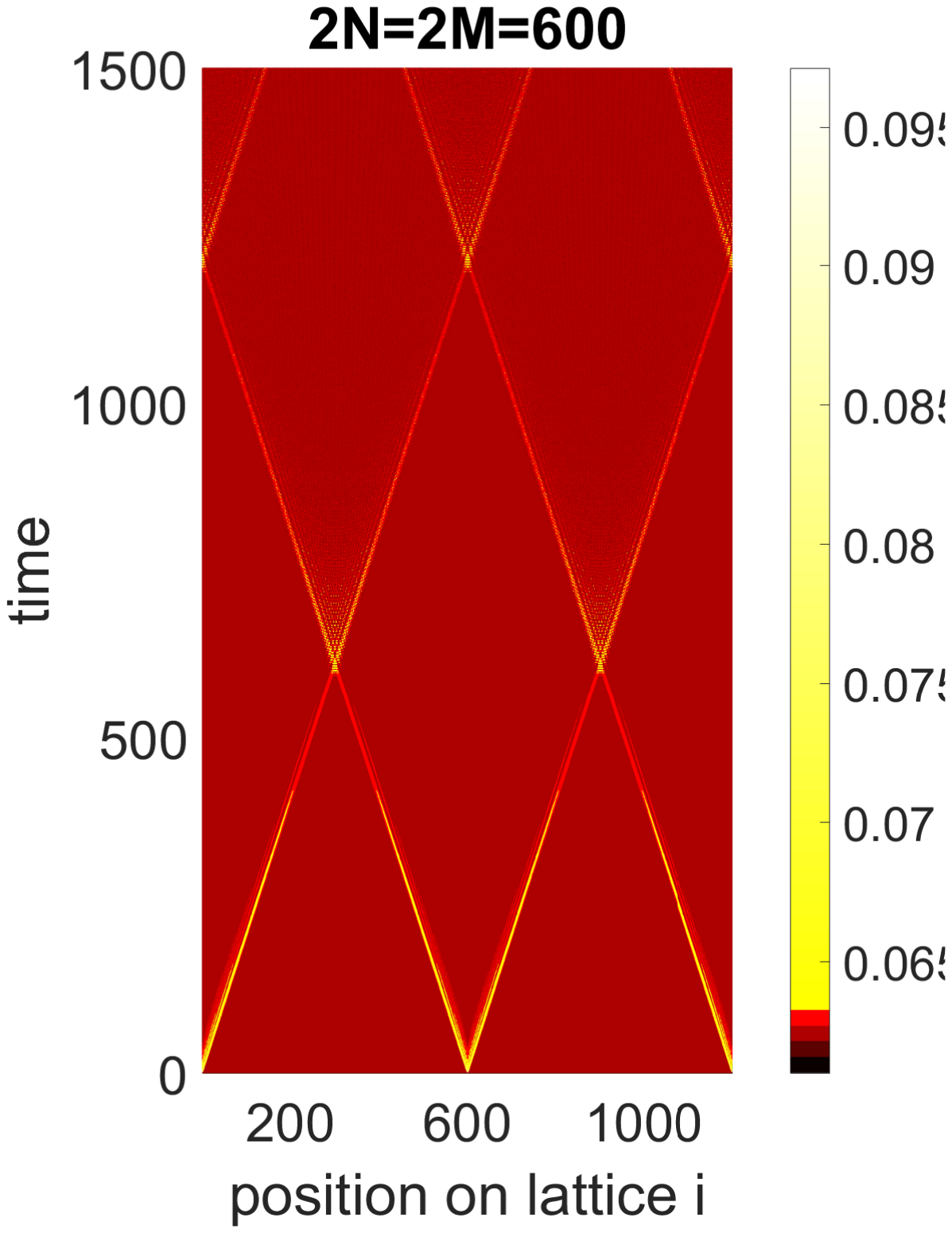}
        }
        \caption[Expectation of number operator(color intensity) and position along the X axis and time along Y-axis.]{This plot shows the expectation of number operator (color intensity) as a function of the position of lattice (x-axis) and time (y-axis) for the four cases: (a) $2N = 2M = 300$, (b) $2N = 2M = 400$, (c) $2N = 2M = 500$, (d) $2N = 2M = 600$.}\label{conebreak}
    \end{figure}
    
    To understand the time-delay for the quench to reach a lattice site, in \ref{conebreak},  we plot the expectation of number operator (as a color intensity) as a function of the lattice position (in the x-axis) and time (in the y-axis). The color intensity map shows how the disturbance travels along the quenched lattice in time. We observe the following features: First, number density peaks propagate along the lattice with constant speed. The figure also shows the formation of a {\it light cone} (in analogy to causal propagation) which marks the existence of maximum speed for the information propagation. This result is consistent with the Lieb-Robinson bound~\cite{lieb1972finite} for short-range interactions which give a theoretical limit for the speed of propagation of information in non-relativistic quantum systems. Second, the fluctuations are caused by the interference between different light cones due to finite size effect and periodic boundary condition. Once the effect of quench reaches the lattice, it goes out of equilibrium and then equilibrates to an almost steady state which matches with the GGE value but later starts to fluctuate due to the finite size effect. As the lattice size increases, the fluctuations from the average value reduces. In the thermodynamic limit, we then expect the system to equilibrate to the GGE value thus verifying generalized relaxation.

    \begin{figure}[h]
        \centering
        \subfloat[]{
            \includegraphics[width=0.5\columnwidth]{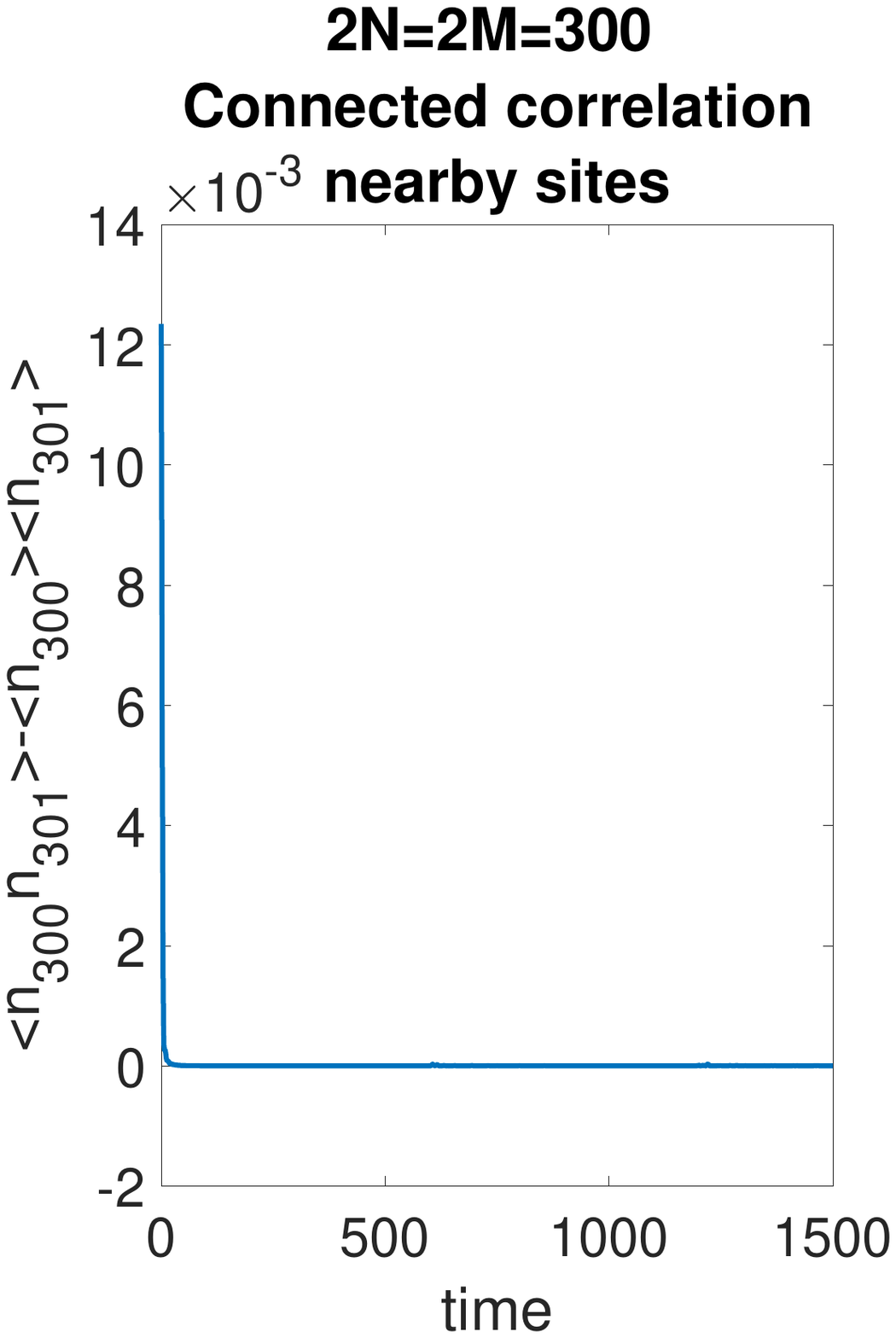}
        }
        \subfloat[]{
            \includegraphics[width=0.5\columnwidth]{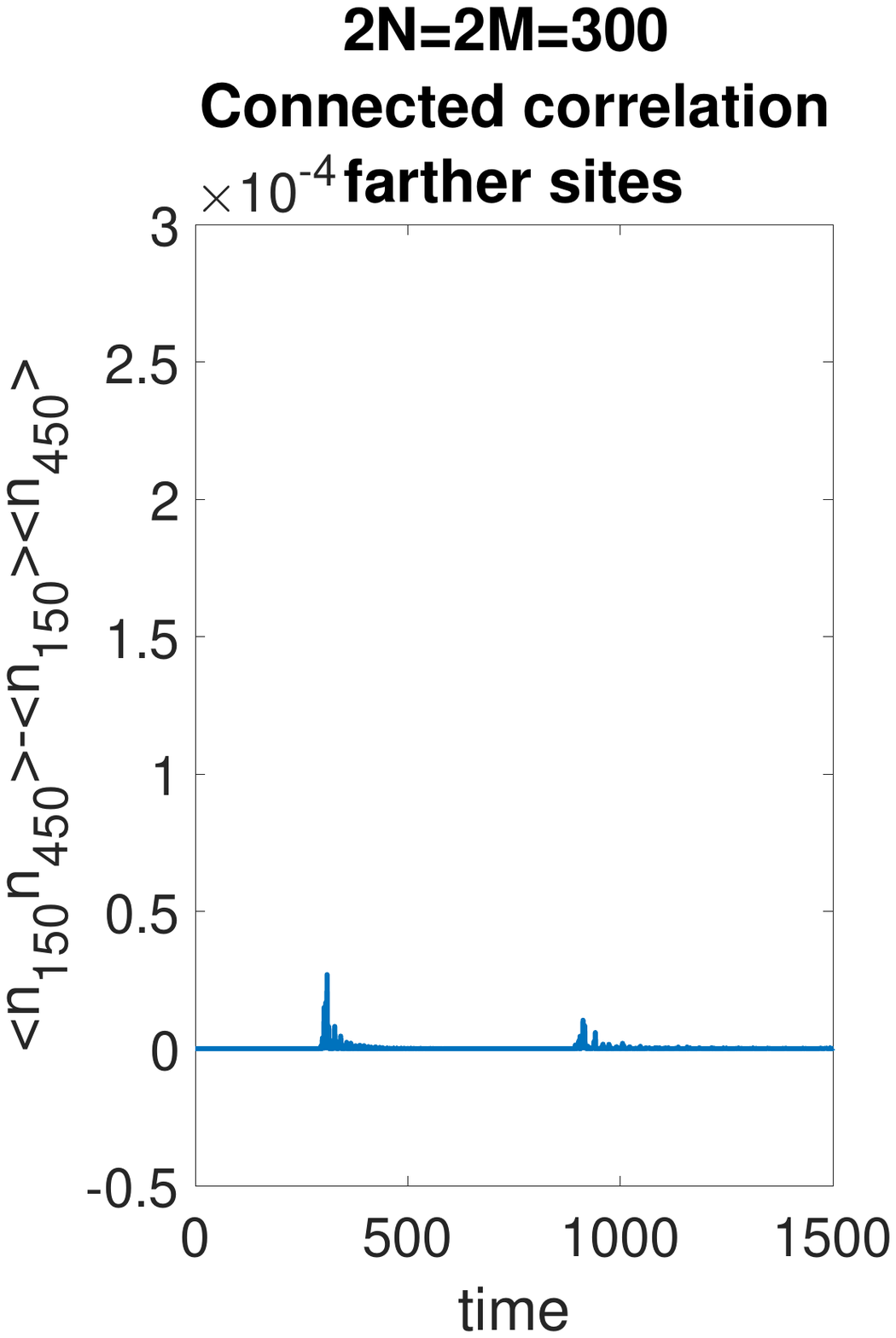}
        }
        \caption[Connected correlation $\langle n_{i} n_{j}\rangle- \langle n_{i}><n_{j}\rangle$ against time]{Connected correlation $\langle n_{i} n_{j}\rangle- \langle n_{i}\rangle\langle n_{j}\rangle$ between two sites each in the disconnected chains after quench. Consider two cases where i and j are the two nearby sites before breaking and the case where they are far apart.}\label{corrbreak}
    \end{figure}

    In \ref{corrbreak}, connected correlation $\langle n_{i} n_{j}\rangle- \langle n_{i}\rangle\langle n_{j}\rangle$ between two sites each in the disconnected chains after 
    quench is plotted. We consider the two cases where $i$ and $j$ are the two nearby sites before breaking and the case where they are far apart. In the first case, the connected correlation is large when the quench happens and then decays to zero since the two sites are in independent chains after the quench. When the two sites are far away from each other and the quench, the connected correlation is minimal initially and then goes to zero after the quench. Any fluctuation which might occur at a later time is due to the finite size.
    
    \begin{figure}[h]
        \includegraphics[width=0.6\columnwidth]{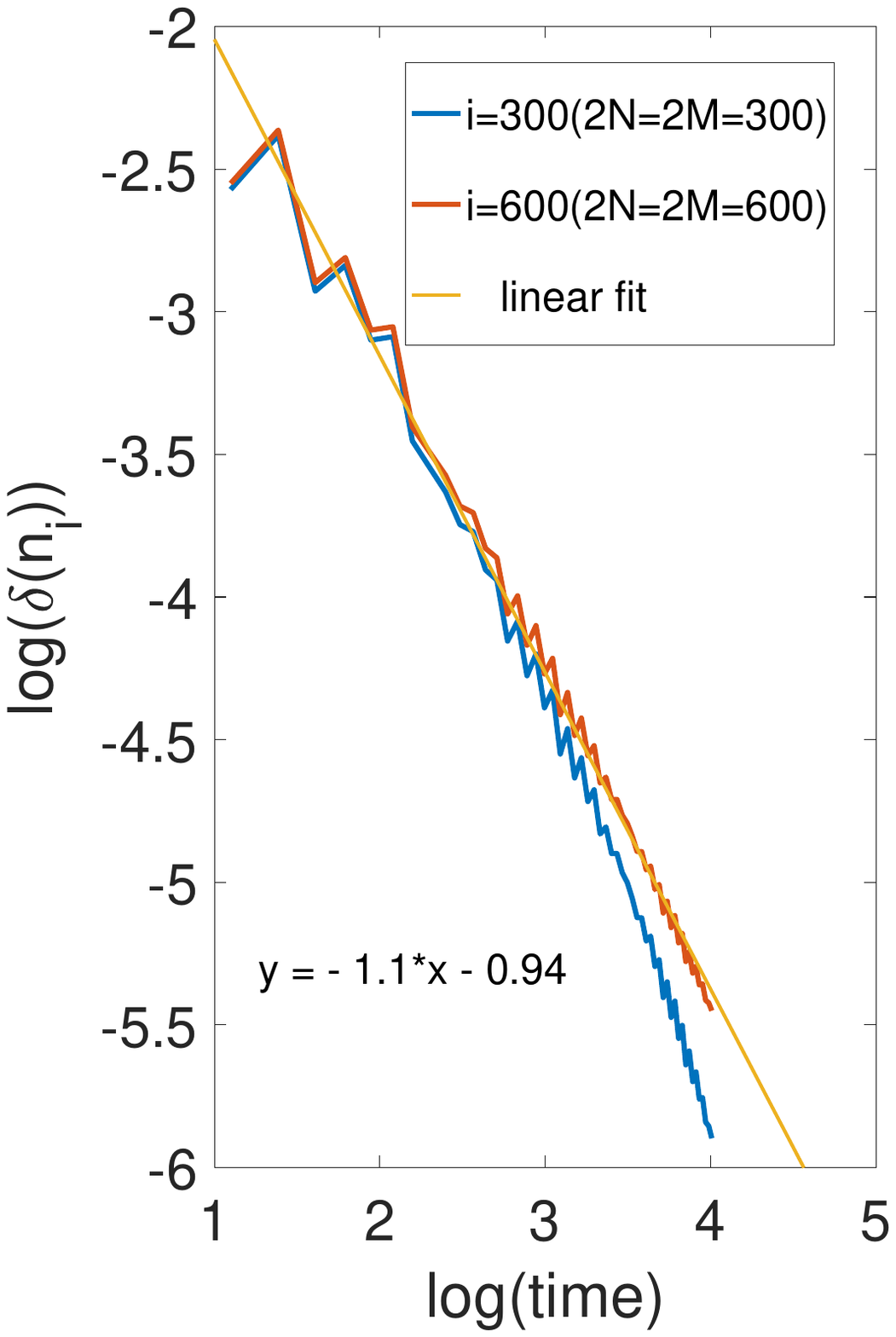}
        \caption[$log(\delta(n_{i}))=log(\langle n_{i}(t)\rangle-\langle n_{i}\rangle_{GGE})$ plotted against time]{The Logarithm of the relative deviation of the lattice occupation number $log(\delta(n_{i}))$ is plotted against time in the log scale for two cases: $2N = 2M = 300$ and $2N = 2M = 600$. We see that the relaxation to GGE goes as power law with exponent approximately $-1$.}\label{decaybreak}
    \end{figure}
    
    To see how the expectation value of the number operator at a site equilibrates to GGE, in \ref{decaybreak}, we plot the logarithm of $\delta n_{i}(t)$ defined in Eq. \ref{eq:rel-deviation} against logarithm of time. For immediate comparison, the figure contains a linear plot with coefficient $-1$. Following points are worth noting regarding the above figure: First, as the lattice size increases, $\delta n_i(t)$ decays as a power-law with exponent close to $-1$. Second, the exponent $-1$ is indicative of ballistic behavior rather than diffusive behavior where the exponent needs to be -0.5~\cite{rademaker2017quantum}. Classically, ballistic behavior arises due to the collisionless transport of particles whereas the system under consideration is composed of non-interacting quasiparticles, mimicking the classical behavior.
    
    \subsection{Nearest neighbor hopping}
    Another local observable that we have studied is the nearest neighbor hopping in the real lattice chain defined as $\langle a_{i}^{\dagger}(t) a_{i+1}(t) + a_{i+1}^{\dagger}(t) a_{i}(t) \rangle$. Here also we have fixed the parameters $h/J=-2$ and the nearest neighbor interaction $J$ to be $0.5$. The initial state is again a thermal state with the lattice chain at temperature $T/J=0.5$. 
    
    \begin{figure}[h]
    	\subfloat[]{
    		\includegraphics[width=0.5\columnwidth]{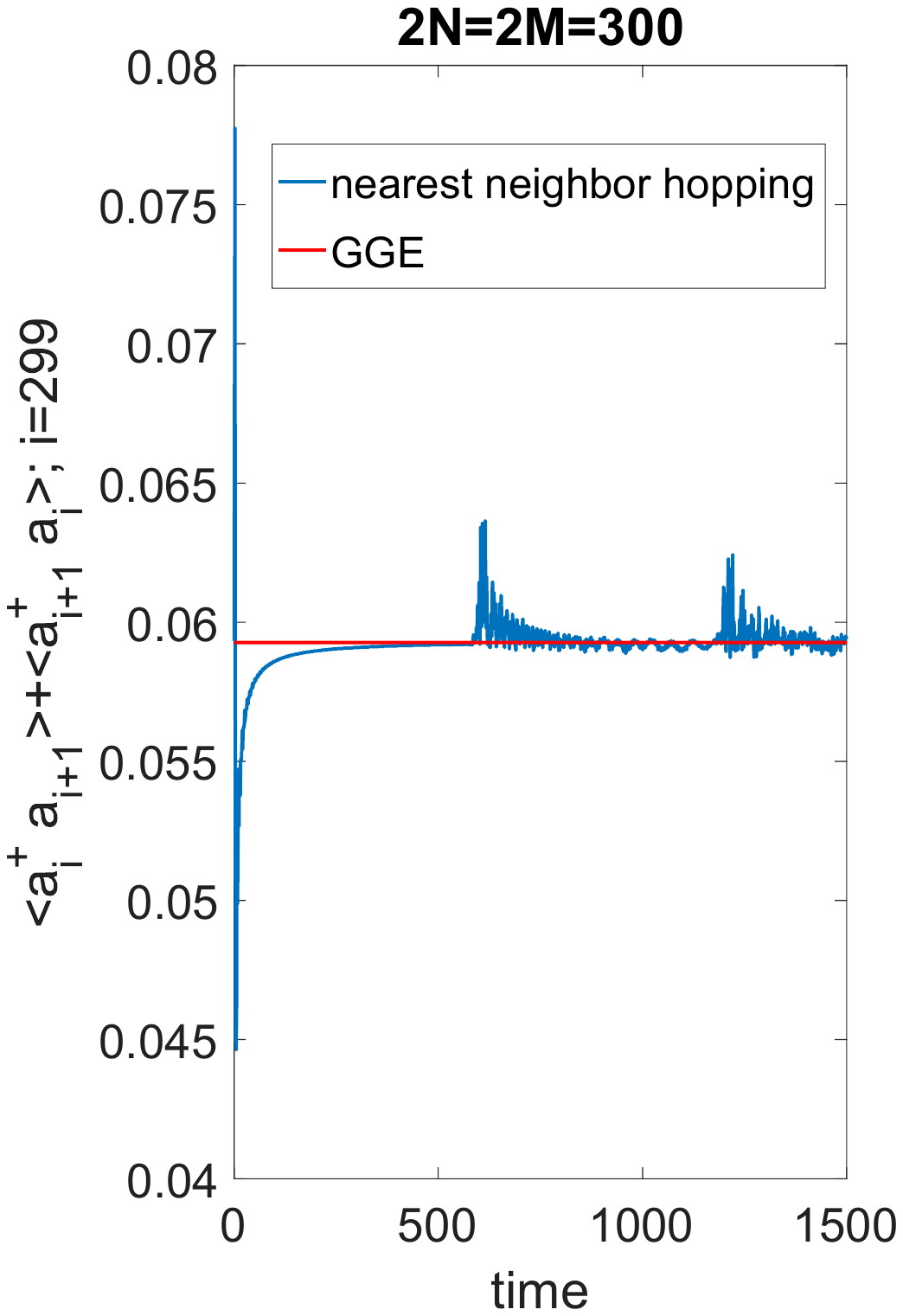}
    	}
    	\subfloat[]{
    		\includegraphics[width=0.5\columnwidth]{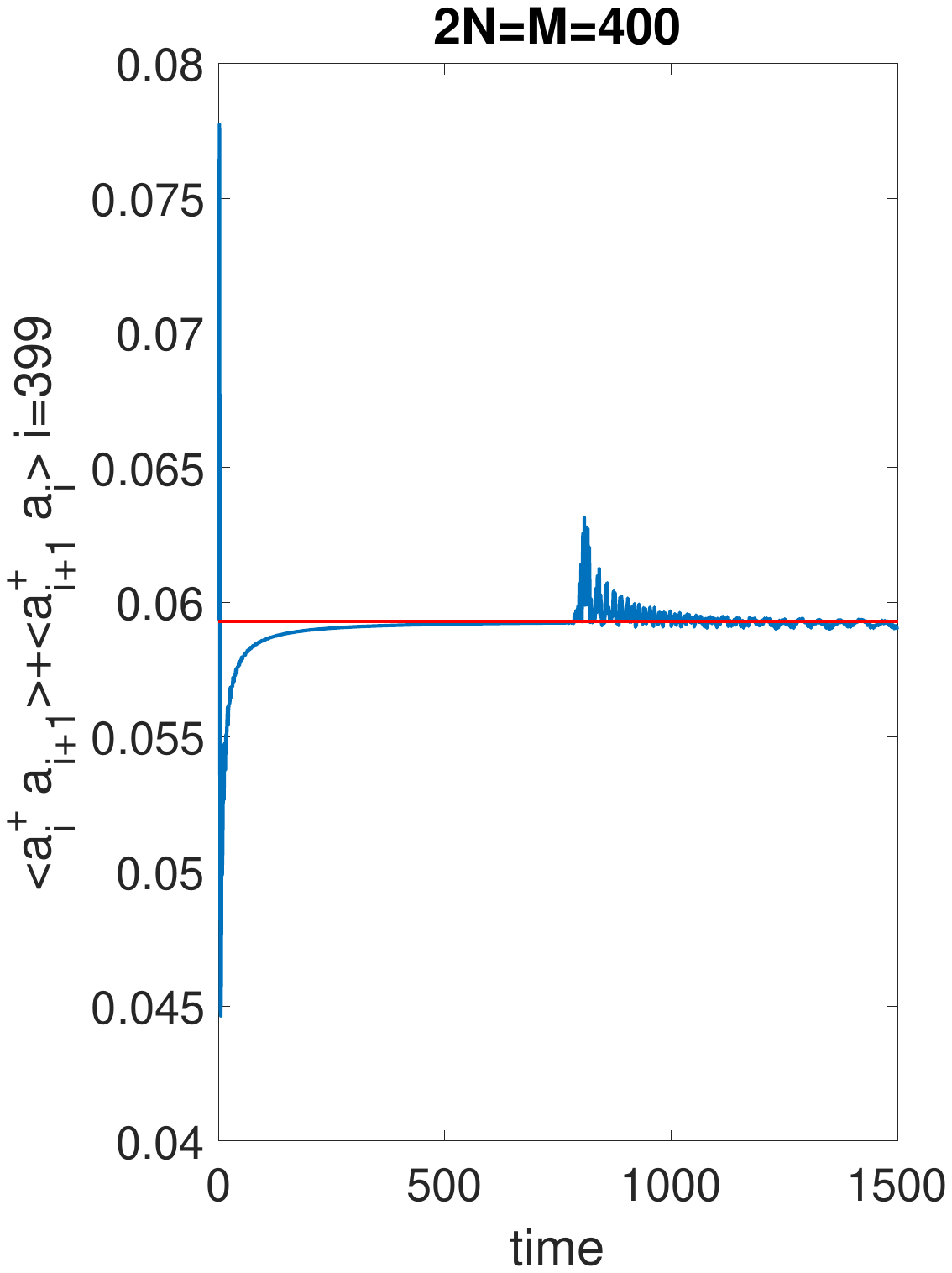}
    	}
    	\hspace{1mm}
    	\subfloat[]{
    		\includegraphics[width=0.5\columnwidth]{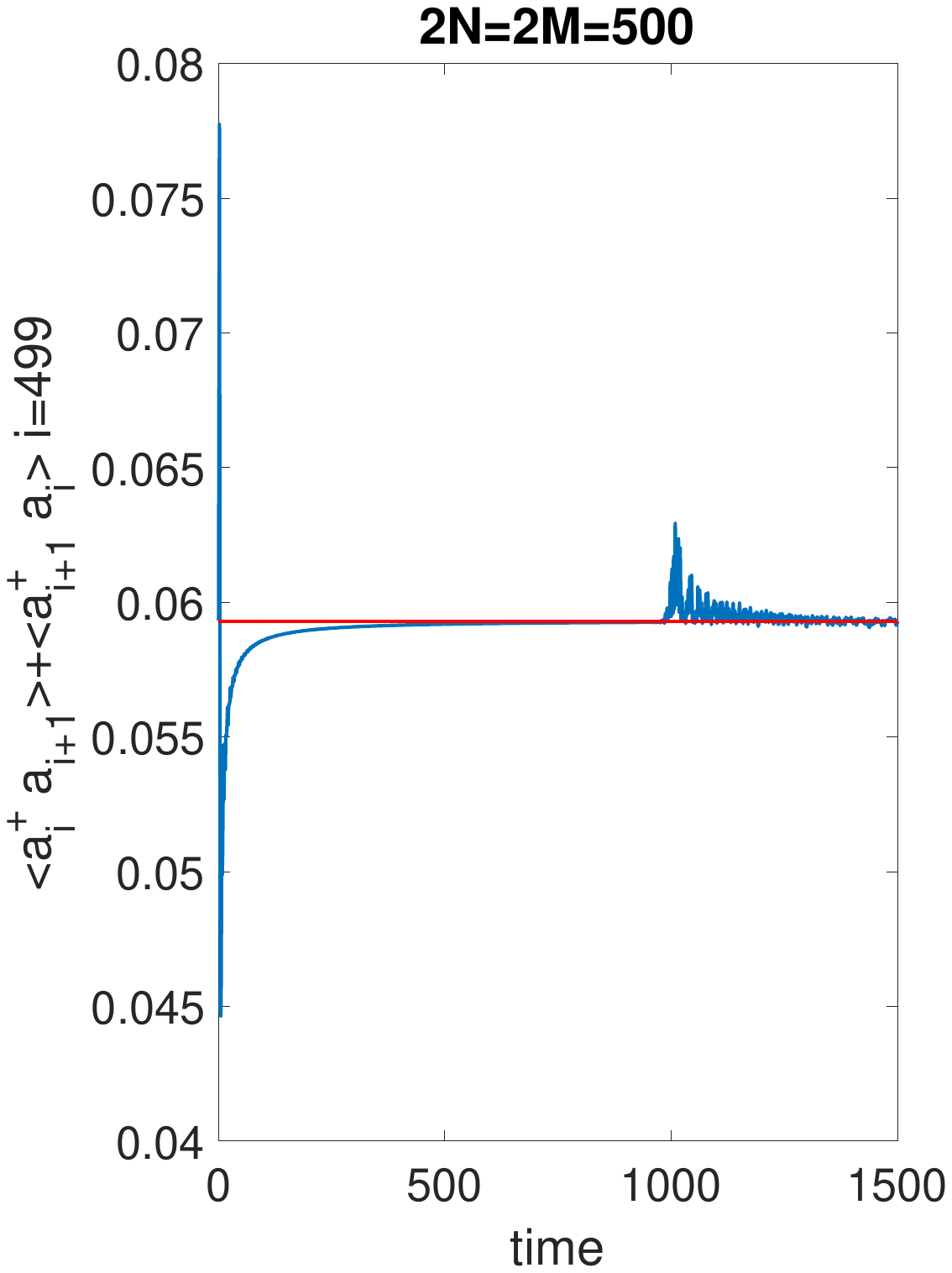}
    	}
    	\subfloat[]{
    		\includegraphics[width=0.5\columnwidth]{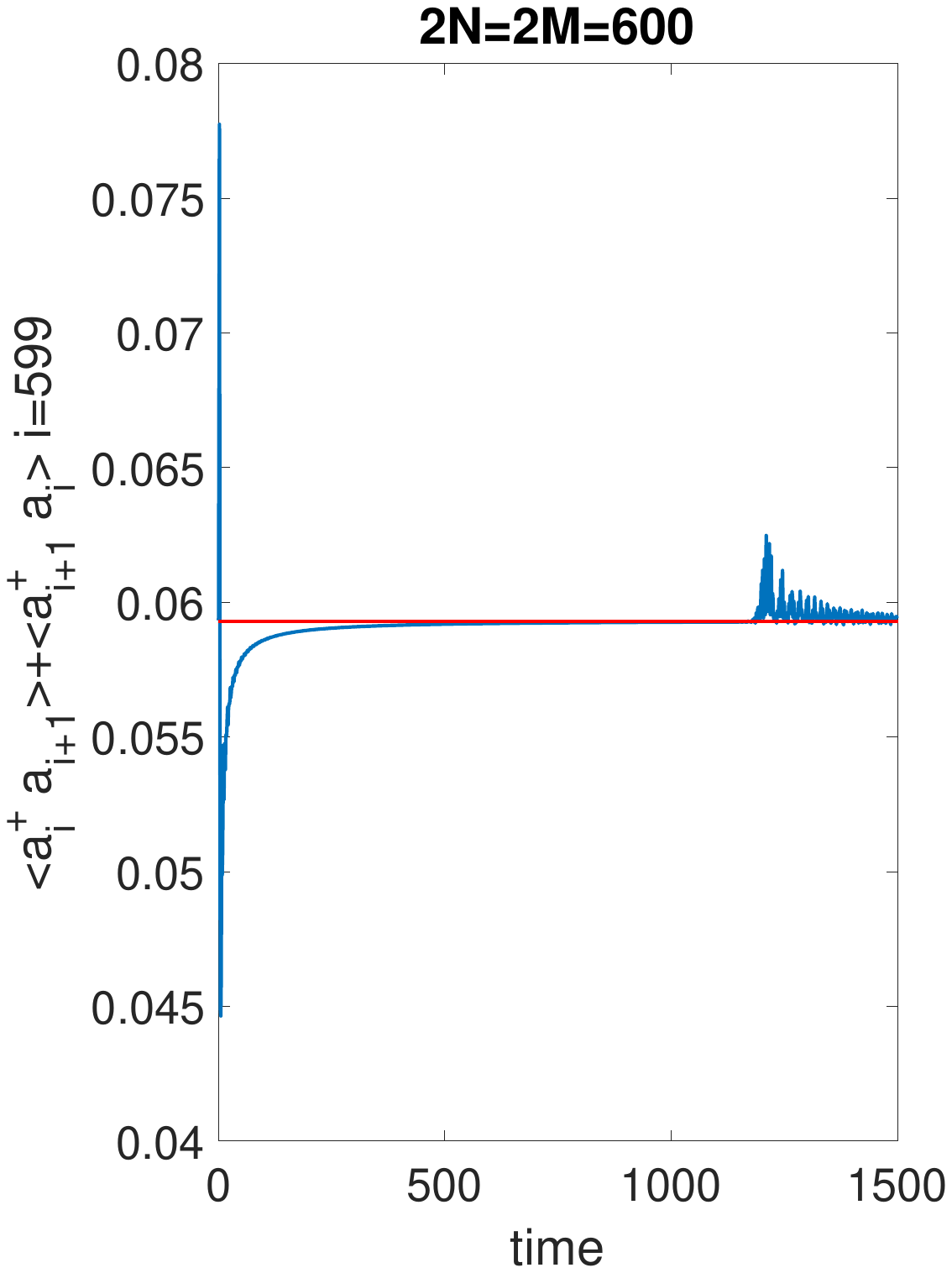}
    	}
    	\caption[Nearest neighbor hopping at a site near quench plotted against time.]{The nearest neighbor hopping at a site near quench is plotted against time where (a) $2N = 2M = 300$, (b) $2N = 2M = 400$, (c) $2N = 2M = 500$, (d) $2N = 2M = 600$. The GGE value is also plotted.}
    	\label{nnh}
    \end{figure}
In \ref{nnh}, we plot the nearest neighbor hopping at a lattice site $i$ near to the site of quench as a function of time. 

As in the previous case of the number operator, we can observe that until the effect of quench reaches the particular site $i$, the system remains in the initial thermal state and as soon as the effect of quench reaches the site, the value fluctuates, and the system goes out of equilibrium. The fluctuations also decay in time, and the nearest neighbor hopping operator equilibrates to the GGE value. Also as the lattice size increases, the fluctuations become smaller, and the system tends to come closer to the GGE value, with sparse recurrences. 

\begin{figure}[h]
	\subfloat[]{
		\includegraphics[width=0.5\columnwidth]{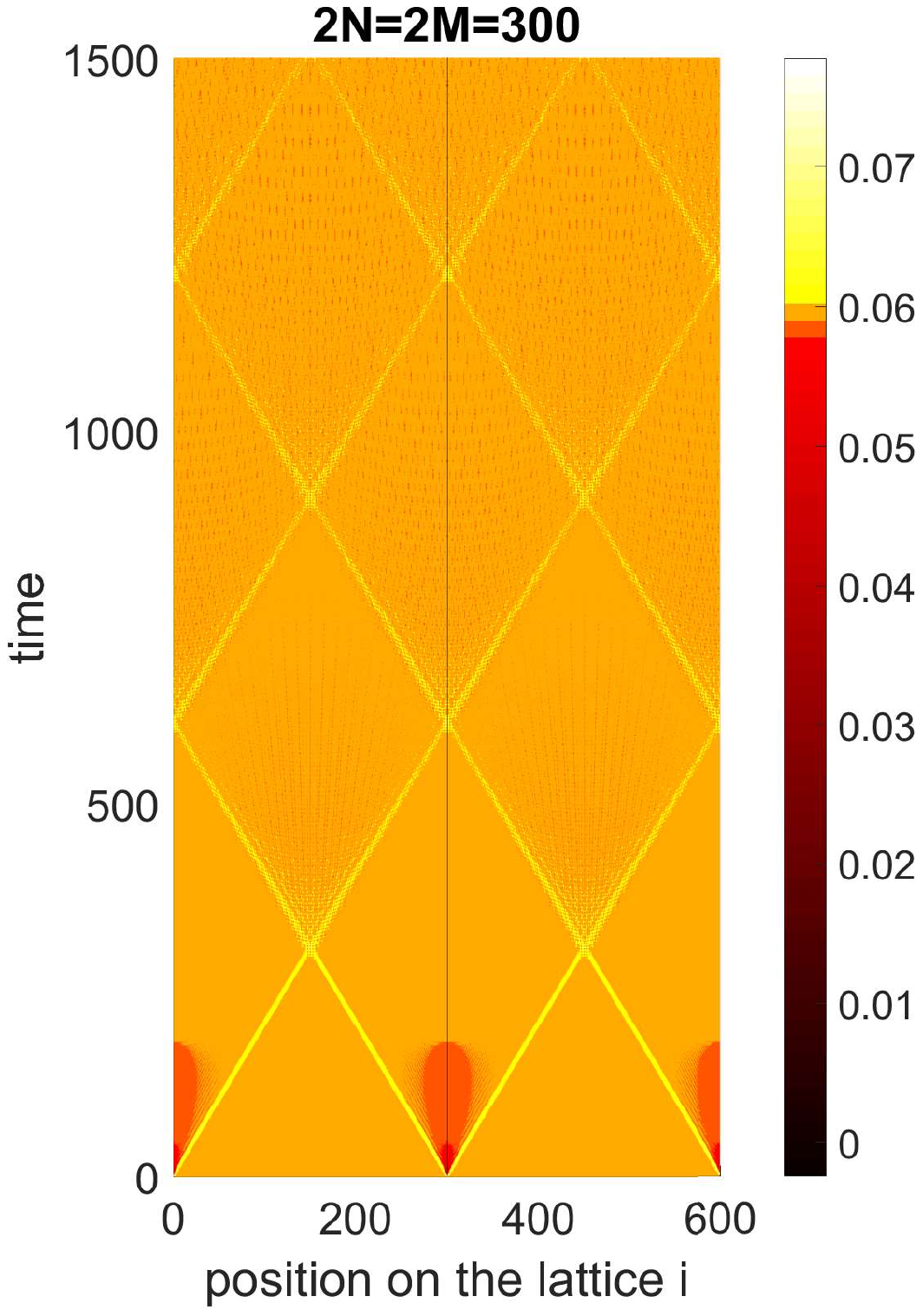}
	}
	\subfloat[]{
		\includegraphics[width=0.5\columnwidth]{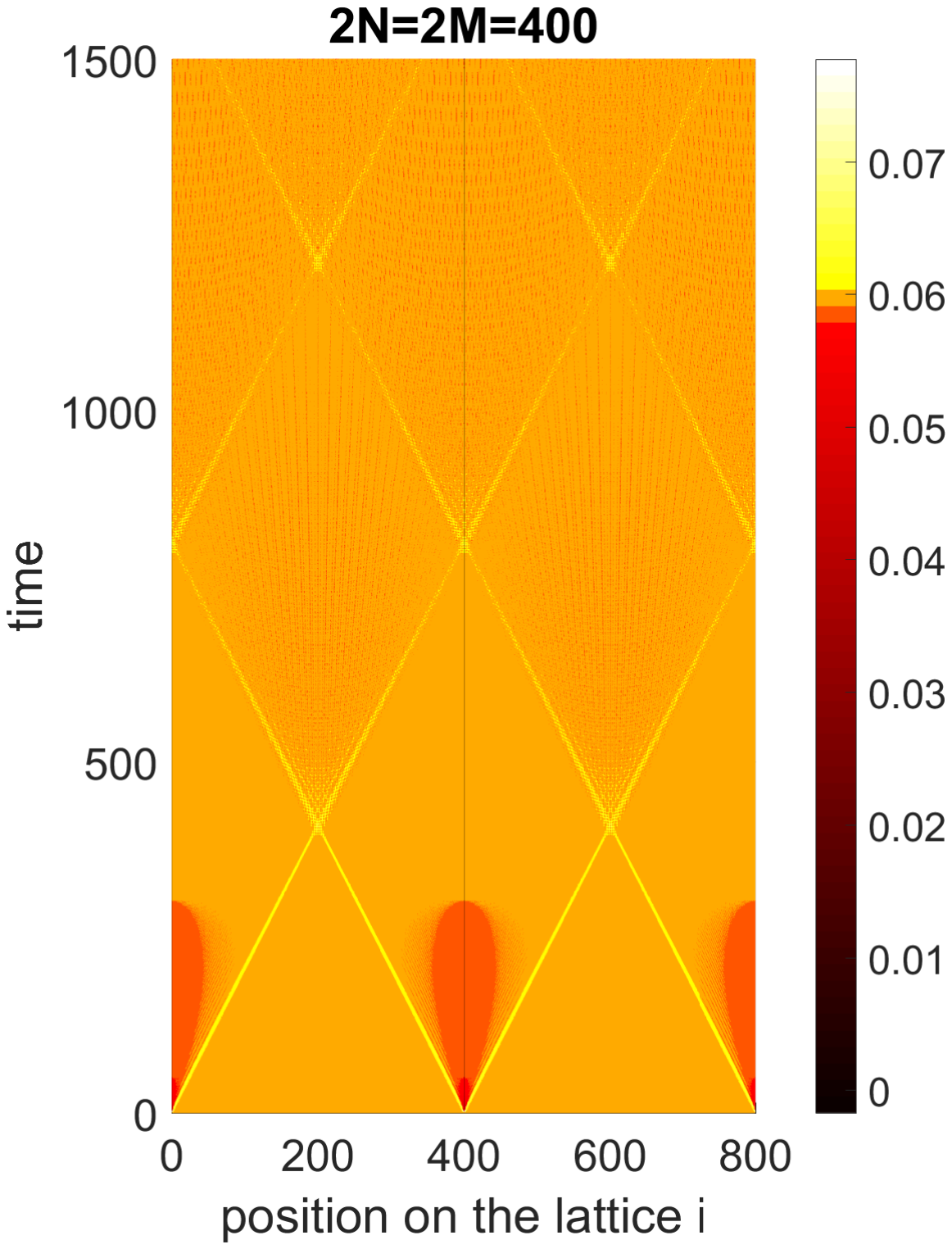}
	}
	\hspace{1mm}
	\subfloat[]{
		\includegraphics[width=0.5\columnwidth]{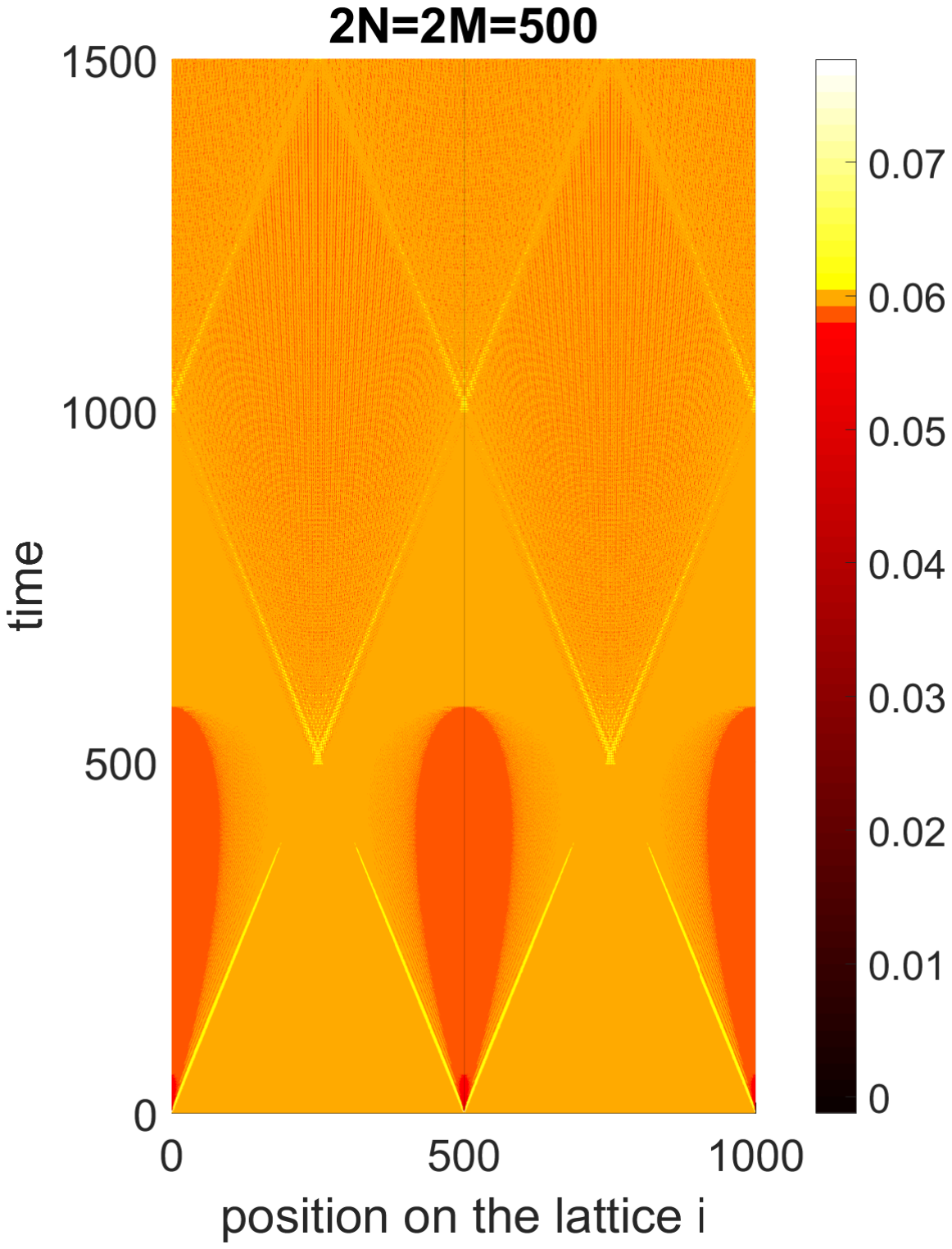}
	}
	\subfloat[]{
		\includegraphics[width=0.5\columnwidth]{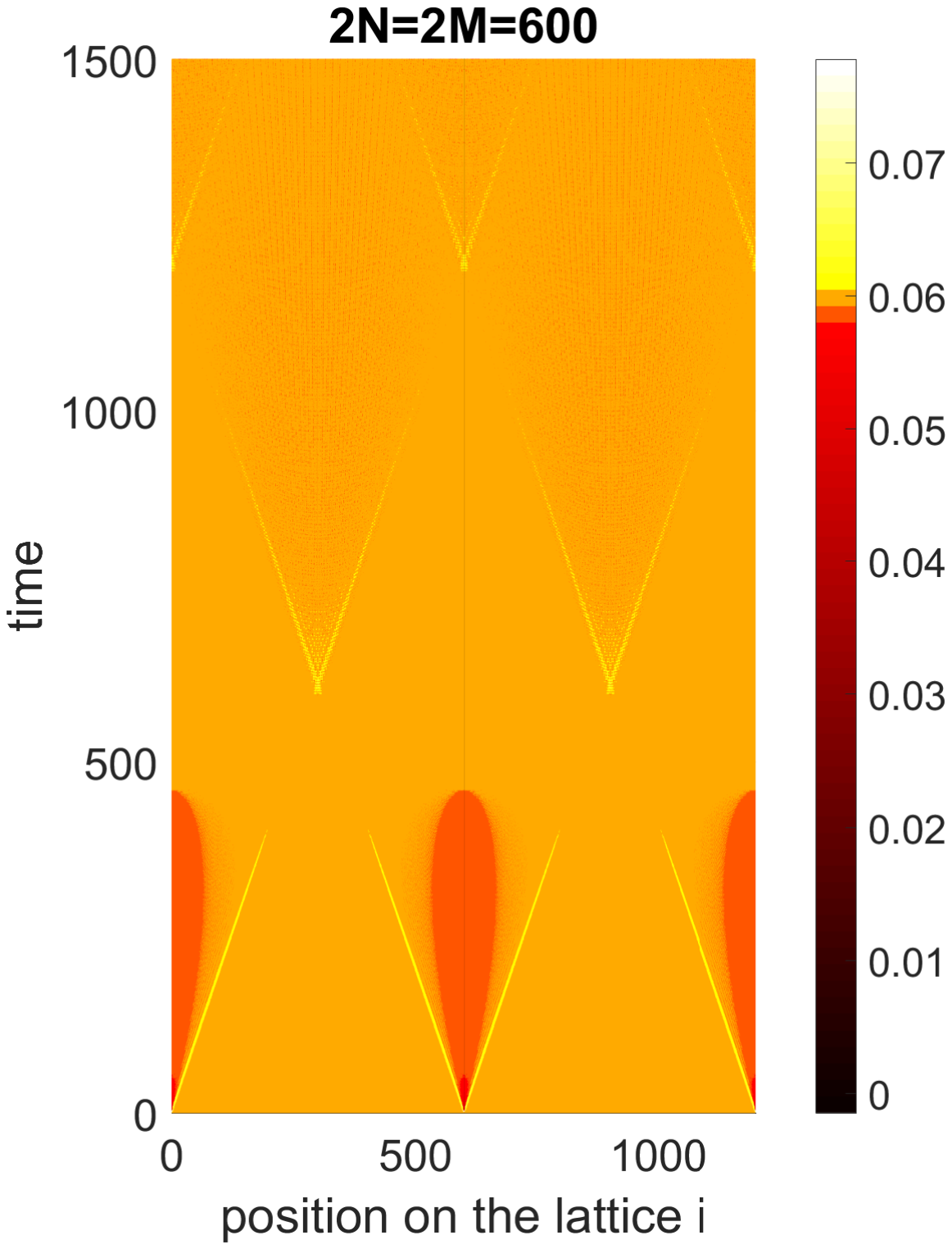}
	}
	\caption[Nearest neighbor hopping(color intensity) and position along the X axis and time along Y-axis.]{This plot shows the nearest neighbor hopping (color intensity) as a function of the position of lattice (x-axis) and time (y-axis) for the four cases: (a) $2N = 2M = 300$, (b) $2N = 2M = 400$, (c) $2N = 2M = 500$, (d) $2N = 2M = 600$.}
	\label{nnhcone}
\end{figure}

In figure \ref{nnhcone} we plot the nearest neighbor hopping (color intensity) as a function of the lattice position (in the x-axis) and time (in the y-axis). Here again we observe all the features observed for the case of expectation value of the number operator and in the thermodynamic limit, we then expect the system to equilibrate to the GGE value thus verifying generalized relaxation.

    \subsection{Information content in bits per Fermion} 
    
    \begin{figure}[h]
        \centering
        \subfloat[]{
            \includegraphics[width=0.5\columnwidth]{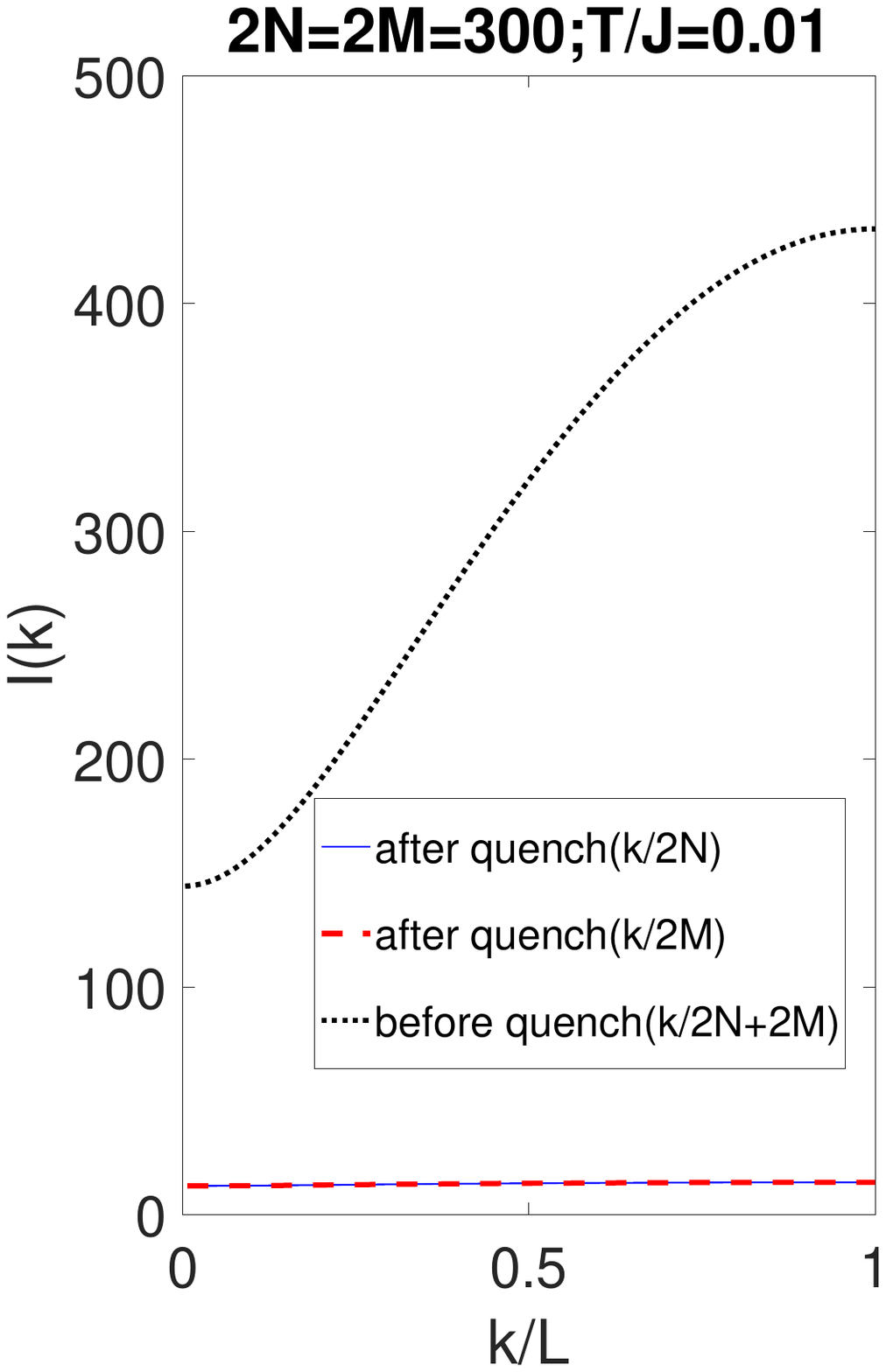}
        }
        \subfloat[]{
            \includegraphics[width=0.5\columnwidth]{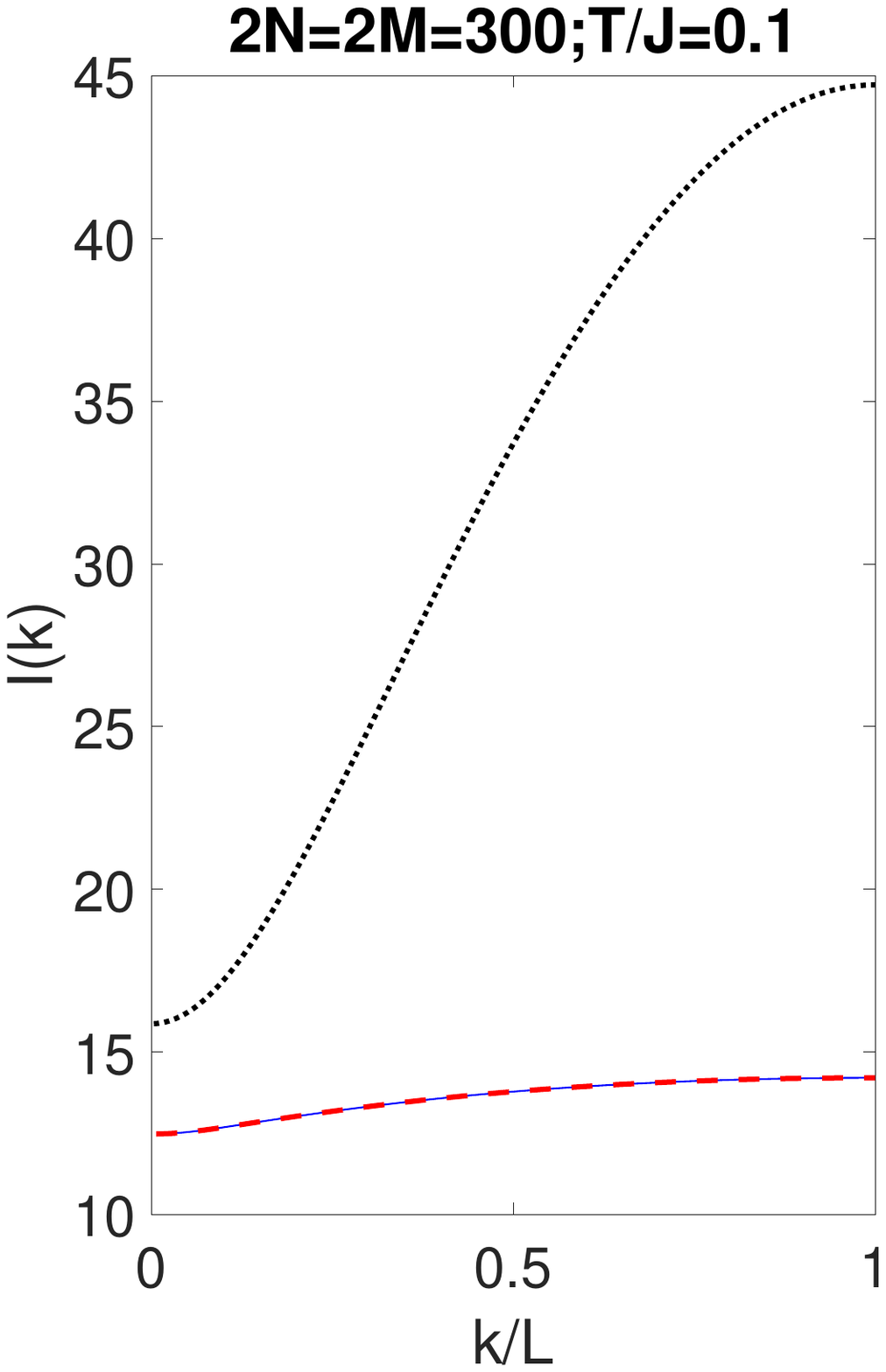}
        }
        \hspace{1mm}
        \subfloat[]{
            \includegraphics[width=0.5\columnwidth]{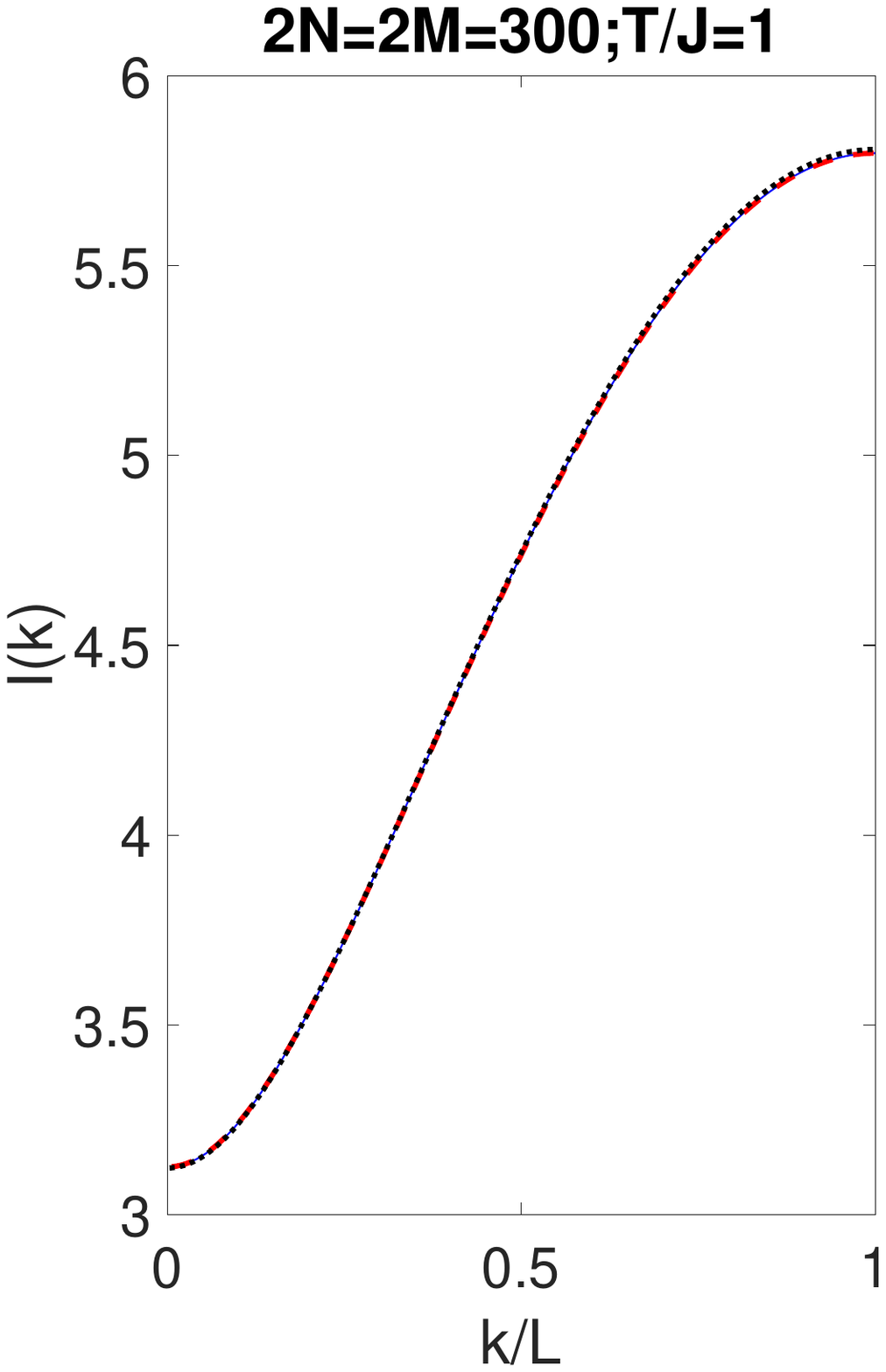}
        }
        \subfloat[]{
            \includegraphics[width=0.5\columnwidth]{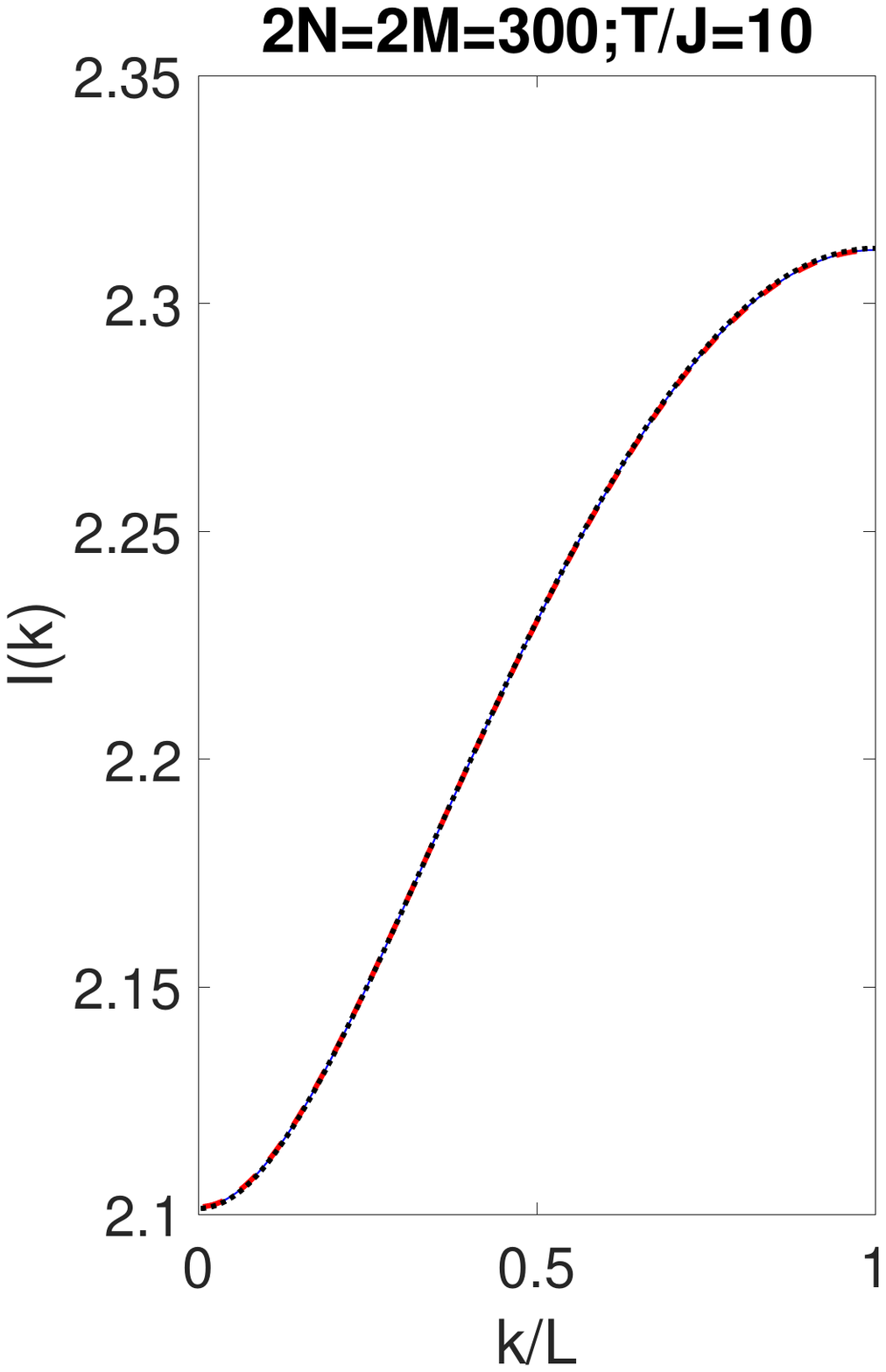}
        }
        \caption{Plot of $I(k)$ against $k/L$ where (a) $2N = 2M = 300$; $T/J=0.01$, (b) $2N = 2M = 300$; $T/J=0.1$, (c) $2N = 2M = 300$; $T/J=1$, (d) $2N = 2M = 300$; $T/J=10$.}\label{bitsbreak}
    \end{figure}

    In the previous subsection, to evaluate the occupation number in a lattice site, we fixed 
    $J = 0.5$ and $\beta_I = 4$. To obtain information about the response of the system for different parameters, in \ref{bitsbreak}, we plot the information content in bits per Fermion $I(k)$ for each normal-mode $k/L$ (where $L$ is the length of the lattice chain and $k = 1,2, \cdots$) before and after the quench for different $T/J$ parameters. 
    Note that we have fixed the values of $J$ and $h/J$ to be $0.5$ and $-2$, respectively. 
    
    We see two distinct features: When $T/J$ is high, corresponding to a high temperature initial thermal state, $I(k)$ before and after the quench almost overlaps. However, when the temperature is low, say at $T/J=0.01$, the profile shows a different trend. We thus infer the following: First, for the low-temperature initial state, the information content per Fermion after the quench is distributed evenly to all the normal modes in contrast to before the quench distribution. Second, the information content per Fermion in each mode after quench is smaller compared to before the quench. However, the total entropy of the system increases after the quench, consistent with the second law of thermodynamics. Third, it implies that the initial state of the system before the quench for small $T/J$ and large $T/J$ is not identical. 
    
    To further investigate this, we calculate fidelity which is a measure of closeness or overlap between two quantum states \cite{Zanardi2006,Vieira2010}. Fidelity between two density matrices $\rho_{1}$ and $\rho_{2}$ can be written as:
    \begin{equation}
    F(\rho_{1},\rho_{2})=\sum_{i} \sqrt{ p_{i} q_{i} }
    \end{equation}
    where $p_i$ and $q_i$ are the eigenvalues of two density matrices, i. e.,
    \begin{equation}
    \rho_{1}=\sum_{i} p_{i} |i><i| \quad \text{and} \quad \rho_{2}=\sum_{i} q_{i} |i><i|
    \end{equation}
    for some orthonormal basis ${|i>}$. Note that we have taken the situation where 
    the two density matrices can be simultaneously diagonalized by unitary matrices.
    
\begin{figure}[t]
        \includegraphics[width=0.7\columnwidth]{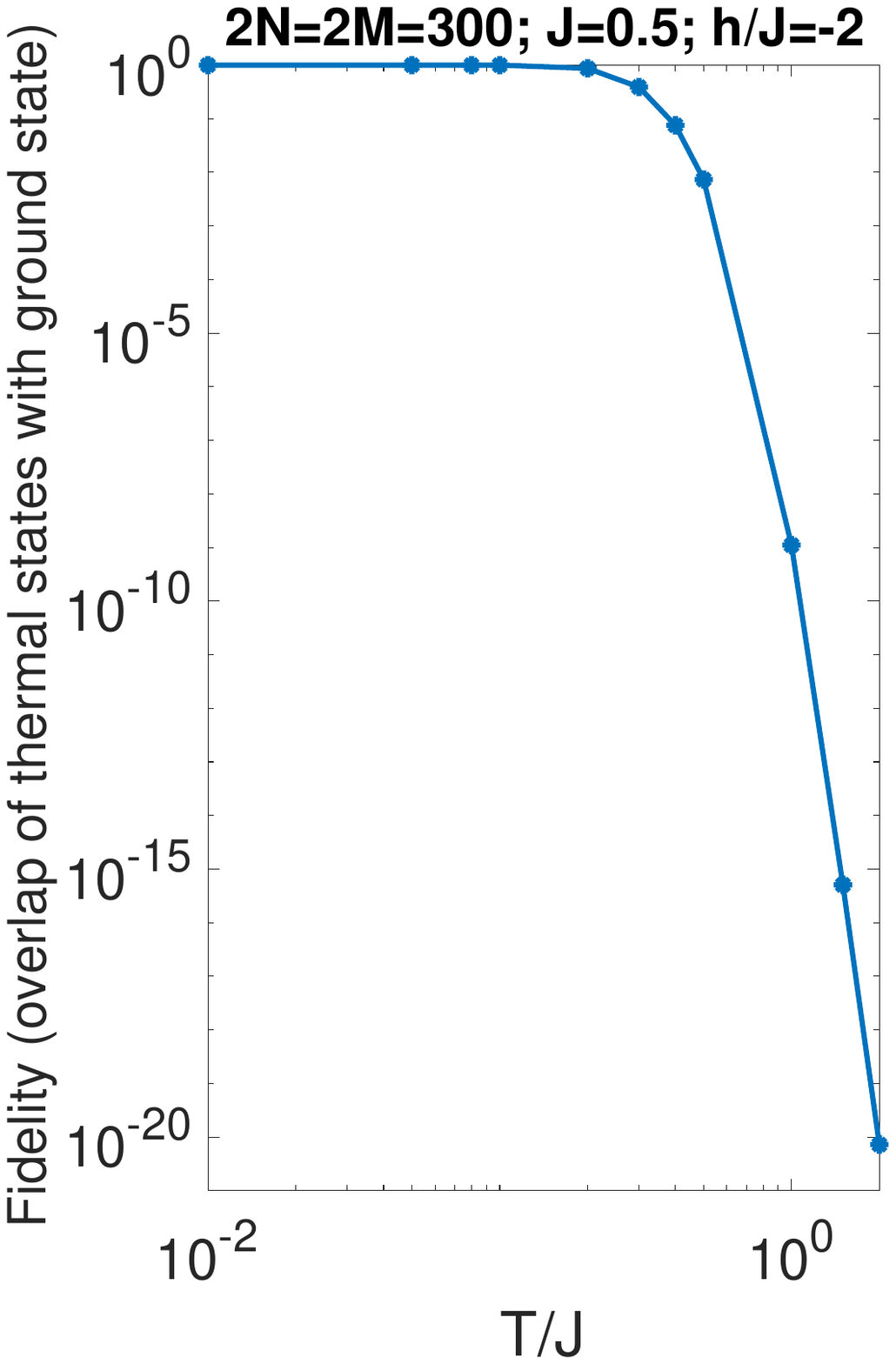}
        \caption{Plot of Fidelity of the initial thermal state (with ground state) as a function of temperature.}\label{fidelity}
    \end{figure}

    \ref{fidelity} contains the plot of the fidelity between $\rho_{T}$ and $\rho_{GS}$ --- between the initial thermal (with non-zero $T/J$) and ground state --- as a function of $T/J$. We see that as $T/J$ decreases, the fidelity goes closer to unity. Thus, the information content in bits per Fermion can be used as an indicator to identify the driving term of the dynamics  in the thermal background \cite{subir,carr2011}.

    \begin{figure}[!h]
        \centering
        \subfloat[]{
            \includegraphics[width=0.5\columnwidth]{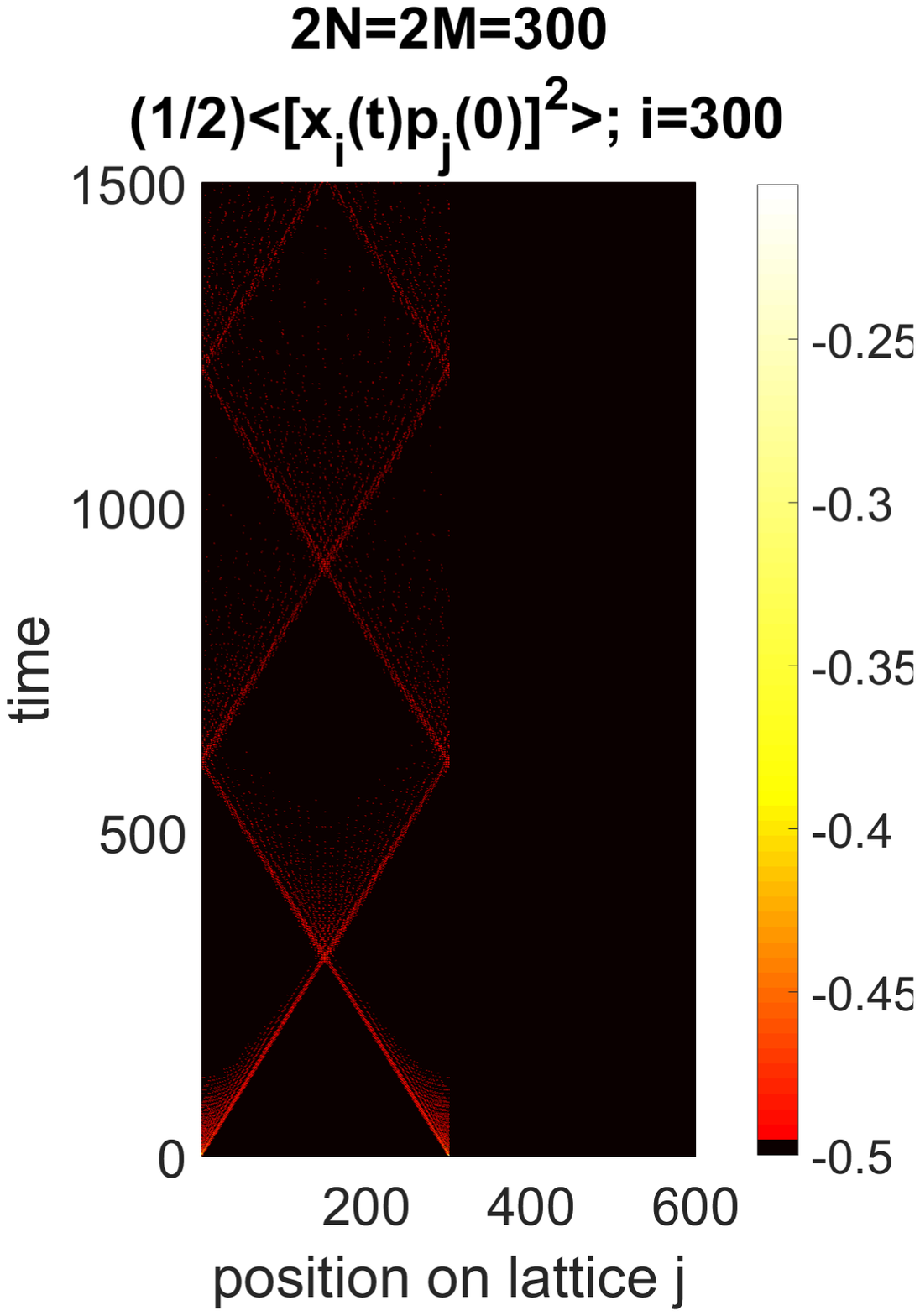}
        }
        \subfloat[]{
            \includegraphics[width=0.5\columnwidth]{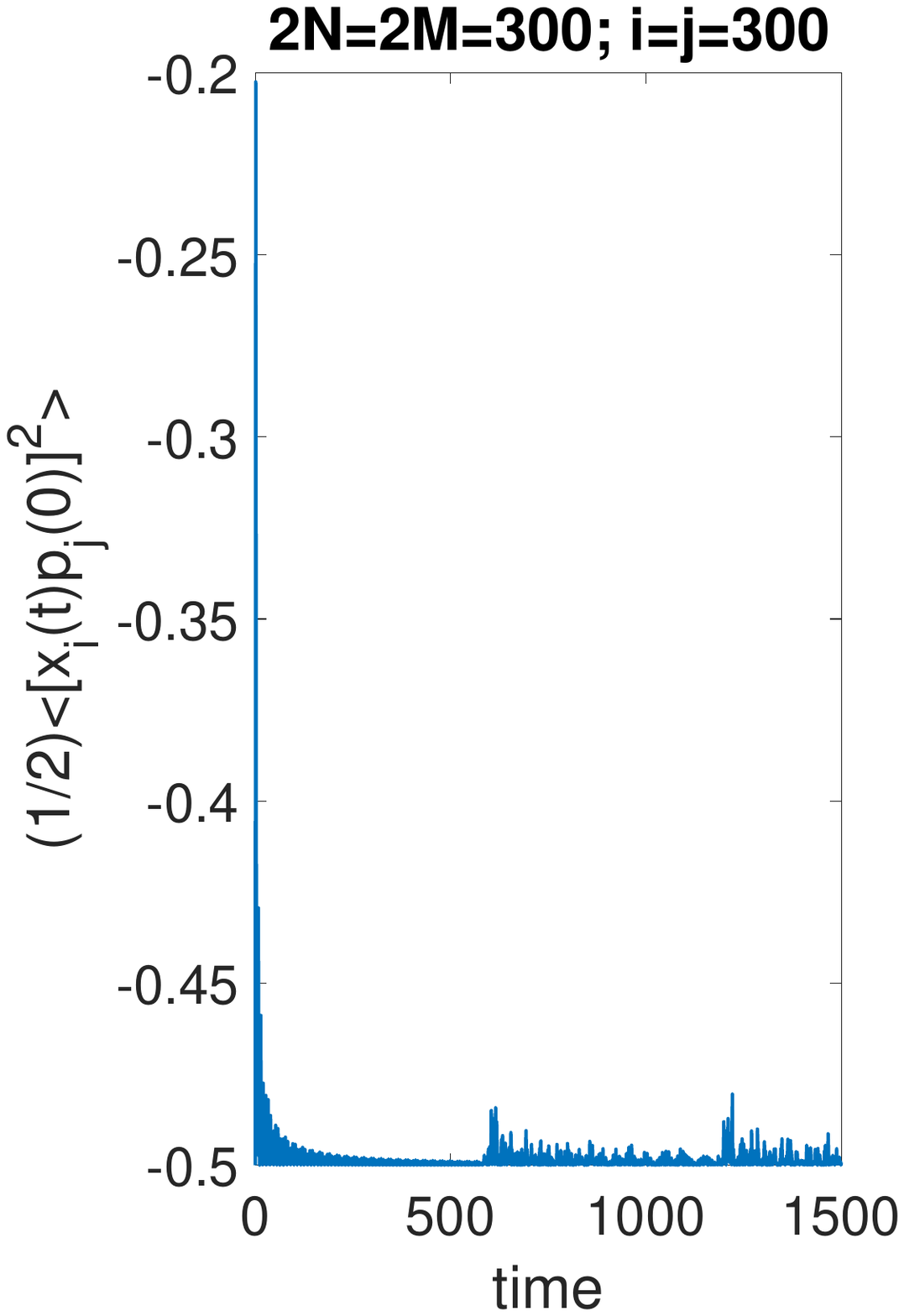}
        }
        \caption{OTOC $F_{ij}(t)=\frac{1}{2}\langle[\hat{x}_i(t),\hat{p}_j(0)]^2\rangle$ is calculated for the chain of size $2N + 2M = 600$. At $t=0$ the quenching (chain breaking) is done, so subequently the two chain segments are causally disconnnected.  A point in one segment will be affected by the initial condition in that segment only. (a) $F(t)$ (color intensity) is plotted as a function of position $j$ (along the x-axis) and time (along y-axis) for $i=300$ which is on the chain of size 2N after the quench. (b) $F_{ij}(t)$ is plotted as a function of time for $i=300$ and $j=300$ ($i=j$).}\label{otoc}
    \end{figure}

\subsection{OTOC}

As mentioned in the previous section, thermalization can be associated with the loss of accessible information (or scrambling) and OTOC is considered a good diagnostic for the strength of scrambling. We evaluate OTOC, given in Eq.~\ref{eq:OTOC-def}, for the model Hamiltonian \ref{ham} .

\ref{otoc} is the plot of OTOC as a function of time and lattice position. Plot (a) contains 
$F(t)$ (color intensity) as a function of position $j$ (along the x-axis) and time (along y-axis) for the lattice chain of size $2N + 2M = 600$ and $i=300$ which is on the disconnected chain size $2N$ after the quench. From the plot, we observe the following:  First, in the case of $i \neq j$, the time evolution of OTOC is present only when $j$ is in the same broken chain as $i$.  Second, the evolution of the light cone is just one half of the entire initial chain of size $2N+2M$. OTOC ascertains that the correlations between the two chains vanish with the quench. Third,  the long-time value of $F_{ij}(t)$ moves towards $-0.5$ which  equals $\frac{1}{2}\langle[\hat{x}_{i}(0),\hat{p}_{j}(0)]^2\rangle$.

The initial value of OTOC i.e., $F_{ij}(0)=-0.5$ for any $i,j$ since the operators are Fermionic in nature. From the OTOC analysis (see  \ref{otoc}) we can see that OTOC parameter, after being affected by the quench,  decays with time. Theoretically, it indicates that due to some initial perturbation at $j-$th site, the effect on the configuration at $i-$th site (recalling $[\hat{x}_i(t),\hat{p}_j(0)]$ measures $\delta x_i(t)/\delta x_j(0)$) settles over time, for any pair $(i,j)$, to the value $-0.5$ indicating  that at long time, the system equilibrates. In the \ref{otoc}, the long time average value obtained is $-0.4983$ which is close to $-0.5$. Therefore, in the thermodynamic limit, the OTOC is expected to settle down to steady value to $-0.5 $ indicating homogeneity.

\section{Analytical understanding}\label{analytic}
The results in the previous section are exact for the model Hamiltonian \ref{ham}.  Since it is cumbersome to obtain an analytical expression for these observables and to have a better understanding of the features of the above model, we now analytically study a qualitatively similar yet simpler model namely the tight-binding model \cite{Mahan2013book}. In the rest of this section, we explicitly write down the analytic expressions for the three observables in this model and discuss the essential features. 

Hamiltonian of the tight-binding model is 
\begin{equation}
H^{\rm TB} =-h \sum_{j=1}^{2N}   a_{j}^{\dagger} a_{j}   -  \dfrac{J}{2} \sum_{j=1}^{2N} 
\left( a_{j+1}^{\dagger}a_{j} + a_{j}^{\dagger} a_{j+1} \right)
\label{eq:Ham-TB}
\end{equation}
where $a_{j}^{\dagger}(a_{j})$ creates(annihilates) Fermion at lattice site $j$. Comparing the above Hamiltonian with \ref{ham}, it is clear that this model is number conserving.

This Hamiltonian with periodic boundary condition can be diagonalized by the Fourier transformation:
{\small
\begin{equation}\label{eq:FT-TB}
\hat{b}_k=\dfrac{1}{\sqrt{2N}}\sum_{j=1}^{2N}\hat{a}_{j} e^{\dfrac{2 \pi i j k}{2N}}\,; \,
\hat{b}^{\dagger}_k=\dfrac{1}{\sqrt{2N}}\sum_{j=1}^{2N}\hat{a}^{\dagger}_{j} e^{\dfrac{-2 \pi i j k}{2N}}
\end{equation}
}

Substituting the Fourier transforms in the Hamiltonain \ref{eq:Ham-TB}, we get
\begin{equation}
H^{\rm TB}=\sum_{k=1}^{2N}\omega_k\hat{b}_k^\dagger \hat{b}_k 
\end{equation}
where
\begin{equation}\label{NormalMode}
\omega_k=-h-J\cos{\dfrac{2 \pi k}{2N}}, \quad k=1,2,...,2N.
\end{equation}

As compared to the Hamiltonian \ref{ham}, the tight-binding model 
does not require one to perform Bogoliubov transformation to diagonalize the Hamiltonian.

\subsection{$\langle n_{j}(t)\rangle $ for the tight-binding model}

Like in the earlier case, we assume the state to be in a thermal state for the initial Hamiltonian, $H^{\rm TB}_{I} = H^{\rm TB}_{2N+2M}$. The density matrix for the initial state is 
\begin{eqnarray}
\hat{\rho_{I}}=\frac{e^{(-\beta_{I} \hat{H}_{I}^{\rm TB})}}{\text{Tr}[e^{(-\beta_{I} \hat{H}_{I}^{\rm TB})}]}
\end{eqnarray}
\begin{equation}
[\hat{\rho_{I}},\hat{H}_{I}]=0
\end{equation}
At $t=0$, we quench the system by splitting into two spin chains of sizes $2N$ and $2M$. (See the illustration in \ref{illustration}.) The Hamiltonian changes to $\hat{H}^{\rm TB}_{F}=H_{2N}^{\rm TB} \oplus H_{2M}^{\rm TB}$.

Our aim is to analytically evalaute the time evolution of average occupation number in real space for the quenched system, i. e., after the chain is broken:
\begin{equation}\label{eq43}
\langle n_{j}(t)\rangle=\langle a_{j}^{\dagger}(t) a_{j}(t)\rangle \qquad j = 1, \cdots, 2N
\end{equation}
Substituting the inverse Fourier transform from \ref{eq:FT-TB} in the above expression 
and using the time-evolution of the operators ($b_{k}(t)=e^{-i\omega_{k}t} b_{k}(0)$), we get,
\begin{eqnarray}\label{eq44}
\langle n_{j}(t)\rangle = \sum_{k=1}^{2N} \sum_{k'=1}^{2N} e^{it(\omega_{k}-\omega_{k'})} e^{\frac{i2\pi j(k-k')}{2N}} \frac{\langle b_{k}^{\dagger}(0) b_{k'}(0)\rangle}{2N},~~~ 
\end{eqnarray}
where $k, k' = 1,\cdots,2N$. From the above expression, we see that the disturbance propagate as a plane wave with the speed %$(v=\omega/k)$
\begin{equation}
v= 2N \frac{(\omega_{k}-\omega_{k'})}{2\pi (k-k')}=2N J \frac{(\cos{\frac{2 \pi k'}{2N}}- \cos{\frac{2 \pi k}{2N}})}{2\pi (k-k')}
\end{equation}
In the limit of $k\rightarrow k'$, the maximum speed of propagation is $J$ when $k = N/2$. Comparing this result with \ref{conebreak}, we see 
that for the Hamiltonian \ref{ham}, the maximum speed of propagation is achieved for $k = N/2$. In other words, the number density peaks propagate along the lattice with constant speed and forms a light cone; implying the existence of maximum speed for the information propagation.

To obtain the time evolution of the occupation number, we substitute the Fourier transform \ref{eq:FT-TB} in equation \ref{eq44}:
\begin{eqnarray}\label{eqa}
    \langle &&b_{k}^{\dagger}(0) b_{k'}(0)\rangle=\nonumber\\
     &&\frac{1}{2N}\sum_{m=1}^{2N}  \sum_{m'=1}^{2N}  (e^{\frac{-i2\pi (km-k'm')}{2N}}\langle a_{m}^{\dagger}(0) a_{m'}(0)\rangle) \, .
\end{eqnarray}
To evaluate $\langle a_{m}^{\dagger}(0) a_{m'}(0)\rangle$, we perform the Fourier 
transform \ref{eq:FT-TB} for the chain of size $2 N + 2M$, we get, 
\begin{eqnarray}\label{eqc}
\langle&& a_{m}^{\dagger}(0) a_{m'}(0)\rangle =\nonumber\\
&&\frac{1}{2N+2M}\sum_{k_{I},k'_{I}=1}^{2N+2M}  e^{\frac{i2\pi (mk_{I}-m'k'_{I})}{2N+2M}} \langle b_{k_{I}}^{\dagger}(0) b_{k'_{I}}(0)\rangle
\end{eqnarray}
where,
\begin{equation}
\label{eq:bkCond1}
\langle b_{k_{I}}^{\dagger}(0) b_{k'_{I}}(0)\rangle=0 \quad\text{for} \quad k_{I} \neq k'_{I}
\end{equation}
\begin{equation}\label{eqb}
\langle b_{k_{I}}^{\dagger}(0) b_{k'_{I}}(0)\rangle=\dfrac{1}{1+e^{\beta E_{K_{I}}}} \quad\text{for}\quad k_{I} = k'_{I} 
\end{equation}
$E_{K_{I}}$ corresponds to the energy Eigenvalues for the chain $2N+2M$ and $k_{I} = 1, \cdots, 2N+2M$. 

{The advantage of working with the number-conserving Hamiltonian is that we can control the total number of particles in the real space since it would be the same as the number of particles in the normal mode. We can study what happens when we change the number of particles in the initial state.

\begin{figure}[h]
    \includegraphics[width=0.7\columnwidth]{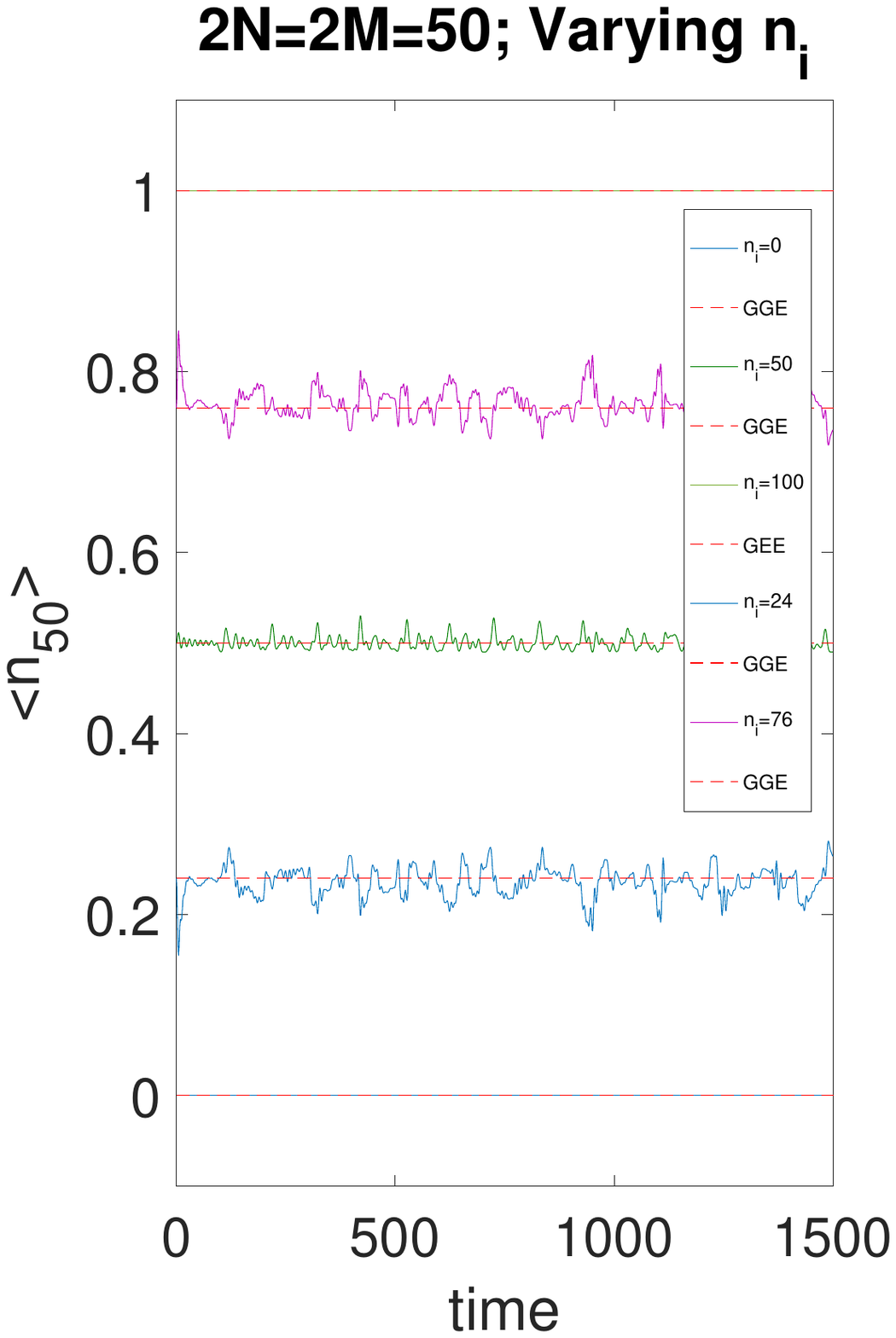}
    \caption[Varying initial number of particles]{The initial state with the particle number  $n_I$ chosen to be $0, 24, 50, 76$ and $100$ for the initial lattice chain of size $2N+2M=100$. The expectation value of the number operator at a site where quench happens is plotted against time. The GGE value is also plotted.}\label{tightbreak}
\end{figure}

In \ref{tightbreak}, we have constructed the initial state with the particle number  $n_I$ chosen to be $0, 24, 50, 76$ and $100$ for the initial lattice chain of size $2N+2M=100$. The expectation value of the number operator at a site where the quench happens is plotted against time. We infer the following: First, the minimum fluctuations are obtained for half the number of sites ($n_I = 50$). Second, when we have particles in half the number of sites, the Hilbert space dimension is $L \choose L/2$, which is the largest for the system to span, resulting in better equilibration to GGE. Third, in Fermionic systems like the ground state, the highest excited state is also unique; therefore, post-quench, there is no freedom left to (re)distribute the population, resulting in a perfect matching with GGE expectation.

For both models, we see that the expectation of the number operator equilibrates to the GGE value in the thermodynamic limit. In the case of the tight-binding model also, there is the evolution of a light cone due to the local quench indicating maximum speed, which we have analytically calculated to be $J$. The relaxation to GGE in the case of the tight-binding model also follows the power law with exponent $-1$. In other words, the GGE behavior of the Hamiltonian \ref{ham} can be inferred from the tight-binding model. We can also populate other lattice sites to change the initial state of the system. Following the discussion in \cite{Sushruth2018}, a well-behaved general state can similarly be argued to land up in GGE configuration for well-behaved Bogoliubov transformations, even under non-periodic boundary conditions.}

\begin{figure}[h]
    \centering
    \subfloat[]{
        \includegraphics[width=0.5\columnwidth]{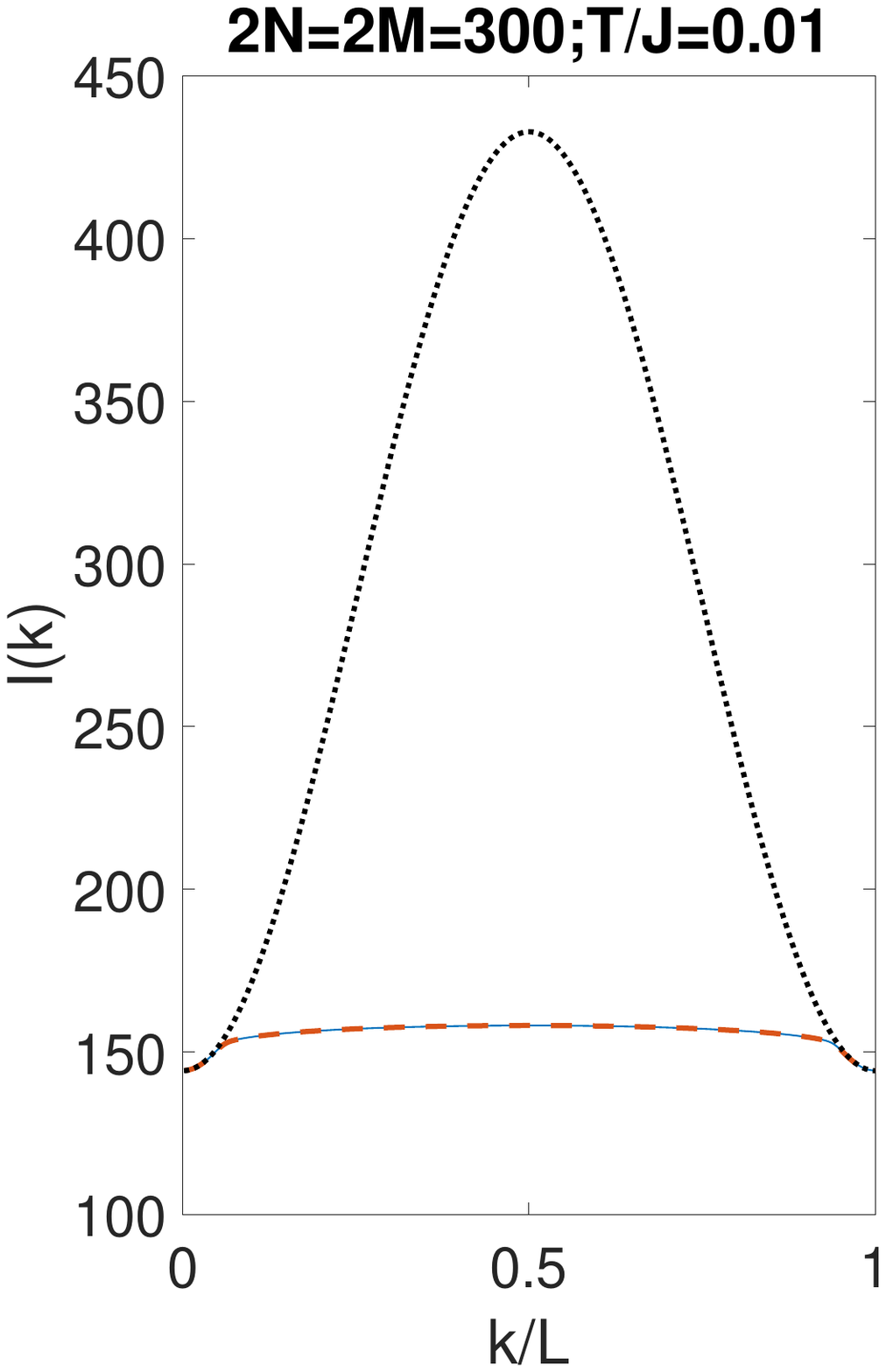}
    }
    \subfloat[]{
        \includegraphics[width=0.5\columnwidth]{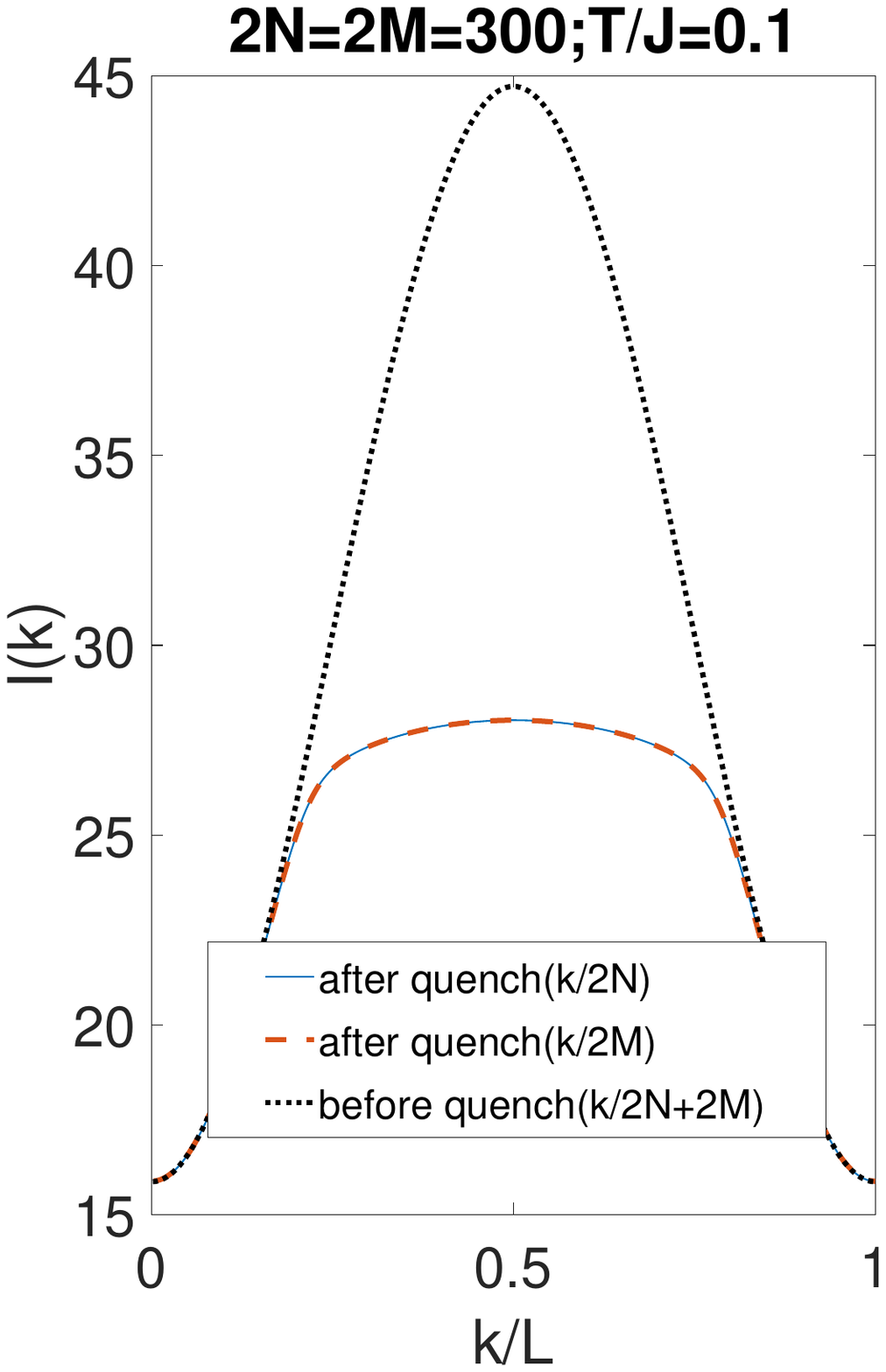}
    }
    \hspace{1mm}
    \subfloat[]{
        \includegraphics[width=0.5\columnwidth]{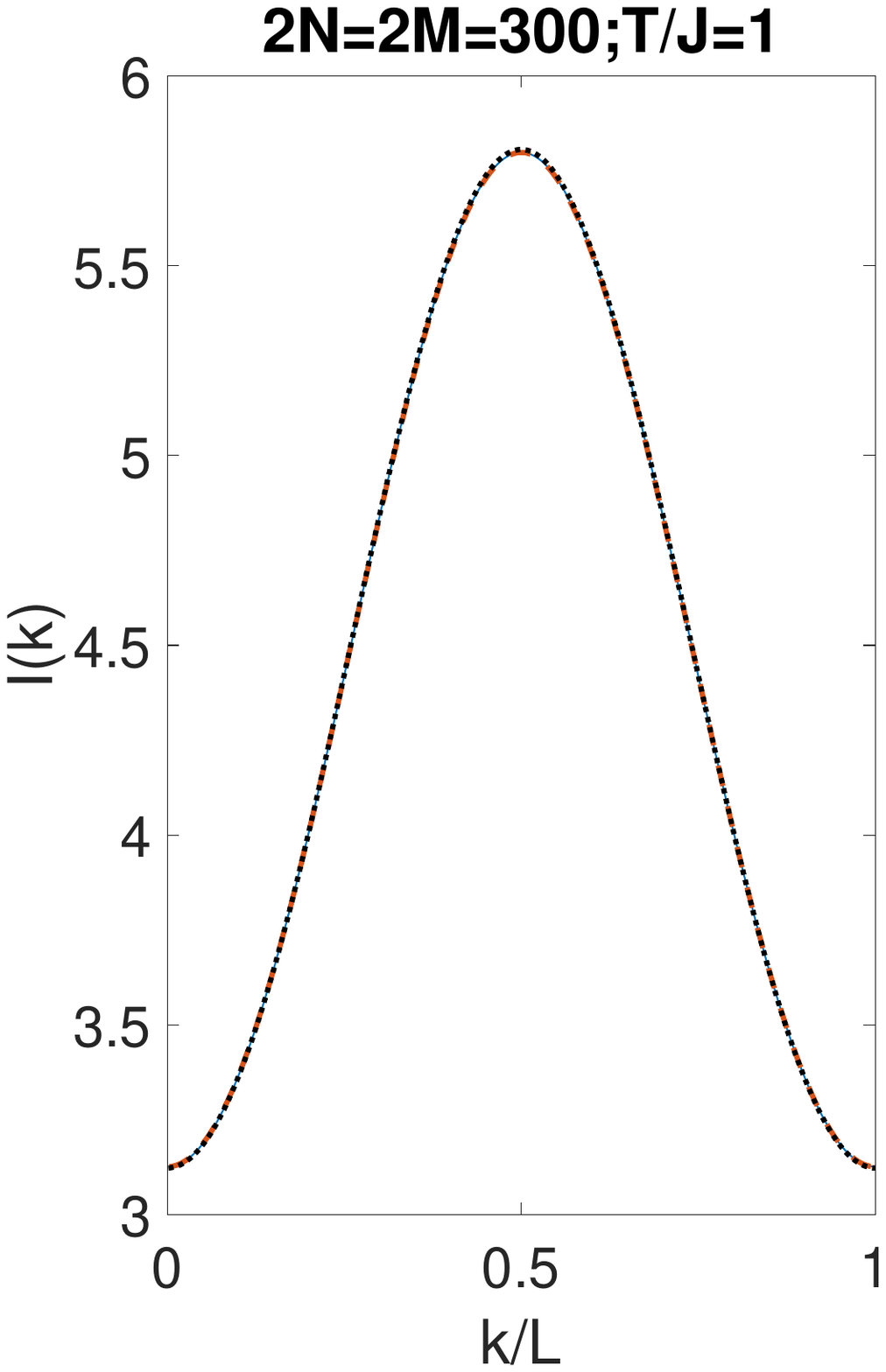}
    }
    \subfloat[]{
        \includegraphics[width=0.5\columnwidth]{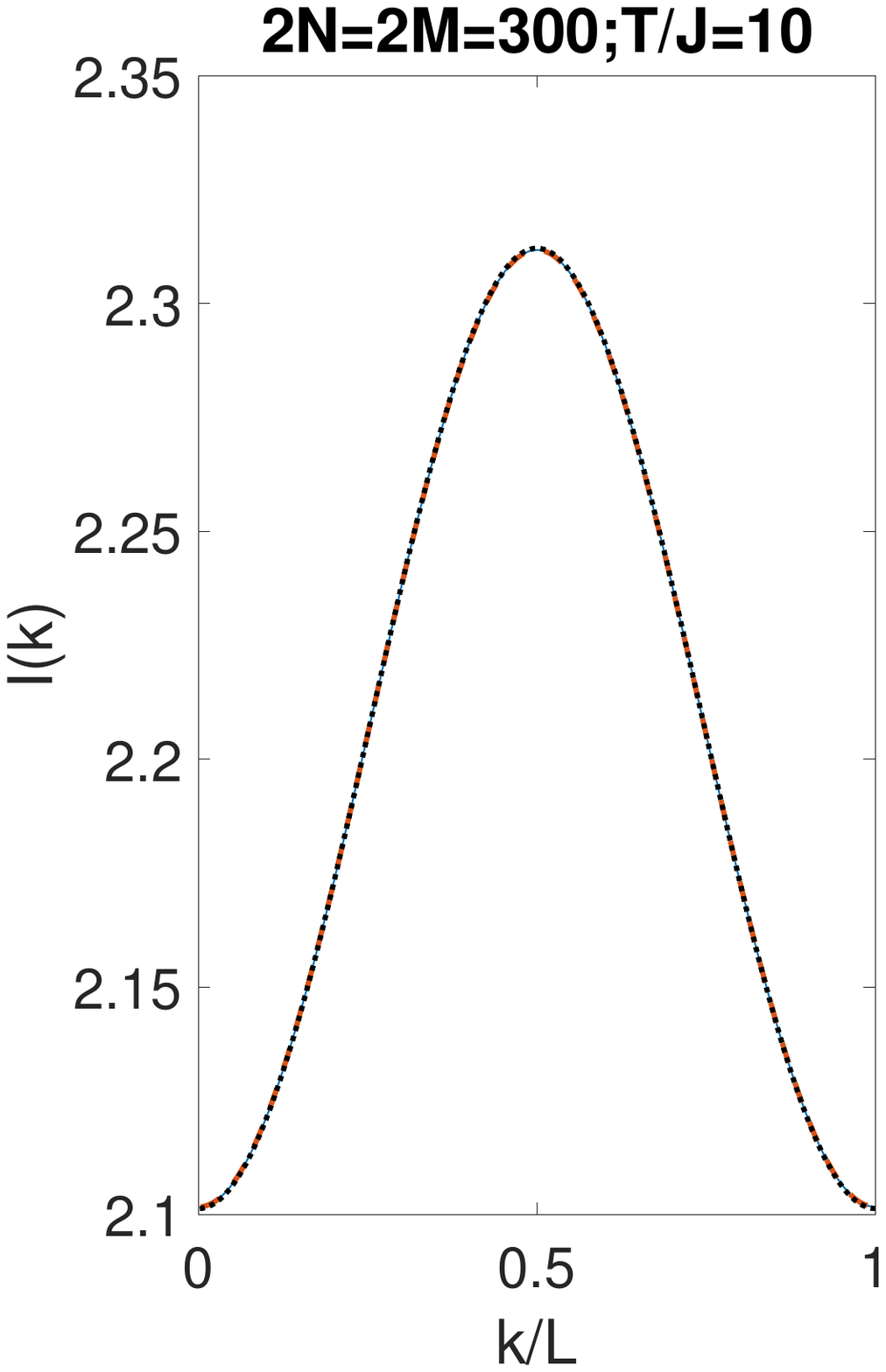}
    }
    \caption{Plot of $I(k)$ against $k/L$ for the tight-binding model for where $2N = 2M = 300$ where (a) $T/J=0.01$, (b) $T/J=0.1$, (c) $T/J=1$, (d) $T/J=10$.}\label{bitsbreaktight}\label{tightbits}
\end{figure}
\subsection{$I(k)$ for the tight-binding model}

The information content in bits per Fermion $I(k)$, for a given normal mode $k$, is given by \ref{eq:BitsperF}. Since the Hamiltonian \ref{eq:Ham-TB} is number conserving, average expectation value of the number operator for a normal mode $\langle \hat{n}_{k} \rangle$ is a conserved quantity. Once we obtain $\langle \hat{n}_{k} \rangle= \langle b_{k}^{\dagger} b_{k}\rangle$, we can calculate $I(k)$. The Fourier transform leads to:
\begin{equation}
\langle b_{k}^{\dagger} b_{k}\rangle=\frac{1}{2N}\sum_{m=1}^{2N}  \sum_{m'=1}^{2N} e^{\frac{-i2\pi k(m-m')}{2N}}\langle a_{m}^{\dagger} a_{m'}\rangle
\end{equation}
where $\langle a_{m}^{\dagger}(0) a_{m'}(0)\rangle$ is given by  \ref{eqc}.

\ref{tightbits} is the plot of $I(k)$ for the tight-binding model. Comparing these plots with the plots of $I(k)$ in \ref{bitsbreak}, we conclude the following:  For both the models, with the lower-temperature initial state, the information content per Fermion after the quench is distributed evenly across all the normal modes in contrast to before the quench distribution. Also, the information content per Fermion in each mode after the quench is smaller compared to before the quench. Like the average expectation value of the number operator,  $I(k)$ behavior of the Hamiltonian \ref{ham} can be inferred from the tight-binding model.

\subsection{OTOC for the tight-binding model}

Out-of-time-order correlator is given by \ref{eq:OTOC-def}. Substituting \ref{eq:OTOC-def2} 
in \ref{eq:OTOC-def} and simplifying the expression, we get,
\begin{eqnarray}
F_{ij}(t)&=&2\left[\text{Re}\left[\frac{i}{2}(\langle a_{j}^{\dagger}(t) a_{l}^{\dagger}(0)\rangle - \langle a_{j}^{\dagger}(t) a_{l}(0)\rangle\right.\right.\nonumber\\
&&- \left.\left.\langle a_{j}(t) a_{l}^{\dagger}(0)\rangle -\langle a_{j}(t) a_{l}(0)\rangle)\right]\right]^{2}-\frac{1}{2}
\end{eqnarray}
In the case of tight binding model, $\langle a_{j}^{\dagger}(t) a_{l}^{\dagger}(0)\rangle =0= \langle a_{j}(t) a_{l}(0)\rangle$. So we are left to calculate the expressions for $\langle a_{j}(t) a_{l}^{\dagger}(0)\rangle$ and $\langle a_{j}^{\dagger}(t) a_{l}(0)\rangle$.

Assuming $j$ and $l$ to be on the broken chain of size $2N$ after the quench, we get,
\begin{equation}
\langle a_{j}(t) a_{l}^{\dagger}(0)\rangle= \frac{1}{2N}\sum_{k,k'=1}^{2N}  e^{-i\omega_{k}t} e^{\frac{-i2\pi (jk-lk')}{2N}} \langle b_{k}(0) b_{k'}^{\dagger}(0)\rangle
\end{equation}
where,
\begin{eqnarray}
    \langle b_{k}(0) b_{k'}^{\dagger}(0)\rangle =\frac{1}{2N}\sum_{m,m'=1}^{2N}   (e^{\frac{i2\pi (km-k'm')}{2N}}\langle a_{m}(0) a_{m'}^{\dagger}(0)\rangle)\nonumber\\
\end{eqnarray}
Since the quench happens at t=0,
\begin{eqnarray}
\langle&& a_{m}(0) a_{m'}^{\dagger}(0)\rangle =\nonumber\\
&&\frac{1}{2N+2M}\sum_{k_{I},k'_{I}=1}^{2N+2M}  e^{\frac{-i2\pi (mk_{I}-m'k'_{I})}{2N+2M}} \langle b_{k_{I}}(0) b_{k'_{I}}^{\dagger}(0)\rangle
\end{eqnarray}
where,
\begin{equation}
\langle b_{k_{I}}^{\dagger}(0) b_{k'_{I}}(0)\rangle=0 \quad\text{for} \quad k_{I} \neq k'_{I}
\end{equation}
\begin{eqnarray}
\langle b_{k_{I}}(0) b_{k'_{I}}^{\dagger}(0)\rangle&=&1-\langle b_{k_{I}}^{\dagger}(0) b_{k'_{I}}(0)\rangle\\
&=&1-(\frac{1}{1+e^{\beta E_{K_{I}}}})\text{ for } k_{I} = k'_{I}
\end{eqnarray}
$E_{K_{I}}$ corresponds to the energy Eigenvalues for the chain $2N+2M$ and $k_{I}=1...2N+2M$

\begin{figure}
    \centering
    \subfloat[]{
        \includegraphics[width=0.5\columnwidth]{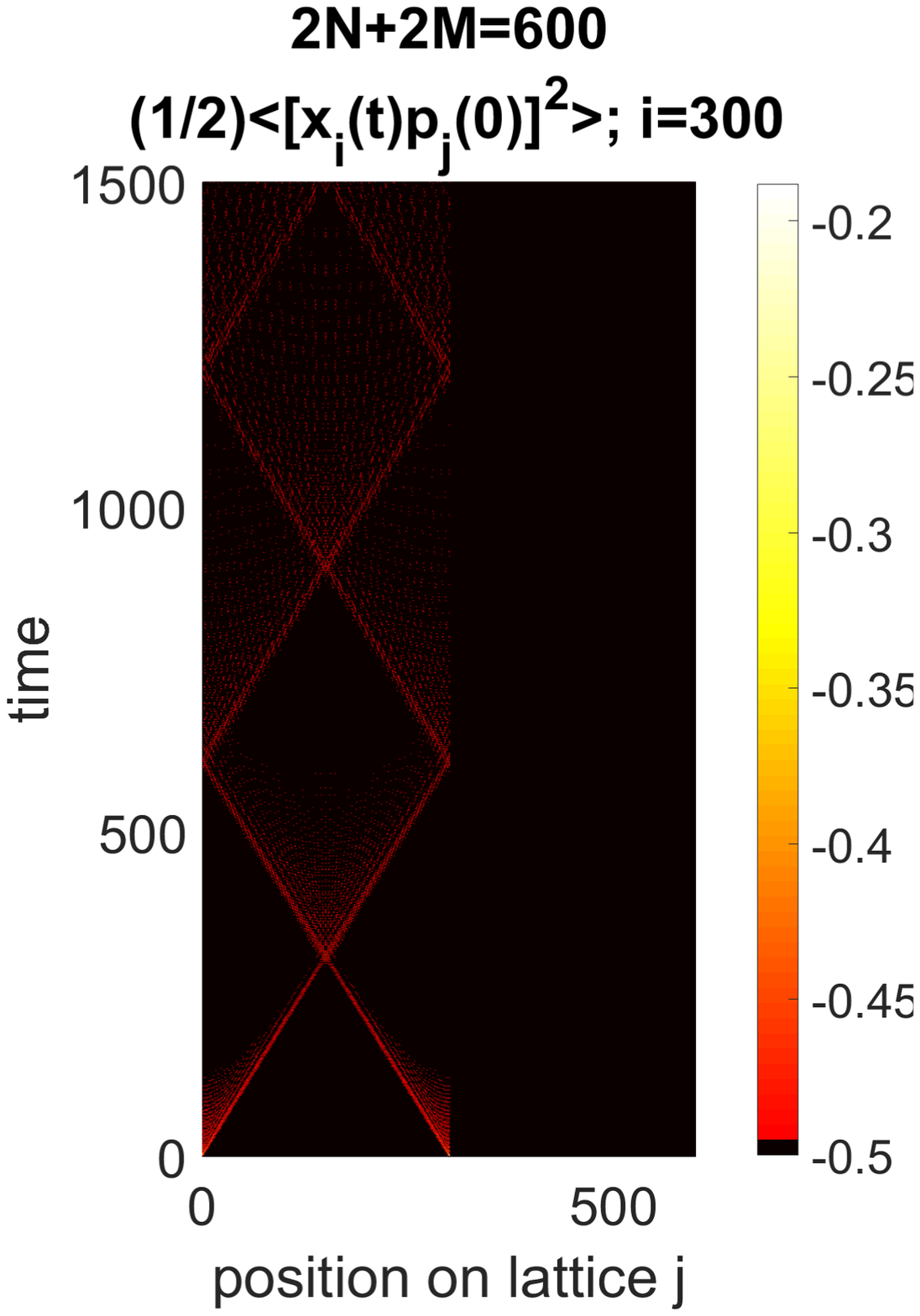}
    }
    \subfloat[]{
        \includegraphics[width=0.5\columnwidth]{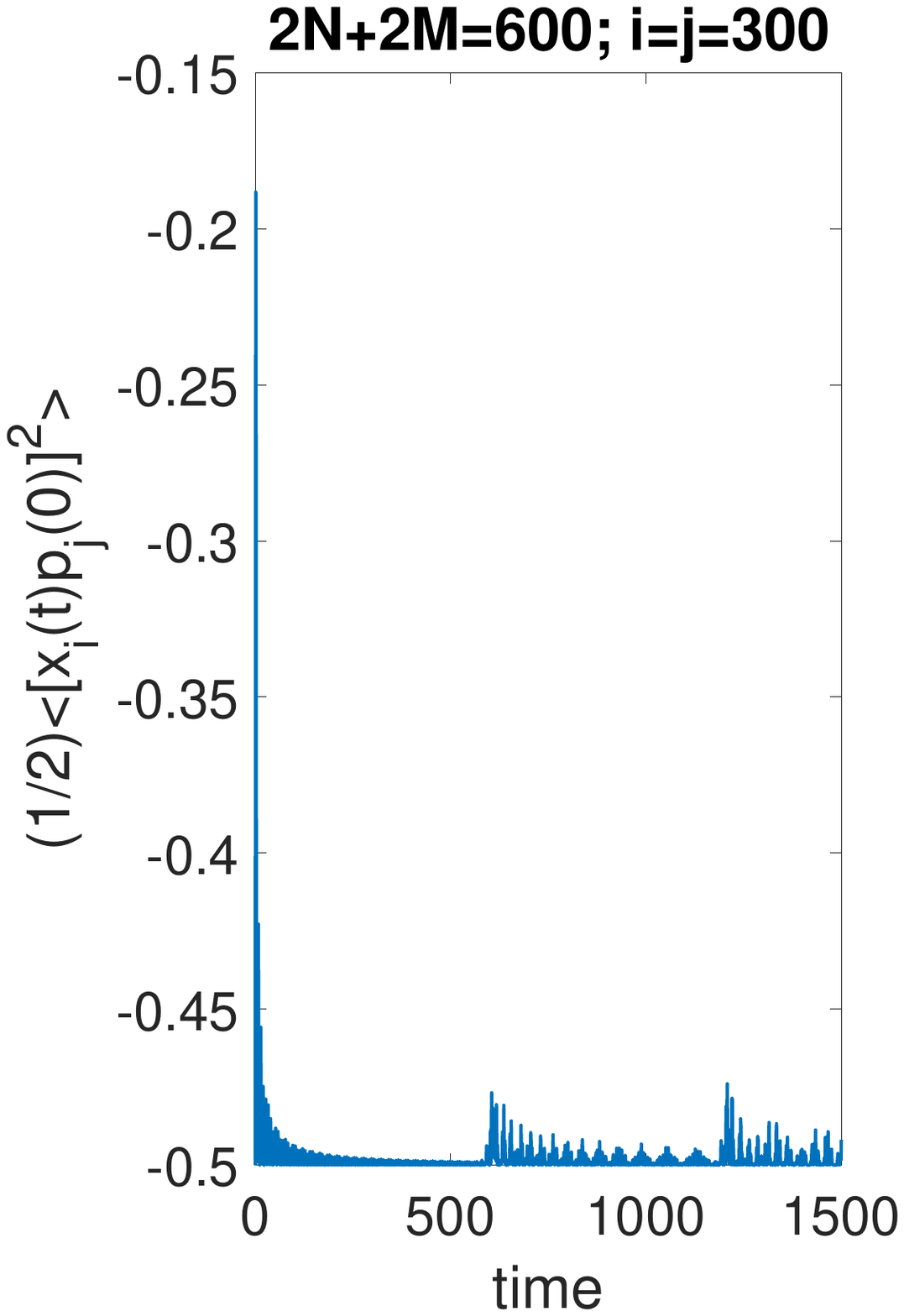}
    }
    \caption{OTOC for tight-binding model for the chain of size $2N + 2M = 600$.  (a) $F_{ij}(t)$ (color intensity) is plotted as a function of position $j$ (along the x-axis) and time (along y-axis) for $i=300$ which is on the chain of size 2N after the quench. (b) $F_{ij}(t)$ is plotted as a function of time for $i=300$ and $j=300$.}\label{otoctight}
\end{figure}
Similarly, we can obtain:
\begin{equation}
\!\!\!\!\! \langle a_{j}^{\dagger}(t) a_{l}(0)\rangle=\frac{1}{2N}\sum_{k,k'=1}^{2N}  \!\!\! e^{i\omega_{k}t} e^{\frac{i2\pi (jk-lk')}{2N}} \langle b_{k}^{\dagger}(0) b_{k'}(0)\rangle
\end{equation}
where $\langle b_{k}^{\dagger}(0) b_{k'}(0)\rangle$ is given by \ref{eqa}. In order to evaluate its thermodynamic limit we use \ref{eqa},\ref{eqc} and \ref{eqb} to obtain after some manipulations,
\begin{eqnarray}\label{OTOCThermodynamic}
\!\!\!\!\! \langle a_{j}^{\dagger}(t) a_{l}(0)\rangle=\frac{1}{2(N+M)} \sum_{K_I=1}^{2(N+M)}
\frac{e^{i \omega_{\frac{N K_I}{N+M}}t}}{e^{\beta E_{K_I}} +1}e^{\frac{\pi i (j-l)}{N+M}K_I},
\end{eqnarray}
which even in thermodynamic limit, can be shown to be bounded from above (using \ref{NormalMode})
\begin{eqnarray}\label{OTOCThermodynamic}
\!\!\!\!\! \langle a_{j}^{\dagger}(t) a_{l}(0)\rangle< 
\frac{1}{e^{-\beta (h +J)} +1}.
\end{eqnarray}
Thus, the system does not turn chaotic even in the thermodynamic limit, hinting at a possible non chaotic behaviour in the continuum limit too \cite{Bombelli_1992}.
\ref{otoctight} contains the plot of OTOC for the tight-binding model. Comparing this with 
\ref{otoc}, we see that both models have similar features. More specifically, the evolution of the light cone in just one half of the entire initial chain of size $2N+2M$ showing the correlations between the two chains vanishes with the quench. Like in the earlier case, at long-times the parameter  $F_{ij}(t)$ approaches $-0.5$.

%%%%%%%%%%%%%%%%%%%%%%%%%%%%%%%%%%%%%%%%%%
\section{Conclusions} \label{Conclusion}
Motivated by the quench like causal sturcture of oppositely moving modes in a collapsing geometry through their eventual causal disruption after the horizon formation, in this work, we consider quenching as a possible mechanism to set up thermality as well as any possible onset of chaos in the modes exterior to the horizon of the black hole. Since complete field theoretic treatment of modes in the collapsing geometry in full generality is tedious, we resort to study an analogue model which captures many conceptual similarities. This model of atomic chain undergoes an action of getting disjoint and one segment post the quench is studied.  The time evolution of local lattice occupation number and nearest neighbor hopping following quench is calculated. Thus, the system jumps between two integrable configurations. The expectation value of the observables is seen to equilibrate to the value given by generalized Gibbs ensemble with the fluctuations vanishing in the thermodynamic limit. The results show that the observables we studied for the non-interacting spinless Fermions do relax to GGE. However this is not true in general as it has been shown that certain one body correlators in such systems do not relax to GGE in the thermodynamic limit ~\cite{wright2014nonequilibrium}. We have verified for the two observables that we considered namely, lattice occupation number and nearest neighbor hopping, that the expectation value of the observable equilibrates to GGE in the thermodynamic limit.

We have obtained a light-cone like evolution with which one can accurately track the evolution of initial data. The light-cone confirms the existence of the maximum limit for the speed of propagation of the information of the quench. This result is consistent with the Lieb-Robinson bound in quantum systems. We have also seen that the relaxation to GGE goes as a power law with exponent approaching  $-1$ indicating ballistic dynamics. We also calculated the connected correlation between two sites each in the disconnected chains after the quench which gradually vanishes.

Further, we calculated the information content in bits per Fermion $I(k)$ per normal mode $k/L$ before and after the quench. We observed exciting trends in the distribution of  $I(k)$ before and after the quench. We see from the plot of $I(k)$ against $k/L$ that for initial low-temperature thermal states, the information content per Fermion after quench is smaller and spreads evenly for all normal modes compared to before the quench, in the spirit of {\it thermalization}. However, the total entropy of the system increases after the quench, consistent with the second law of thermodynamics.

Another measure of  predictability of evolution is OTOC. If OTOC grows in time, the evolution becomes chaotic and final state cannot be ascribed with accuracy. The dynamics of the systems demonstrates that post-quench the OTOC parameter decays once the quench hits the system and the gradually settles to a value close to $-0.5$.

We also study a closely resembling model with analytically more tractable equations, which captures the essential features of the first model. In this model we also analytically demonstrated the march of the system towards a GGE configuration, strongly suggesting that internal interactions within the system do not remain of much importance once the quench is sufficiently strong.

Thus the analogue model remains integrable even after undergoing causal disruption and also lands up in GGE configuration, notwithstanding with the conflict between apparent thermality and onset of chaos. In the black hole setting the apparent conflict arises due to maximality of entanglement between the interior, near horizon and asymptotic modes. However, the system remains integrable throughout without any chaotic behaviour setting up a GGE like distribution can well suggest the apparent thermality, in the spirit of \cite{Sonner:2017hxc}. Since these results appear largely irrespective of the details of interaction a full field theoretic calculation can be expected to inherit these features, probably demonstrable in certain analogue system undergoing similar kind of quench action,  will be pursued elsewhere. 

%%%%%%%%%%%%%%%%%%%%%%%%%%%%%%%%%%%%%%%%%%

%%%%%%%%%%%%%%%%%%%%%%%%%%%%%%%%%%%%%%%%%%
\vspace{6pt} 

\section{Acknowledgement}
%{\it Acknowledgement : }
The authors thank Soumya Bera and Swastik Bhattacharya for discussions.
The authors thank Nicholas Sedlmayr for email communications and comment 
about the previous draft of the manuscript. SB  was supported by Inspire fellowship of DST, Government
of India. Research of KL is supported by DST-INSPIRE
Faculty Fellowship of the Government of India. Research of SS is funded by DST-Max Planck Partner Group on Cosmology and Gravity and SERB CORE-Research Grant.

%\bibliographystyle{apsrev4-1}
%\bibliography{ref-new}

\end{document}